\documentclass[12pt]{article}
\usepackage[latin1]{inputenc}

\usepackage{amsmath}
\usepackage{color}
\usepackage{amsfonts}
\usepackage{amssymb}
\usepackage{graphicx}
\usepackage{geometry}
\usepackage{amssymb,epsfig}
\usepackage{hyperref}
\usepackage{comment}
\usepackage{tabularx}
\usepackage{bm}
\usepackage{euscript}
\usepackage{graphicx}
\usepackage{color}
\usepackage{amsfonts}
\usepackage{exscale}
\usepackage{amsbsy}
\usepackage{textcomp}
\usepackage{comment}
\usepackage{hyperref}
\usepackage{slashed}
\usepackage{authblk}
\usepackage{mathtools}
\usepackage{tabularx}
\usepackage{bm}
\usepackage{euscript}
\usepackage{graphicx}
\usepackage{color}
\usepackage{amsfonts}
\usepackage{exscale}
\usepackage{amsbsy}
\usepackage{textcomp}
\usepackage{comment}
\usepackage{hyperref}
\usepackage[numbers,sort&compress]{natbib}

\usepackage{subcaption}

\def\simge{
  \mathrel{\rlap{\raise 0.511ex
      \hbox{$>$}}{\lower 0.511ex \hbox{$\sim$}}}}

\def\simle{
  \mathrel{\rlap{\raise 0.511ex
      \hbox{$<$}}{\lower 0.511ex \hbox{$\sim$}}}}

\makeatletter
\renewcommand\section{\@startsection {section}{1}{\z@}%
  {-3.5ex \@plus -1ex \@minus -.2ex}
  {2.3ex \@plus.2ex}%
  {\normalfont\large\bfseries}}
\renewcommand\subsection{\@startsection{subsection}{2}{\z@}%
  {-3.25ex\@plus -1ex \@minus -.2ex}%
  {1.5ex \@plus .2ex}%
  {\normalfont\bfseries}}
\renewcommand\subsubsection{\@startsection{subsubsection}{3}{\z@}%
  {-3.25ex\@plus -1ex \@minus -.2ex}%
  {1.5ex \@plus .2ex}%
  {\normalfont\itshape}}
\makeatother

\def\pplogo{\vbox{\kern-\headheight\kern -29pt
    \halign{##&##\hfil\cr&{\ppnumber}\cr\rule{0pt}{2.5ex}&\ppdate\cr}}}
\makeatletter
\def\ps@firstpage{\ps@empty \def\@oddhead{\hss\pplogo}%
  \let\@evenhead\@oddhead 
}
\def\maketitle{\par
  \begingroup
  \def\thefootnote{\fnsymbol{footnote}}
  \def\@makefnmark{\hbox{$^{\@thefnmark}$\hss}}
  \if@twocolumn
  \twocolumn[\@maketitle]
  \else \newpage
  \global\@topnum\z@ \@maketitle \fi\thispagestyle{firstpage}\@thanks
  \endgroup
  \setcounter{footnote}{0}
  \let\maketitle\relax
  \let\@maketitle\relax
  \gdef\@thanks{}\gdef\@author{}\gdef\@title{}\let\thanks\relax}
\makeatother

\numberwithin{equation}{section}

\newcommand{\expval}[1]{\ensuremath{ \left\langle #1 \right\rangle }}

\newcommand{\deriv}[2]{\ensuremath{ \frac{d #1 }{d #2 } } }

\newcommand{\di}{\ensuremath{ \mathrm{d} }}

\newcommand{\hamil}{\ensuremath{\EuScript{H}}}

\newcommand{\expon}[1]{\ensuremath{  \operatorname{e}^{#1} }  }

\begin{document}

\setcounter{page}0
\def\ppnumber{\vbox{\baselineskip14pt
  }}

\def\ppdate{
} \date{\today}

\title{\bf Partial equilibration of anti-Pfaffian edge modes at $\nu=5/2$
  \vskip 0.5cm}
\author{Hamed Asasi}
\author{Michael Mulligan}
\affil{\small \it Department of Physics and Astronomy, University of California,
  Riverside, CA 92511, USA}

\bigskip

\maketitle

\begin{abstract}
  The thermal Hall conductance $K$ of the fractional quantum Hall
  state at filling fraction $\nu=5/2$ has recently been measured to be
  $K=2.5 \pi^2k_B^2T/3h$ [M. Banerjee et al., Nature {\bf 559}, 205 (2018)].
  The half-integer value of this result (in units of $\pi^2k_B^2T/3h$) provides strong evidence for the presence of a Majorana
  edge mode and a corresponding quantum Hall state hosting
  quasiparticles with non-Abelian statistics.
  Whether this measurement
  points to the realization of the PH-Pfaffian or the
  anti-Pfaffian state has been the subject of debate.
  Here we consider the implications of this measurement for anti-Pfaffian edge-state transport.
  We show that in the limit of a strong Coulomb interaction and an approximate spin degeneracy in the lowest Landau level, the anti-Pfaffian state admits low-temperature edge phases that are consistent with the Hall conductance measurements.
  These edge phases can exhibit fully-equilibrated electrical transport coexisting with partially-equilibrated heat transport over a range of temperatures.
  Through a study of the kinetic equations describing low-temperature electrical and heat transport of these edge states, we
  determine regimes of parameter space, controlling the interactions
  between the different edge modes, that agree with experiment.
\end{abstract}
\bigskip

\newpage

\tableofcontents

\newpage

\vskip 1cm

\section{Introduction}

Since the discovery of the quantum Hall plateau at filling fraction
$\nu=5/2$ \cite{willet-1987}, the nature of the ground state of this
system has been the subject of debate.  Numerical studies point toward
either the Moore-Read Pfaffian state \cite{moore-1991,read-2000} or
its particle-hole conjugate, the anti-Pfaffian state \cite{levin-2007,
  lee-2007}, as the true ground state of the system
\cite{haldane-num-1988,macdonald-num-1989,pakrouski-num-2015,rezayi-num-2017,
  rezayi-2011}.  Both of these states host quasiparticles with
non-Abelian statistics.  On the other hand, quantum point contact
tunneling experiments
\cite{lin-tun-2012,radu-2008,5halv-exp-review-2014,yang-feldman-review,
  PhysRevB.90.161306} support either the anti-Pfaffian state, the
$SU(2)_2$ state, or the Abelian $331$ or $113$ states.  Observation of
upstream neutral modes \cite{bid-2010,gross-2012} only hints at the
realization of a non-Abelian state.

As first pointed out by Kane and Fisher \cite{KF-Qthermal}, the
thermal Hall conductance $K$ provides a sensitive probe of the
topological order of a fractional quantum Hall (FQH) state.  $K/T$
equals the difference in the number of right and left moving chiral
edge modes (in units of $\kappa_0=\frac{\pi^2k_B^2}{3h}$ at
temperature $T$) for an Abelian quantum Hall state; more generally,
$K$ is determined by the chiral central charge $c_- = c_R - c_L$ of
the edge states \cite{2002NuPhB.636..568C} ($c_{R/L}$ is the sum of
the right/left, i.e., holomorphic/anti-holomorphic, central charges of
the edge modes).  Consequently, the remarkable measurement of Banerjee
et al.~\cite{banerj-half} that finds $K=2.5\kappa_0T$ at $\nu=5/2$
provides strong evidence for a non-Abelian quantum Hall state.  Taken
at face value, this result suggests the recently proposed topological
order, the particle-hole symmetric Pfaffian state (PH-Pfaffian)
\cite{son-2015} which has chiral central charge $c_- = 5/2$, is
realized.  One explanation \cite{zucker-feldman-2016} for the apparent
contradiction between this experimental result and prior numerical
work invokes disorder and Landau Level mixing, which are inevitably
present in any real sample, but difficult to include in numerics.
Another scenario is that long-range disorder results in puddles of
Pfaffian and anti-Pfaffian states, which (intuitively) contribute
$(c_-^{\rm Pfaffian} + c_-^{\rm anti-Pfaffian})/2 = (7/2 + 3/2)/2$ to
the thermal Hall conductance. The resulting state can exhibit the
thermal Hall conductance of $K=2.5 \kappa_0T$ in some parameter
regimes \cite{wang-2018,mross-2018, PhysRevB.97.165124}.  However, the
conditions for this observation were found to be rather restrictive.

Simon \cite{simon-interp} has proposed an alternative interpretation:
The experimental measurement may not directly reflect the bulk
topological order; instead $K=2.5\kappa_0T$ may be due to suppressed
thermal equilibration relative to charge equilibration of
anti-Pfaffian edge modes.  This partial equilibration is believed to
occur at $\nu=2/3$ and potentially $\nu=8/3$ \cite{Banerjee:2017aa,
  banerj-half}.  The distinction between various candidate $\nu=5/2$
states, based on the thermal Hall conductance, is clear only if the
different edge channels of the quantum Hall sample are well
equilibrated with each other.  If instead there's no equilibration
between edge modes, the thermal Hall conductance is proportional to
the total central charge $c = c_R + c_L$ of the edge state.  If the
edge modes only partially equilibrate, the thermal Hall conductance
can in principle take any value between the fully-equilibrated
conductance and the non-equilibrated one.  (These statements are true
only if ideal contacts are assumed \cite{KF-contacts,chamon-fradkin}.)
The Pfaffian state has three bosonic modes and one Majorana mode (with
central charge $c^{\rm Majorana} = 1/2$), all moving ``downstream"
along the edge.  Assuming the contacts are ideal, the resulting
thermal Hall conductance equals $K=\frac{7}{2}\kappa_0T$, which is not
consistent with the measurement $K=2.5\kappa_0T$.  The anti-Pfaffian
state has three ``downstream" bosonic modes, one ``upstream" bosonic
mode, and one ``upstream" Majorana mode.  Therefore, depending on the
degree of equilibration, the thermal Hall conductance of the
anti-Pfaffian state can take any value between
$K={3 \over 2} \kappa_0T$ and $K={9 \over 2} \kappa_0T$.  Since the
electrical Hall conductance $G = {5 \over 2} \sigma_0$
\cite{banerj-half} ($\sigma_0 = e^2/h$), realization of this idea
requires partial thermal equilibration simultaneous with full charge
equilibration, at least of the edge modes belonging to the first
Landau level.

There have been a variety of different scenarios proposed for partial
equilibration of the anti-Pfaffian edge state. Simon
\cite{simon-interp} originally suggested that the low velocity of the
Majorana mode combined with long-range disorder might hinder the
equilibration of the Majorana mode with the rest of the edge
modes. However, it has been argued that the parameter regime required
by this interpretation is not realized experimentally
\cite{feldman-comment,banerj-half}.  Partial equilibration in the
anti-Pfaffian state can also occur if the modes in the lowest Landau
level do not equilibrate with modes in the first Landau level.  One
possible realization was described by Ma and Feldman
\cite{ma-feldman-2019}.  Another mechanism whereby equilibration of
the Majorana mode is suppressed was proposed by Simon and Rosenow
\cite{simon-rosenow-2019}. There, equilibration between edge modes was
assumed to be dominated by scattering via intermediate tunneling to
Majorana zero modes localized in the bulk, rather than charge
tunneling along the edge, considered in
\cite{simon-interp,feldman-comment,ma-feldman-2019}.

In this note, we continue the study of the role of equilibration in
anti-Pfaffian edge-state transport.  In contrast to
\cite{simon-interp,simon-rosenow-2019}, we assume that electron
tunneling, induced by short-range disorder, serves to equilibrate the
edge modes.  Tunneling between spin-up and spin-down edge modes of the
lowest Landau level plays a prominent role in our scenario.  These
tunnelings were not considered in the previous analysis
\cite{ma-feldman-2019} of transport in the anti-Pfaffian state, as it
was argued that weak spin-orbit coupling suppresses such tunnelings.
The effective theories we consider are driven by such spin-flip
interactions.  The resulting low-energy edge states have an
approximate spin symmetry in the lowest Landau level that we show can
serve to suppress thermal equilibration while simultaneously allowing
complete charge equilibration over a range of experimentally-relevant
temperatures, in the presence of a strong Coulomb interaction.

We analyze charge and heat equilibration when the edge modes are
biased at different temperatures and voltages. While we consider the
effects of bias in temperature and voltage only to linear order,
including higher order contributions can have interesting
consequences. It has been argued that the interplay between the
electrical and the thermal transport can generate distinct shot noise
profiles along the Hall bar edge
\cite{PhysRevLett.123.137701,PhysRevB.99.161302,PhysRevB.101.075308,park2020noise}.
These noise profiles fall into three universality classes depending on
the chirality structure of the edge modes.  Specifically, it has been
suggested that the universality class for the noise profile of the
anti-Pfaffian state is different from that of the Pfaffian and the
PH-Pfaffian states, and so the measurement of shot noise along the
edge of the quantum Hall system at filling fraction $\nu=5/2$ is
another tool that can be used to distinguish between the different
candidates.

We start in Section \ref{kineq-section} with a review of the general
framework that we use in order to examine the transport of charge and
heat in the anti-Pfaffian state. Starting from the effective field
theory of a general quantum Hall edge state, we derive the equations
that describe charge and heat transport in the ohmic regime.  In
Section \ref{nu2-transport-section} we discuss the simple example of
transport along the $\nu=2$ quantum Hall edge; this example
illustrates the possible importance of an approximate spin symmetry in
the lowest Landau level and helps to motivate the anti-Pfaffian edge
phases considered in the remainder of the paper.  In Section
\ref{conductance-section} we describe how we model the contacts and
calculate the electrical and thermal conductance.  In Section
\ref{antipf-theory-section} we discuss the edge theory of the
anti-Pfaffian state. We identify low-temperature fixed points of this
theory that we argue to be relevant to experiment and discuss two of
the fixed points that are driven by spin-flip tunneling.  In Section
\ref{antipf-transport-section} we apply the framework presented in
Section \ref{kineq-section} to these low-energy fixed points.  We
calculate the electrical and thermal conductances for each of these
theories and discuss the regime of parameters consistent with the
measured electrical $G=2.5 \sigma_0$ and thermal $K=2.5\kappa_0T$
conductances.  We discuss the degree to which such parameter regimes
are realistic.  Finally, in Section \ref{sec-QPC-tunn} we examine
quantum point contact tunneling in the anti-Pfaffian state in the
vicinity of these low-energy edge states.

Throughout this paper, we use the notation $\sigma_0=\frac{e^{2}}{h}$ and
$\kappa_0=\frac{\pi^2k_B^2}{3h}$.
However, for our calculations we use units where $e=\hbar=k_B=1$ so that
$\sigma_0=\frac{1}{2\pi}$ and $\kappa_0=\frac{\pi}{6}$.

\section{Edge-state transport for an Abelian quantum Hall state}
\label{gen-transport-section}

\subsection{Hydrodynamic kinetic equations }\label{kineq-section}

In this section we derive the kinetic equations that describe the
low-temperature dc transport of charge and heat along the edge of an
Abelian quantum Hall state, closely following
\cite{KF-imp,incoherent-23,wen-1991-edge,KF-contacts,KF-thermLL}.  We
highlight the dependence of these equations on the low-temperature
state of the edge modes.  These equations are readily generalized to
the anti-Pfaffian edge theory, which includes a chiral Majorana
fermion.

Consider a layered quantum Hall state with filling fraction $\nu_i$
for each layer $i = 1, \ldots, N$.  The action for the chiral boson
edge modes $\phi_i$ is $S_{\text{edge}} = S_0 + S_{\text{tunneling}}$
where
\begin{subequations}\label{bos-gen-action}
  \begin{align}
    S_0 &= -\frac{1}{4\pi} \int_{t,x} \left[ \sum_i \frac{1}{\nu_{i}} \partial_x \phi_i(\eta_i\partial_t\phi_i +
          v_i\partial_x\phi_i) + \sum_{i \neq j} v_{ij}
          \partial_x\phi_i\partial_x\phi_j\right], \\
    S_{\text{tunneling}}&=- \int_{t,x} \sum_{p\in P}\left[\xi_p(x) \expon{i\sum_j^{N} m^{(p)}_j
                          \phi_j}+ \text{h.c.} \right].
  \end{align}
\end{subequations}
Here, $v_{ij}$ parameterizes the short-ranged Coulomb interaction
coupling the edge-mode charge densities
${1 \over 2\pi} \partial_x \phi_i$; the velocities $v_i$ are
non-negative; $\eta_i = \pm 1$ denotes the chirality of the edge mode
($\eta_i = +1$ is a right-moving or ``downstream" mode, while
$\eta_i = - 1$ is a left-moving or ``upstream" mode);
$\int_{t,x} = \int dt dx$; $P$ is the set of charge-conserving
processes that tunnel $\nu_j m_j^{(p)}$ electrons/bosons between the
edge channels; and $\xi_p$ is a Gaussian random field with statistical
average
$\overline{\xi_p(x) \xi^{*}_{p'}(x')}=\delta_{pp'}W_p\delta(x-x')$.  To
study the transport properties of $S_{\text{edge}}$ it's convenient to
diagonalize $S_0$ using the transformation
$\phi_i=\Lambda_{i\alpha} \tilde{\phi}_{\alpha}$:
\begin{align}\label{gen-action-diag}
  S_0 &=-\frac{1}{4\pi} \int_{t,x}\left[   \sum_{\alpha}
        \partial_x \tilde{\phi}_{\alpha}( \tilde{\eta}_{\alpha}\partial_t
        \tilde{\phi}_{\alpha} - \tilde{v}_\alpha\partial_x
        \tilde{\phi}_{\alpha}) \right].
\end{align}
This transformation is of the form
$\Lambda_{i\alpha}=\sqrt{\nu_i}\tilde{\Lambda}_{i\alpha}$ where
$\tilde{\Lambda}_{i\alpha}$ satisfies
$\tilde{\Lambda}^T\eta \tilde{\Lambda}=\eta$
($\eta_{ij}=\delta_{ij}\eta_i$).  Note that $\tilde v_\alpha \geq 0$,
$\eta_\alpha = \pm 1$, and $\sum_i \eta_i = \sum_\alpha
\eta_\alpha$. Throughout this paper, we will use Latin indices $i,j$ for
the fractional modes and Greek indices $\alpha,\beta,\gamma$ for the
bosonic modes that diagonalize the action.

The leading order renormalization group equation for the variance $W_p$ is
\begin{align}
  \deriv{W_p}{l} = (3-2\Delta_p)W_p
\end{align}
with $\Delta_p$ the scaling dimension of the tunneling operator
${\cal O}_p = \expon{i \sum_j m^{(p)}_j \phi_j} = \expon{i
  \sum_{j,\alpha} m^{(p)}_j \Lambda_{j \alpha} \tilde \phi_\alpha}$.
When all tunneling operators appearing in $S_{\text{tunneling}}$ are
irrelevant, $\Delta_p> \frac{3}{2}$, the fixed point action is $S_0$.
At zero temperature, the currents
$\tilde{I}_{\alpha}=-\frac{1}{2\pi}\partial_t\tilde{\phi}_{\alpha}$
and
$\tilde{J}_\alpha(x)=\frac{\eta_{\alpha}}{4\pi}(\partial_t\tilde{\phi}_{\alpha})^2$
(no sum over $\alpha$) associated to each mode are separately
conserved.  In particular, the static components of these currents
satisfy for each $\alpha$,
\begin{align}
  \label{chargecurrent-conserv}
  \partial_x \tilde{I}_{\alpha}(x,\omega=0) & = 0, \\
  \label{heatcurrent-conserv}
  \partial_x \tilde{J}_{\alpha}(x,\omega=0) & = 0.
\end{align}
At low temperatures, the irrelevant terms in $S_{\text{tunneling}}$
perturbatively correct these continuity equations to allow
equilibration between the different edge channels; only the total
charge and heat currents (related to $\tilde I_\alpha$ and
$\tilde J_\alpha$ via the $\Lambda_{i \alpha}$ transformation---see
below) remain conserved.

First consider the correction to Eq.~\eqref{chargecurrent-conserv} in
the presence of the chemical potential bias
$H_{\mu}=-\frac{1}{2\pi}\int \di x \ \sum_{\alpha} \tilde \mu_{\alpha}
\partial_x \tilde{\phi}_{\alpha}$ and uniform temperature $T$.
In the ohmic regime $\tilde \mu_{\alpha}\ll T$ we have (using Eq.~\eqref{elect-kin-final-app}):
\begin{align}
  \partial_x\expval{ \tilde{I}_{\alpha}(x) }
  &=-\sigma_0 \eta_{\alpha}  \sum_{p\in P}\left[ g_p (\sum_{i} m_i^{(p)}\Lambda_{i\alpha})
    \left(\sum_{j} \eta_{\beta} m_j^{(p)} \Lambda_{j\beta}
    \tilde \mu_{\beta}(x) \right) \right], \qquad
    g_p \propto W_p T^{2\Delta_p-2}.
\end{align}
We assume local equilibrium so that
$\expval{\tilde{I}_{\alpha}(x) }=\eta_{\alpha} \sigma_0 \tilde
\mu_{\alpha}(x)$.
These equations are more transparent physically in
the original basis where $I_i=\Lambda_{i\alpha}\tilde{I}_{\alpha}$ and
$ \tilde \mu_\alpha  =\mu_i  \Lambda_{i \alpha} $:
\begin{align}
  \partial_x \expval{I_i(x) }
  &= -\eta_{i}\nu_i \sum_{p\in P} g_p  m_i^{(p)}
    \left(\sum_{\j}  m_j^{(p)}
    \expval{I_{j}(x) } \right) 
\end{align}
or in matrix form (dropping expectation value signs),
\begin{subequations}\label{elec-eq-matrix}
  \begin{align}
    \partial_x \bm{I} &= G^e \bm{I}, \\
    G^e_{ij} &= -\eta_i\nu_i \sum_{p\in P} g_p m_{i}^{(p)} m_{j}^{(p)}.
  \end{align}
\end{subequations}
These equations constitute the kinetic equations for dc charge transport about the $S_0$ fixed point.
Equilibration of charge is parameterized by the charge matrices $G_{ij}^e$.

If a tunneling operator is relevant, $\Delta_p \leq \frac{3}{2}$, we
have to determine the resulting low-energy fixed point in order to
derive the appropriate transport equations.  There exists a similar
set of conserved charge and heat currents and we treat the leading
irrelevant terms (with respect to the corresponding disordered fixed
point) perturbatively.  The kinetic equations for charge transport are
similar to Eq.~\eqref{elec-eq-matrix}
. The difference lies in the set
of processes that drive inter-mode equilibration and, consequently,
the precise expressions for $G^e_{ij}$.  A simple example of a
disordered fixed point---relevant to our later analysis of
anti-Pfaffian edge transport---occurs along the edge of the integer
quantum Hall state at filling fraction $\nu=2$.  Charge transport
about this fixed point is discussed in Section
\ref{nu2-transport-section}; details of this analysis are given in
Appendix \ref{sec-rand-dd}.

We treat the effects of irrelevant interactions on the
$\tilde J_\alpha$ continuity equations \eqref{heatcurrent-conserv}
similarly with the details relegated to Appendix
\ref{therm-cond-apndx}. These interactions induce heat exchange
between the edge modes when these modes are at different local temperatures
$T_{\alpha}(x)$. To linear order in
$(T_{\alpha}-T_{\beta})/T_{\alpha}$, we find
\begin{align}
  \partial_x\expval{ \tilde{J}_{\alpha}(x)}
  &=\kappa_0 \sum_{\beta \neq \alpha}
    g^{Q}_{\alpha \beta}\frac{T_{\beta}^{2}(x)-T_{\alpha}^{2}(x)}{2} ,\qquad
    g^Q_{\alpha \beta}=\sum_{p\in P} g_p \frac{ 12d_{\alpha}^{(p)} d_{\beta}^{(p)}}{1+2\Delta_p}.
\end{align}
The constants
$d_\alpha^{(p)} = {1 \over 2} (\sum_i m_i^{(p)} \Lambda_{i
  \alpha})^2$.  Similar to charge transport, the set of processes $P$
and conductivity coefficients $g^Q_{\alpha \beta}$ depend on the
low-temperature fixed point of the theory.  Assuming local equilibrium
we express the local currents $\tilde{J}_{\alpha}(x)$ in terms of
local temperatures $T_\alpha(x)$ as ($\tilde{c}_{\alpha}$ is the
central charge of mode $\alpha$)
\begin{align}
  \expval{ \tilde{J}_{\alpha}(x) }&=\frac{1}{2}\kappa_0\tilde{\eta}_{\alpha}\tilde{c}_{\alpha}T^2_{\alpha}(x).
\end{align}
The resulting kinetic equations take the form (again dropping the expectation value signs):
\begin{subequations}\label{therm-eq-matrix}
  \begin{align}
    \partial_x\tilde{\bm{J}} &= G^{Q} \tilde{ \bm{J} }, \\
    G^{Q}_{\alpha \beta}
                             &= \frac{\eta_{\beta}}{c_{\beta}}(g^Q_{\alpha
                               \beta} - \delta_{\alpha
                               \beta}\sum_{\gamma}g^Q_{\alpha \gamma}).
  \end{align}
\end{subequations}
Similar to the charge kinetic equations, equilibration of heat is
controlled by $G_{\alpha \beta}^Q$.  (We precisely relate the kinetic
equations for the $\tilde J_\alpha$ currents to heat transport later.)
Note that $G^e_{\alpha \beta}$ and $G^Q_{\alpha \beta}$ need not
coincide.

\subsection{Edge-state transport at $\nu=2$}\label{nu2-transport-section}

We now illustrate some aspects of the previous discussion for the case of $\nu=2$ edge-state transport.  This
allows us to offer an alternative explanation for the large
equilibration lengths reported in \cite{muller-1992,wurtz-2002},
relevant to our study of the anti-Pfaffian edge-state transport.

Consider the action for the edge modes of the integer quantum Hall state at $\nu=2$.
Ignoring possible edge reconstruction, $S =S_0 +S_{\text{tunneling}}$:
\begin{subequations}
  \begin{align}
    \label{nu2-action}
    S_0 &= -\frac{1}{4\pi} \int_{t,x}
          \Big[ \sum_{i =1}^2 \partial_x\phi_i(\partial_t\phi_i+v_i \partial_x\phi_i)
          + 2 v_{12} \partial_x \phi_1 \partial_x \phi_2 \Big], \\
    S_{\text{tunneling}}&=- \int_{t,x} \left[\xi_{12}(x) \expon{i(\phi_1-\phi_2)
                          }+ \text{h.c.} \right], \quad
                          \overline{\xi_{12}(x) \xi^{*}_{12}(x') } = W_{12}\delta(x-x')  .
  \end{align}
\end{subequations}
Here, the most relevant tunneling term transfers a spin-up electron of
the first edge channel $\phi_1$ into a spin-down electron of the
second edge channel $\phi_2$.  Because $\expon{i (\phi_1 - \phi_2)}$
has scaling dimension $\Delta_{12}=1$ (for any value of $v_{12}$) and
is therefore relevant, it drives the system to an IR fixed point
(different from the clean fixed point $S_0$) described by
$S = S_{\Delta_{12}=1} + S_{\text{int}}$ \cite{KF-imp}(see Appendix
\ref{D12-apndx}):
\begin{subequations}
  \begin{align}
    \label{nu2-finaction}
    S_{\Delta_{12}=1} &= -\frac{1}{4\pi} \int_{t,x}
                        \left[\partial_x\phi_{\rho_{12}}(\partial_t\phi_{\rho_{12}}+v_{\rho_{12}} \partial_x\phi_{\rho_{12}})
                        \right]
                        -\frac{1}{4\pi}\int_{t,x} \left[
                        \partial_x
                        \tilde{\phi}_{\sigma_{12}}(\partial_t\tilde{\phi}_{\sigma_{12}}+v_{\sigma_{12}}\partial_x\tilde{\phi}_{\sigma_{12}}\right], \\
    S_{\text{int}} &=  -\frac{2v_{\sigma_{12},\rho_{12}}}{4\pi}\int_{t,x} \ \partial_x \phi_{\rho_{12}} \left(
                     \frac{\sqrt{2}}{a}O^{zx}\cos(\sqrt{2}
                     \tilde{\phi}_{\sigma_{12}})+\frac{\sqrt{2}}{a}O^{zy}\sin(\sqrt{2}
                     \tilde{\phi}_{\sigma_{12}})+O^{zz}\partial_x\tilde{\phi}_{\sigma_{12}}
                     \right),
  \end{align}
\end{subequations}
where $\phi_{\rho_{12}}$ is the total charge mode,
$\tilde \phi_{\sigma_{12}}$ is the gauge-transformed spin mode and
\begin{align}
  v_{\rho_{12}}=\frac{v_{1}+v_2}{2}+v_{12},\quad v_{\sigma_{12}}=\frac{v_{1}+v_2}{2}-v_{12},
  \quad v_{\sigma_{12},\rho_{12}} =\frac{v_{1}-v_2}{2}.
\end{align}
Following Section \eqref{kineq-section} (see Appendix
\ref{sec-rand-dd} for details) we write down the kinetic equation for
the charge current $I_{\rho}=-\partial_t\phi_{\rho}/2\pi$ and the
gauge-transformed neutral current $\tilde I_{\sigma}=-\partial_t\tilde
\phi_{\sigma}/2\pi$ in the vicinity of $S_{\Delta_{12}=1}$.
In the linear regime we find it more convenient to express the kinetic
equations in terms of a basis similar to the original fractional
modes. We define the ``slow'' fractional basis as
\begin{subequations}\label{eq-strmix-basis}
  \begin{align}
    I_1' &= \frac{1}{\sqrt{2}}(I_{\rho}+\tilde I_{\sigma})
    \\
    I_2' &= \frac{1}{\sqrt{2}}(I_{\rho}-\tilde I_{\sigma}).
  \end{align}
\end{subequations}
In this basis the kinetic equation is
\begin{subequations}
  \begin{align}
    \partial_x
    \begin{pmatrix}
      I_1'(x) \\
      I_2'(x)
    \end{pmatrix}
    &= G^e \begin{pmatrix}
      I_1'(x) \\
      I_2'(x)
    \end{pmatrix} \\
    G^e
    &=-g
      \begin{pmatrix}
        1 & -1 \\
        -1 & 1
      \end{pmatrix}
  \end{align}
\end{subequations}

where the conductivity coefficient is (see Appendix \ref{elect-cond-apndx})
\begin{align}\label{eq-nu2-g}
  g = \frac{2v_{\sigma_{12},\rho_{12}}^2 T^2}{3v_{\rho_{12}}^2W_{12}} = \frac{2
  T^2}{3W_{12}}\left( \frac{v_1-v_2}{v_1+v_2+2v_{12}} \right)^2.
\end{align}
If $|v_1 - v_2| \ll |v_1 + v_2 + 2v_{12}|$, $g \approx 0$ and charge
equilibration is weak.
We can write $v_i = v_i^{(0)} + w$ and
$v_{12} = w$, where $v_{i}^{(0)} > 0$ parametrizes the edge confining
potential and $w > 0$ is the magnitude of the short-ranged Coulomb
interaction (see the discussion following Eq.~\eqref{eq-gen-bos-action}).
The above inequality translates to
$|v^{(0)}_1 - v^{(0)}_2| \ll |v^{(0)}_1 + v^{(0)}_2 + 4w|$.
There are two reasons why this
inequality might be satisfied.
(1) Based on the measurements of the velocities of the
charge $\phi_{\rho_{12}}$ and neutral $\phi_{\sigma_{12}}$ modes \cite{bocq-nature-2013,kuma-prb-2011}, we
infer that $v^{(0)}_i\ll w$.
(2) If there exists approximate degeneracy between the spin-up and spin-down modes we have
$v_1^{(0)}\approx v^{(0)}_2$.

We can estimate
$|v^{(0)}_1 - v^{(0)}_2|$ using a simple model of the confining
potential $V(x)$.
Assume a potential of the form $V(x)=Ax^2$
which is slowly varying on the scale of the magnetic length. Then the
velocity of mode $\phi_i$ in the absence of the short-ranged Coulomb
potential is
\begin{align}\label{eq-v0-model}
  v^{(0)}_i = \frac{1}{B}\partial_xV(x)|_{E_i+V(x)=E_F}= \frac{\sqrt{2A(E_{F}-E_i) }}{B}
\end{align}
where $B$ is the magnetic field, $E_F$ is the bulk Fermi energy, and
$E_i$ is the energy of the Landau level corresponding to mode
$\phi_i$, deep within the bulk of the sample and away from any defect.
When $E_F$ sits in the middle of Landau levels
$E_F-E_i\sim \hbar \omega_c$.  From an experimental study of
equilibration between Landau level edge modes \cite{wurtz-2002}, we
infer that the Zeeman gap $\Delta E_Z=E_2-E_1$ is much smaller than
the cylotron gap $\hbar \omega_c$ by about an order of magnitude.
Therefore, for the difference in velocities we can write
\begin{align}\label{eq-dv0-estim}
  \frac{v^{(0)}_1-v^{(0)}_2}{(v^{(0)}_1+v^{(0)}_2)/2} \approx \frac{
  E_{2}-E_1 }{2(E_F-(E_1+E_2)/2 )}\approx \frac{
  \Delta E_Z }{ \hbar \omega_{c}}
\end{align}
and so $|v^{(0)}_1 - v^{(0)}_2|$ is also much smaller than the typical (average)
velocity $(v^{(0)}_1 + v^{(0)}_2)/2$.

To summarize, these estimates show that the conductivity
coefficient $g$ between the
spin-up and spin-down modes can be small even in the strong tunneling (large
$W_{12}$) regime.

\subsection{Electrical and thermal Hall conductance: overview}\label{conductance-section}

\begin{figure}[h!] \centering
  \includegraphics[width=0.5\textwidth]{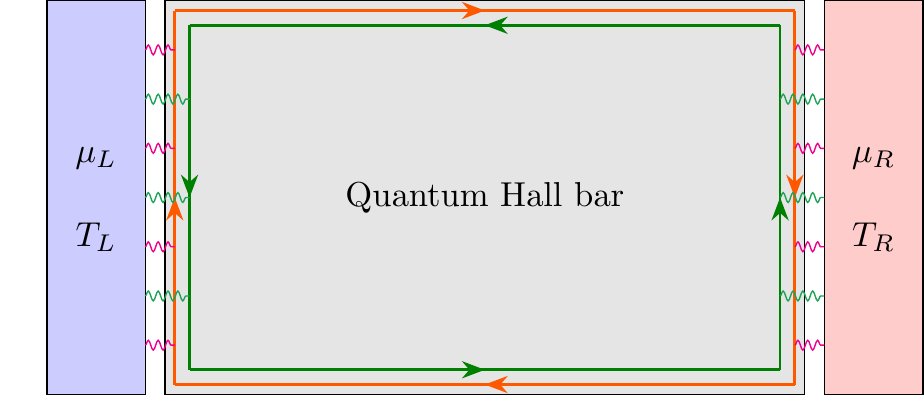}
  \caption{Quantum Hall bar geometry. Two counter-propagating modes
    (red and green directed lines) are shown. The wiggly lines
    represent tunnelings from the contacts to the edge modes along
    left and right line junctions.  Interactions between the edge
    modes are taken to occur along the top and bottom edges.}
  \label{setup-1}
\end{figure}

We are interested in transport in the two-terminal Hall bar geometry depicted in
Fig.~\ref{setup-1}.
The left and right edges of the Hall bar are coupled to leads held at chemical potentials $\mu_L$ and
$\mu_R$ and temperatures $T_L$ and $T_R$.
In order to find the electrical and thermal conductance we assume
ideal contacts in the following sense: a mode $\phi_i$
($\phi_i'$ for the slow modes) carries charge current
$I_i=\eta_i\nu_i\sigma_0\mu_c$ ($I_i'=\eta_i\nu_i\sigma_0\mu_c$)
upon leaving the $c \in \{L, R \}$ contact region, while the mode
$\tilde{\phi}_{\alpha}$ (refer to \eqref{gen-action-diag}) carries
heat current
$\tilde{J}^Q_\alpha=\eta_\alpha c_\alpha \kappa_0\frac{T^{2}_c}{2}$
upon leaving contact $c$.

Given this setup, we use the following procedure to calculate the
electrical and thermal conductances of the edge modes.  In order to
solve for the electrical conductance, we first solve the linear
differential equations in \eqref{elec-eq-matrix}. Taking
$\bm{\mathcal{I}}_n$ to be the eigenvectors of the matrix $G^e$ with
eigenvalue $\mathfrak{g}_n$, the general solution to the charge
transport equations is
\begin{align}
  \bm{I}(x) = \sum_n a_n \bm{ \mathcal{I} }_n \expon{\mathfrak{g}_n x}
\end{align}
for arbitrary coefficients $a_{n}$.
We then impose the above ``ideal contact" boundary
conditions to determine the $a_n$ for the top/bottom
edges of the Hall bar. We use a similar procedure to solve the heat transport
equations \eqref{therm-eq-matrix}.
From these solutions we find the total charge and heat
currents moving along the top/bottom edge of the Hall bar:
\begin{subequations}
  \begin{align}
    I_{\text{total, top/bottom}} & = \sum_{i} I_i(x), \\
    J^Q_{\text{total, top/bottom}} & = \sum_{\alpha} \tilde J^Q_\alpha(x),
  \end{align}
\end{subequations}
where $x$ is restricted to either the top/bottom edge of the Hall
bar. In the case where some of the modes are strongly mixed (for
example the edge modes of the $\nu=2$ quantum Hall state near the
$\Delta_{12}=1$ fixed point, as described in section
\ref{nu2-transport-section}) we use the slow modes basis to write
\begin{align}
  I_{\text{total, top/bottom}} & = \sum_{i\notin \text{strongly mixed}}
                                 I_i(x)+\sum_{i\in \text{strongly
                                 mixed}} I_i'(x).
\end{align}
Note that this expression still represents the total charge current,
since the gauge transformations that eliminate the strong-disorder
tunnelings (See appendix \ref{D12-apndx}) only rotate the neutral
currents.

The two-terminal charge and heat Hall conductances are then:
\begin{align}
  G = \frac{I_{\text{total, top}} + I_{\text{total, bottom}}}{\mu_L-\mu_R},
  \quad K = \frac{J^{Q}_{\text{total, top}}+J^Q_{\text{total, bottom}}}{T_L-T_R} .
\end{align}
Depending on the degree of inter-mode equilibration along the top and
bottom edges, the two-terminal conductance takes values between the
fully-equilibrated and non--equilibrated values.  For the electrical
conductance, $G_{\text{fully--eq}}=\sigma_0\sum_i\eta_i\nu_i$ while
$G_{\text{non--eq}}=\sigma_0\sum_i\nu_i$.  For the thermal
conductance,
$K_{\text{fully-eq}}=\kappa_0T\sum_{\alpha}\tilde{\eta}_{\alpha}\tilde{c}_{\alpha}=\kappa_0T\sum_i\eta_ic_i$
while
$K_{\text{non-eq.}}=\kappa_0T\sum_{\alpha}\tilde{c}_{\alpha}=\kappa_0T\sum_ic_i$,
where $c_i, c_\alpha$ are the central charges of the various edge
modes.  (A chiral boson has central charge equal to $+1$; a chiral
Majorana fermion has central charge equal to $+1/2$.)

\section{Low-temperature theories of anti-Pfaffian edge states at $\nu=5/2$}\label{antipf-theory-section}

\subsection{Setup and assumptions}\label{antipf-setup-section}

\begin{figure}[h!] \centering
  \includegraphics[width=0.6\textwidth]{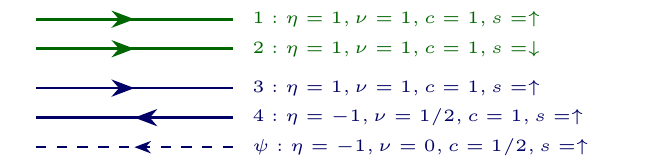}
  \caption{Edge modes of the anti-Pfaffian state at $\nu = 5/2$ in the
    absence of edge reconstruction: $\eta_i= \pm 1$ denotes the
    chirality of the edge mode; $\nu_i$ is the charge carried by the
    edge mode; $c_i$ is the central charge of the edge mode; and $s_i$
    is the spin of the Landau level associated to a particular edge
    mode.  Subscripts labeling the different edge modes are suppressed
    in the figure.}
  \label{antipf-edge}
\end{figure}

In the absence of edge reconstruction, the anti-Pfaffian state at $\nu = 5/2$ hosts a
total of five edge modes (Fig.~\ref{antipf-edge}) \cite{levin-2007,lee-2007}.
The lowest Landau level contributes (1) a spin-up integer mode and (2) a spin-down integer mode, both moving downstream.
From the first Landau level we have (3) a
downstream spin-up integer mode, (4) an upstream spin-up
$\nu=\frac{1}{2}$ bosonic mode, and an upstream Majorana mode $\psi$.

Charge tunneling between the edge channels requires broken translation
symmetry since the edge modes generally lie at different Fermi
momenta.  Quenched disorder is effective in tunneling electrons only
if it can absorb this momentum mismatch.  Estimates based on
experimental parameters (see \cite{feldman-comment}) suggest that
disorder satisfies this requirement.  We take the disorder to be
short-ranged.  Relaxation of such an assumption, however, has
interesting consequences for the equilibration of edge modes, as
suggested by Simon \cite{simon-interp}.

With these considerations, the low-energy effective theory for the
anti-Pfaffian edge state at $\nu = 5/2$ takes the form
$S = \sum_{i=1}^4 S_i + S_{\psi} + \sum_{i \neq j}
S_{ij}+S_{\text{tunneling}}$ \cite{levin-2007,lee-2007}:
\begin{subequations}
  \label{antipf-action}
  \begin{align}
    S_i &= -\frac{1}{4\pi} \int_{t,x}
          \left[\partial_x\phi_i({\eta_i \over \nu_i}\partial_t\phi_i+v_i \partial_x\phi_i)
          \right], \\
    S_{\psi} &=\frac{1}{4} \int_{t,x} \ i\psi (\partial_t \psi - u
               \partial_x \psi ), \\
    \label{coulombpfaffian}
    \sum_{i \neq j} S_{ij} & =  -\sum_{i \neq j}\frac{v_{ij}}{4\pi} \int_{t,x} \
                             \partial_x \phi_i \partial_x\phi_j, \\
    S_{\text{tunneling}}&=- \int_{t,x} \sum_{p\in P}\left[\xi_p(x) \expon{i\sum_j m^{(p)}_j
                          \phi_j} \psi^{m^{(p)}_{\psi}}+ \text{h.c.} \right].
  \end{align}
\end{subequations}
Similar to before, $P$ is the set of charge-conserving processes
defined by the integers $(m_j^{(p)}, m_\psi^{(p)})$ that tunnel
electrons between the edge modes, and $\xi_p$ is a Gaussian random
field with statistical average
$\overline{\xi_p(x) \xi^{*}_{p'}(x')}=\delta_{pp'}W_p\delta(x-x')$.

Unless the Coulomb interaction between edge modes of the different
Landau levels can be ignored, it's not obvious what tunneling
operators are most relevant.  In principle, multi-electron tunneling
operators can be more relevant than those that only involve a
single-electron tunneling process.  However, the largeness of the
Landau-gap compared to the electrochemical potential difference
between the edge modes, present in the experiments
\cite{banerj-half,banerj-any}, and the large equilibration lengths
reported for modes in different Landau levels
\cite{wurtz-2002,devyatov-2007} suggest that tunneling between edge
channels belonging to different Landau levels is generally suppressed.

Experiments also report a large equilibration length between spin-up
and spin-down modes \cite{muller-1992,devyatov-2007}.  This has been
attributed to suppressed tunneling between these modes due to weak
spin-orbit coupling.  This is the assumption made in
Ref.~\cite{ma-feldman-2019}; we relax this assumption in this paper.
Following the analysis in Section \ref{nu2-transport-section}, where
we provided an alternative explanation for the large equilibration
length between spin-up and spin-down integer modes, we assume strong
tunneling between spin-up and spin-down electrons of the lowest Landau
level.  Therefore, the most relevant tunnelings to include in
$S_{\text{tunneling}}$ are
\begin{subequations}
  \label{tunn-terms}
  \begin{align}
    S_{\text{tunneling},12} &= - \int_{t,x} \left[ \xi_{12}(x)
                              \expon{i(\phi_1-\phi_2)}  +\text{h.c.} \right], \\
    S_{\text{tunneling},34\psi} &= - \int_{t,x} \left[ \xi_{34}(x)
                                  \expon{i(\phi_3+2\phi_4)}\psi  +\text{h.c.} \right].
  \end{align}
\end{subequations}

If the Coulomb interaction between edge modes of different Landau
levels is ignored the term $S_{\text{tunneling},12}$ is always
relevant; $S_{\text{tunneling},34\psi}$ is relevant if the Coulomb
interaction between edge modes of the first Landau level interaction
is sufficiently strong.  If the modes in the lowest Landau level are
decoupled from the modes in the first Landau level (and if
equilibration of the first Landau level edge modes occurs via
$S_{\text{tunneling},34\psi}$), the low-temperature thermal Hall
conductance is the sum of the contributions from the lowest Landau
level and the first Landau level
$K=K_{\text{LLL}}+K_{\text{1LL}}=\frac{5}{2}\kappa_0T$.

However, we aren't aware of any reason that the Coulomb interaction
between the Landau levels is suppressed.  Consequently, either of the
two tunneling terms in \ref{tunn-terms} can be relevant or irrelevant,
depending on the specific nature of the Coulomb interaction, i.e., the
values of the $v_{ij}$ in Eq.~\eqref{coulombpfaffian}; even strong
Coulomb repulsion between all the modes doesn't uniquely specify an IR
fixed point.  We identify four possible IR fixed points:
\begin{enumerate}
\item $W_{12}=0$ and $W_{34}=0$ while $\Delta_{12}> \frac{3}{2}$ and
  $\Delta_{34}>\frac{3}{2}$
\item $W_{12}=0$ ($\Delta_{12}> \frac{3}{2}$) and $\Delta_{34}=1$
\item $\Delta_{12}=1$ and $ W_{34}=0$ ($\Delta_{34}> \frac{3}{2}$)
\item $\Delta_{12}=1$ and $\Delta_{34}=1.$
\end{enumerate}
Above, $\Delta_{12}$ and $\Delta_{34}$ are the scaling dimensions of $\expon{i (\phi_1 - \phi_2)}$ and $\expon{i (\phi_3 + 2 \phi_4)} \psi$.
The second case was analyzed in \cite{ma-feldman-2019}, where it was argued that $K=2.5\kappa_0T$ requires fine-tuning.
The first case is similar to the second one in this
regard so we won't discuss it.
In this paper, we investigate the third and the fourth low-temperature fixed points.
In Section \ref{antipf-transport-section}, we describe the conditions under which $K=2.5 \kappa_0T$ is consistent with either of these fixed points.

\subsection{$\Delta_{12}=1,W_{34}=0$ disordered fixed point}\label{D12-section}

In order to study this fixed point we change variables to charge $\phi_{\rho_{\tiny{12}}}=\frac{1}{\sqrt{2}} (\phi_1+\phi_{2})$ and spin $\phi_{\sigma_{12}}=\frac{1}{\sqrt{2}} (\phi_1-\phi_{2})$ modes
\cite{KF-imp}.
For $v_{ij}$ such that there is no coupling between $\partial_x \phi_{\sigma_{12}}$ and $\partial_x \phi_{4}$ the theory has an
emergent $SO(3)$ symmetry \cite{KFP,KF-imp,mirlin} that acts on the $\phi_{\sigma_{12}}$ sector.
In Appendix \ref{D12-apndx} we show how this symmetry can be used to eliminate $S_{\text{tunneling},12}$, after which an $SO(3)$ transformed spin mode $\tilde \phi_{12}$ is introduced.
The resulting action becomes $S = S_{\Delta_{12}=1} + S_{\text{int}}$ where
\begin{subequations}
  \begin{align}
    \label{D12-finaction}
    S_{\Delta_{12} = 1} & = S_{\rho_{12}} + S_{\sigma_{12}} + S_{3} + S_{4} + S_\psi + \sum_{i \neq j \in \{\rho_{12}, 3, 4 \}}S_{ij}, \\
    S_{\rho_{12}} &= -\frac{1}{4\pi} \int_{t,x}
                    \left[\partial_x\phi_{\rho_{12}}(\partial_t\phi_{\rho_{12}}+v_{\rho_{12}} \partial_x\phi_{\rho_{12}})
                    \right], \\
    S_{\sigma_{12}} &= -\frac{1}{4\pi}\int_{t,x} \left[
                      \partial_x
                      \tilde{\phi}_{\sigma_{12}}(\partial_t\tilde{\phi}_{\sigma_{12}}+v_{\sigma_{12}}\partial_x\tilde{\phi}_{\sigma_{12}}\right],
  \end{align}
\end{subequations}
and
\begin{subequations}
  \begin{align}\label{D12-finaction-irrel}
    S_{\text{int}} &= \sum_{i \in \{3,4,\rho_{12}\}} S_{\sigma_{12}, i} +
                     S_{\text{tunneling},34\psi}, \\
    S_{\sigma_{12},i} &= -\frac{2v_{\sigma_{12}, i}}{4\pi}\int_{t,x} \ \partial_x \phi_{i} \left(
                        \frac{\sqrt{2}}{a}O^{zx}\cos(\sqrt{2}
                        \tilde{\phi}_{\sigma_{12}})+\frac{\sqrt{2}}{a}O^{zy}\sin(\sqrt{2}
                        \tilde{\phi}_{\sigma_{12}})+O^{zz}\partial_x\tilde{\phi}_{\sigma_{12}}
                        \right).
  \end{align}
\end{subequations}
$O^{ab}(x)$ are matrix elements of the $SO(3)$ rotation that we use to
eliminate the $\xi_{12}(x)$ tunneling term.  The $S_{ij}$ and $v_{ij}$
with $i,j \in \{\rho_{12}, \sigma_{12}, 3, 4\}$ obtain from the
$S_{ij}$ and $v_{ij}$ with $i,j \in \{1, 2, 3, 4\}$ after the above
field redefinition.  $S_{\Delta_{12}=1}$ describes the $\Delta_{12}=1$
fixed point at which the terms in $S_{\text{int}}$ vanish:
$v_{\sigma_{12}, \rho_{12}}=v_{\sigma_{12}, 3}=v_{\sigma_{12},
  4}=W_{34}=0$. The density-density interactions in
$S_{\sigma_{12}, i}$ are irrelevant near the $\Delta_{12} = 1$ fixed
point. We assume $S_{\text{tunneling},34\psi}$ is irrelevant at this
fixed point, \textit{i.e.} $\Delta_{34} > \frac{3}{2}$, so that
$S_{\Delta_{12}=1}$ describes the low energy behavior of the
anti-Pfaffian edge.
When $S_{\text{tunneling},34\psi}$ is
relevant, the low-energy theory might be described by one of the other
fixed points in \ref{antipf-setup-section}. In Appendix
\ref{apndx-FP-validity} we discuss the domain of validity of
describing the low-temperature physics using perturbation theory
around the fixed point action \ref{D12-finaction}.

In order to analyze the finite-temperature transport in the vicinity of the $\Delta_{12}=1$ fixed point, the terms in $S_{\text{int}}$ must be included.
Consequently, we need to make a choice for the short-ranged Coulomb interaction $v_{ij}$ and diagonalize $S_{\Delta_{12} = 1}$.
The choice of the Coulomb interaction is non-universal.

Denote by $S_B = \sum_i S_i + \sum_{i \neq j} S_{ij}$, the quadratic part of
\eqref{antipf-action} that describes the chiral bosons, and write it as
\begin{align}\label{eq-gen-bos-action}
  S_{B} &= -\frac{1}{4\pi} \int_{t,x}
          \left[\sum_i\frac{\eta_{i}}{\nu_{i}}\partial_x\phi_i \partial_t\phi_i+ \sum_{i,j}V_{ij}\partial_x\phi_i \partial_x\phi_j
          \right].
\end{align}
We model the ``velocity matrix" $V_{ij}$ following \cite{chamon-wen}.
In the absence of a short-ranged Coulomb interaction, the action for the bosonic modes is
\begin{align}
  S_{0,B} &= -\frac{1}{4\pi}\sum_i \int_{t,x} \frac{1}{\nu_{i}}
            \left[\partial_x\phi_i(\eta_i\partial_t\phi_i+v^{(0)}_i \partial_x\phi_i)
            \right].
\end{align}
Thus, $v^{(0)}_i$ is the velocity of $\phi_i$ when the
Coulomb interaction is ignored.
We include the short-ranged Coulomb interaction via the ansatz,
\begin{align}
  S_{\text{Coulomb}} = -\pi w \int_{t,x}\ n_{\text{tot}}(x)^2 =
  -\frac{w}{4\pi}\int_{t,x} \ (\sum_i\partial_x\phi_i)^2,
\end{align}
where $n_{\text{tot}}=\frac{1}{2\pi}\sum_i\partial_x\phi_i$ is the
total charge density and $w$ is the strength of the Coulomb interaction.
The Hamiltonian for the bosonic modes is $H_{B}=H_{0,B}+H_{\text{Coulomb}} = \frac{1}{4\pi}\int \di x \sum_{ij}V_{ij}\partial_x\phi_i\partial_x\phi_j $, where the ``velocity matrix" is
\begin{align}\label{V-model}
  V_{ij}=
  \begin{cases}
    \frac{1}{\nu_{i}}v^{(0)}_i + w \quad & i=j, \\
    w &i\neq j.
  \end{cases}
\end{align}
First consider the limit $v_i^{(0)}=0$ for all $i$ at which the total
Hamiltonian is given by the Coulomb term only.
Here, the action is diagonalized using a charge-neutral basis.
One such basis choice, that is consistent with our earlier treatment of the relevant $\expon{i (\phi_1 - \phi_2)}$ term, is
\begin{align}\label{chaneut-basis}
  \begin{pmatrix}
    \phi_{\rho}\\
    \phi_{\sigma_{1}}\\
    \phi_{\sigma_2} \\
    \phi_{\sigma_3}
  \end{pmatrix}
  =
  \begin{pmatrix}
    \sqrt{\frac{2}{5}} & \sqrt{\frac{2}{5}} & \sqrt{\frac{2}{5}}
    &\sqrt{\frac{2}{5}}\\
    \frac{1}{\sqrt{2}}&-\frac{1}{\sqrt{2}}&0&0\\
    \frac{1}{\sqrt{6}}&\frac{1}{\sqrt{6}}&-\frac{2}{\sqrt{6}}&0\\
    \frac{1}{\sqrt{15}}&\frac{1}{\sqrt{15}}&\frac{1}{\sqrt{15}}&\frac{6}{\sqrt{15}}
  \end{pmatrix}
                                                                 \begin{pmatrix}
                                                                   \phi_{1}\\
                                                                   \phi_{2}\\
                                                                   \phi_{3} \\
                                                                   \phi_{4}
                                                                 \end{pmatrix}.
\end{align}
Notice that $\phi_{\sigma_1} = \phi_{\sigma_{12}}$.  When
$v_i^{(0)} = 0$, the velocity of the charge mode $\phi_{\rho}$ is
$\nu w$ ($\nu=\frac{5}{2}$), while the velocities of the neutral modes
$\phi_{\sigma_{\alpha}}$ are zero.  This three-fold degeneracy in the
velocity matrix exists because there is a freedom in choosing the
neutral basis given by
$\phi_{\sigma_i}=\Lambda^{\sigma}_{ij}\phi_{\tilde{\sigma}_j}$ where
$\Lambda^{\sigma}$ is an arbitrary $SO(2,1)$ rotation.

Experiments \cite{bocq-nature-2013,kuma-prb-2011} suggest
the velocity of the charge mode is generally about an order of
magnitude larger than the velocity of a neutral mode.  This was
predicted earlier in \cite{aleiner-1994}.
Thus, we assume small, but finite $v^{(0)}_{i}\ll w$.
The modes that diagonalize
$S_{\Delta_{12}=1}$ when $v_i^{(0)} \neq 0$ are not exactly the charge and neutral basis in
Eq.~\eqref{chaneut-basis}.
We denote the diagonal modes as
$\phi_r,\phi_{\sigma_{12}},\phi_{s_2},\phi_{s_3}$; in the small
$v^{(0)}_i$ limit, the $\phi_r$ mode is ``close" to the total charge
mode while $\phi_{s_2}$ and $\phi_{s_3}$ are ``almost neutral''
modes.
Based on \eqref{eq-v0-model}, we expect all the $v^{(0)}_i$ as
well as the Majorana velocity $u$, to have the same order of
magnitude, which we denote by $v^{(0)}$.
Therefore, to leading order in $v^{(0)}/w$, the velocities for the
modes $\phi_r,\phi_{\sigma_{12}},\phi_{s_2},\phi_{s_3}$ are
\begin{align}\label{D12-vs}
  v_r=\nu w +O(v^{(0)}),\quad v_{\beta}=O(v^{(0)})\ \text{for}\
  \beta=\sigma_{12},s_2,s_3.
\end{align}

The density-density interactions between the $\phi_{\sigma_{12}}$ mode
and the other modes (the first term in \eqref{D12-finaction-irrel})
become irrelevant on scales larger than $v_{\sigma_{12}}^2/W_{12}$.
In Section \ref{antipf-transport-section} we include the effects of
such interactions on charge and heat transport near the
$\Delta_{12} = 1$ fixed point.  The couplings for these interactions,
$v_{\sigma_{12},r}, v_{\sigma_{12},s_2}$, and $v_{\sigma_{12},s_3}$,
vanish in the limit where there's a degeneracy between the up and down
spin electrons in the lowest Landau level.  To see this, consider a
general $SO(3,1)$ transformation $\Lambda_{i\alpha}$ from the
fractional modes $\phi_i,i=1,2,3,4$ to some new modes $\phi_{\alpha}$
with $\alpha=\sigma_{12},\tilde{2},\tilde{3},\tilde{4}$, such that one
of the modes is the spin mode $\phi_{\sigma_{12}}$.  From the
definition of the spin mode we see that
\begin{subequations}\label{s12-transf-eq}
  \begin{align}
    \phi_1 &= \frac{1}{\sqrt{2}}\phi_{\sigma_{12}}+\sum_{\alpha\neq \sigma_{12}
             } \Lambda_{1\alpha} \phi_{\alpha} \\
    \phi_2 &= -\frac{1}{\sqrt{2}}\phi_{\sigma_{12}}+\sum_{\alpha\neq \sigma_{12}
             } \Lambda_{2\alpha} \phi_{\alpha} 
  \end{align}
\end{subequations}
with $\Lambda_{1\alpha}=\Lambda_{2\alpha}$ for $\alpha\neq
\sigma_{12}$ while $\Lambda_{3,\sigma_{12}}=\Lambda_{4,\sigma_{12}}=0$. The
velocity matrix tranforms as $v_{\alpha
  \beta}=\sum_{ij}V_{ij}\Lambda_{i\alpha}\Lambda_{j\beta}$. So for
$v_{\sigma_{12},\alpha}$ we have
\begin{align}
  \beta \neq \sigma_{12}: v_{\sigma_{12},\beta}
  =& \sum_{ij}V_{ij}\Lambda_{i
     \sigma_{12}}\Lambda_{j\beta} \\
  \nonumber =& V_{11}  \Lambda_{1,\sigma_{12}} \Lambda_{1,\beta} + V_{22}
               \Lambda_{2,\sigma_{12}} \Lambda_{2,\beta}
               + V_{12} \left(\Lambda_{1,\sigma_{12}}
               \Lambda_{2,\beta}+\Lambda_{2,\sigma_{12}}\Lambda_{1,\beta}\right) \\
  \nonumber &+ \sum_{j \neq 1,2} V_{1 j} ( \Lambda_{1,\sigma_{12}}
              \Lambda_{j,\beta}+\Lambda_{j,\sigma_{12}}\Lambda_{1,\beta})
              +V_{2 j}( \Lambda_{2,\sigma_{12}}
              \Lambda_{j,\beta}+\Lambda_{j,\sigma_{12}}\Lambda_{2,\beta}) \\
  \nonumber &+ \sum_{i,j\neq 1,2} V_{ij} \Lambda_{i,\sigma_{12}}
              \Lambda_{j,\beta}.
\end{align}
Using \eqref{s12-transf-eq} we get
\begin{align}\label{eq-vsa-final}
  \beta \neq \sigma_{12}: v_{\sigma_{12},\beta}
  =&  \frac{1}{\sqrt{2}} \Lambda_{1,\beta}( V_{11} - V_{22})
     + \frac{1}{\sqrt{2}} \sum_{j \neq 1,2}
     \Lambda_{j,\beta} (V_{1j}-V_{2j})
\end{align}
which vanishes when $V_{11}=V_{22}$ and $V_{1i}=V_{2i},i=3,4$,
i.e., when there exists symmetry between the spin-up and
spin-down modes.  Note that this result is independent of our specific
modeling of the velocity matrix.

\subsection{$\Delta_{12}=\Delta_{34}=1$ disordered fixed
  point} \label{D12D34-section}

Here, in addition to the field redefinition of the edge modes arising from the lowest Landau level considered in the previous section, we introduce the charge
$\phi_{\rho_{34}}=\sqrt{2}(\phi_3+\phi_4)$ and neutral
$\phi_{\sigma_{34}}=\phi_3+2\phi_4$ fields \cite{levin-2007,lee-2007}.
We also define the Majorana vector $\bm{\psi}^T=(\psi_1,\psi_2,\psi_3)$
with Majorana fermions
$\psi_1=\expon{i\phi_{\sigma_{34}}}+\expon{-i\phi_{\sigma_{34}}},
\psi_2=i(\expon{i\phi_{\sigma_{34}}}-\expon{-i\phi_{\sigma_{34}}}),\psi_3=\psi$. In
terms of these fields the action is
\begin{subequations}
  \begin{align}
    S &= \sum_{i \in \{\sigma_{12},\rho_{12},\rho_{34} \} }  S_i + \sum_{i \neq j \in \{
        \sigma_{12},\rho_{12},\sigma_{34},\rho_{34} \}} S_{ij}  +
        S_{\text{neutral}},
    \\
    S_{\rho_{34}} &= -\frac{1}{4\pi} \int_{t,x} \
                    \partial_x\phi_{\rho_{34}}(\partial_t\phi_{\rho_{34}}+v_{\rho_{34}}
                    \partial_x\phi_{\rho_{34}} ),
    \\
    S_{\text{neutral}} &= S_{\text{sym}} + S_{\text{anis} }, \\
    S_{\text{sym.}} &= \frac{1}{4}\int_{t,x} \
                      i\bm{\psi}^{T}(\partial_t\bm{\psi}-\overline{v}\partial_x\bm{\psi}
                      -\frac{\bm{\xi}_{34}.\bm{L}}{2}\bm{\psi}), \quad
                      \bm{\xi}_{34}=\Big(\frac{\xi_{34}+\xi_{34}^{*}}{2}, \frac{\xi_{34}-\xi_{34}^{*}}{2i},0 \Big),
    \\
    S_{\text{anis.}} &= -\frac{1}{4}\int_{t,x} \
                       i\bm{\psi}^{T}\delta v \partial_x\bm{\psi}, \\
    S_{i j  } &= -\frac{2v_{i j }}{4\pi} \int_{t,x} \
                \partial_x\phi_{i}
                \partial_x\phi_{j},
  \end{align}
\end{subequations}
where the average velocity
$\overline{v}\equiv \frac{2v_{\sigma_{34}}+u}{3}$ and the anisotropic
velocity matrix
$\delta v \equiv
\operatorname{diag}(v_{\sigma_{34}}-\overline{v},v_{\sigma_{34}}-\overline{v},u-\overline{v})$. $\bm{L}=(L^x,L^y,L^z)$
is the vector composed of the three generators of $SO(3)$.
$S_{\text{sym}}$ has an $SO(3)$ gauge symmetry
$\bm{\psi}(x,t)=O(x)\tilde{\bm{\psi}}(x,t)$ provided the disorder
vector also transforms as
\begin{align}
  \tilde{\xi}_{34}^a = \frac{1}{2} \epsilon^{abc}
  \left(O^T(\bm{\xi}_{34}.\bm{L})O\right)^{bc} + \overline{v}\epsilon^{abc} (O^T\partial_xO)^{bc}.
\end{align}
However under this transformation, the term
$\tilde{\bm{\psi}}^T (O^T\delta v
\partial_xO)\tilde{\bm{\psi}}$ shows up in $S_{\text{anis}}$.
In order to get rid of such a term we instead require $\xi_{34}$ to transform as
\begin{align}
  \tilde{\xi}_{34}^a = \frac{1}{2} \epsilon^{abc}
  \left(O^T(\bm{\xi}_{34}.\bm{L})O\right)^{bc} + \epsilon^{abc}
  (O^T v\partial_xO)^{bc},
\end{align}
with velocity matrix
$v=\operatorname{diag}(v_{\sigma_{34}},v_{\sigma_{34}},u)$.  Requiring
$\tilde{\bm{\xi}}_{34}=0$, the transformed action becomes
$S = S_{\Delta_{12}=\Delta_{34}=1}+S_{\text{int}}$ where
\begin{subequations}
  \label{D12D34-finaction}
  \begin{align}
    S_{\Delta_{12}=\Delta_{34}=1} &= \sum_{i \in \{\sigma_{12},\rho_{12},\rho_{34} \} }  S_i +
                                    S_{\text{neutral sym}}+ S_{ \rho_{12}, \rho_{34} },
    \\
    S_{\text{neutral sym}} &= \frac{1}{4}\int_{t,x} \
                             i\tilde{\bm{\psi}}^{T}(\partial_t\tilde{\bm{\psi}}-\overline{v}\partial_x\tilde{\bm{\psi}}
                             ),
    \\
    S_{\rho_{12}, \rho_{34}} &= -\frac{v_{\rho_{12}, \rho_{34}}}{8\pi} \int_{t,x} \
                               \partial_x\phi_{\rho_{12}} \partial_x\phi_{\rho_{34}},
  \end{align}
\end{subequations}
and
\begin{subequations}\label{D12D34-finaction-irrel}
  \begin{align}
    S_{\text{int}} &= \sum_{i \in \{\rho_{12},\rho_{34} \}}(S_{
                     \sigma_{34},i}+S_{\sigma_{12},i
                     })+S_{\sigma_{12}, \sigma_{34}}+S_{\text{neutral int}}, \\
    S_{\text{neutral int} } &=  -\int_{t,x} \
                              i\tilde{\bm{\psi}}^{T} \widetilde{\delta v} \partial_x\tilde{\bm{\psi}},
    \\
    S_{\sigma_{34},i} &= -\frac{v_{i .\sigma_{34}}}{8\pi} \int_{t,x} \
                        \partial_x\phi_{\sigma_{34}}
                        \left(i \tilde{\bm{\psi}}^{T}L^z(x)\tilde{\bm{\psi}}\right), \\
    S_{\sigma_{12},\sigma_{34}} &=
                                  -\frac{2v_{\sigma_{12},\sigma_{34}}}{4\pi}\int_{t,x}
                                  \left(i
                                  \tilde{\bm{\psi}}^{T}L^z(x)\tilde{\bm{\psi}}\right)
    \\
    \nonumber & \qquad \qquad \times \left(
                \frac{\sqrt{2}}{a}O^{zx}\cos(\sqrt{2}
                \tilde{\phi}_{\sigma_{12}})+\frac{\sqrt{2}}{a}O^{zy}\sin(\sqrt{2}
                \tilde{\phi}_{\sigma_{12}})+O^{zz}\partial_x\tilde{\phi}_{\sigma_{12}}
                \right),
  \end{align}
\end{subequations}
with $\widetilde{\delta v}(x)\equiv O^T(x) \delta v O(x)$ and
$L^z(x)\equiv O^T(x)L^zO(x)$.
The $\Delta_{12}=\Delta_{34}=1$ fixed point is described by $S_{\Delta_{12}=\Delta_{34}=1}$
about which the terms in $S_{\sigma_{34},i}$ and $S_{\sigma_{34},i}$ are
irrelevant.
Here, the auto-correlation of elements
of matrices  $L^z(x)$ and $\widetilde{\delta v}(x)$ decay on length
scales $\sim \overline{v}^2/W_{34}$.

We model the short-ranged Coulomb interaction as in the previous section.
Here, the diagonal modes are
$\phi_r,\phi_{\sigma_{12}},\phi_{s_2},\phi_{\sigma_{34}}$, where
$\phi_{s_2}$ is some ``almost neutral'' mode.
To leading order in $v^{(0)}/w$ the velocities for these modes  are
\begin{align}\label{D12D34-vs}
  v_r=\nu w +O(v^{(0)}),\quad
  v_{\beta}=O(v^{(0)}) \ \text{for}\  \beta=\sigma_{12},s_2,\sigma_{34}.
\end{align}
Since $u=O(v^{(0)})$, we can write $\overline{v}\approx
O(v^{(0)})$. As for the magnitude of couplings in
\eqref{D12D34-finaction-irrel}, we have
$v_{\sigma_{34},\beta}=O(v^{(0)})$ for $\beta=r,s_2$ while
$v_{\sigma_{12},\beta}$ vanish for $\beta=r,\sigma_{12},s_2$ in the
spin-degenerate limit as demonstrated in the previous section.

\section{Transport and equilibration along the $\nu=5/2$ edge}\label{antipf-transport-section}

In this section, we analyze the low-temperature transport properties
of the effective theories of the $\nu = 5/2$ anti-Pfaffian state
described in Sections \ref{D12-section} and \ref{D12D34-section}.  We
will apply charge and heat kinetic equations introduced in Section
\ref{gen-transport-section} to each of these fixed points, calculate
the expressions for conductivity coefficients, and eventually solve
for the electrical and thermal Hall conductances.  We estimate the
parameter regime that describes the experimental observation of
$\kappa = 2.5 \kappa_0 T$ so as to determine the experimental
relevance of each fixed point.

\subsection{$\Delta_{12}=1$ fixed point}

\subsubsection{Charge transport}
At this fixed point, the processes that cause equilibration are the
irrelevant terms in \eqref{D12-finaction-irrel}. Using
\eqref{elec-eq-matrix} (see appendix \ref{elect-cond-apndx} for details) we
write down the equations describing charge transport resulting from
such interactions. In the basis $(I'_{1} ,I'_2,I_3,I_4)$ the
matrix $G^e$ is
\begin{align}\label{eq-D12-Ge}
  G^{e}
  &=-(\sum_{\beta=r,s_2,s_3}g_{V_{\sigma_{12}, \beta}})
    \begin{pmatrix}
      1 & -1 & 0 & 0\\
      -1 & 1 & 0 & 0 \\
      0 & 0 & 0 & 0 \\
      0 & 0 & 0& 0
    \end{pmatrix}
                 -g_{V_{34}} \begin{pmatrix}
                   0 & 0 & 0 & 0\\
                   0 & 0 & 0 & 0 \\
                   0 & 0 & 1 & 2 \\
                   0 & 0 & -1& -2
                 \end{pmatrix}
\end{align}
with
\begin{subequations}\label{D12-gs}
  \begin{align}
    \beta = r, s_2,s_3:
    g_{V_{\sigma_{12}, \beta}}&
                                =  \frac{2\pi^2 v_{\sigma_{12},\beta}^2
                                T^2}{3v_{\beta}^2W_{12}} \\
    g_{V_{34}} &= \frac{\Gamma(\Delta_{34})^{2}}{\Gamma(2\Delta_{34})} \frac{W_{34}}{\overline{v}_{V_{34}}^{2\Delta_{34}}}\left(2\pi
                 aT\right)^{2\Delta_{34}-2}, \quad
                 \overline{v}_{V_{34}}= O( v^{(0)}).
  \end{align}
\end{subequations}
The velocities are defined in \eqref{D12-vs}, and $a$ is the
short-distance cutoff \cite{wen-1991-edge}.

The last term in $G^e$ couples the downstream and upstream
charge modes. Therefore largeness of $g_{V_{34}}$ (see
below) 
is required for
the proper quantization of the electrical conductance at $G=2.5
\sigma_0$. To quantify this we solve for the electrical conductance using
\eqref{eq-D12-Ge} and boundary conditions specified in Section
\ref{conductance-section}. We find
\begin{align}\label{eq-D12-Gsol}
  G =  \sigma_0 \left(2 + \frac{2+\expon{-g_{V_{34}}L}}{2(2-\expon{-g_{V_{34}}L})} \right),
\end{align}
where $L$ is the effective length on the sample's top/bottom edge along which equilibration takes place.
If the electrical conductance is measured to be $G=2.50 \sigma_0$
within the uncertainty $\Delta G = 0.01 \sigma_0$ we
find the bound $g_{V_{34}}L \gtrsim 4$.

\subsubsection{Heat transport}

Based on \eqref{therm-eq-matrix}, the heat transport matrix
$G^Q$ in the basis $(r,\phi_{\sigma_{12}},\phi_{s_2},\phi_{s_{3}},\psi)$ is
\begin{align}\label{eq-D12-GQ}
  G^Q=& \frac{12 g_{V_{\sigma_{12}, r}}}{5}
        \begin{pmatrix}
          -1 & 1 & 0 & 0 & 0 \\
          1 & -1  & 0 & 0 & 0  \\
          0 & 0 & 0 & 0 & 0 \\
          0 & 0 & 0 & 0  & 0 \\
          0 & 0 & 0 & 0 & 0
        \end{pmatrix}
                          +\frac{12 g_{V_{\sigma_{12}, s_2}}}{5}
                          \begin{pmatrix}
                            0 & 0 & 0 & 0 & 0 \\
                            0 & -1  & 1 & 0 & 0  \\
                            0 & 1 & -1 & 0 & 0 \\
                            0 & 0 & 0 & 0  & 0 \\
                            0 & 0 & 0 & 0 & 0
                          \end{pmatrix}
                                            +\frac{12 g_{V_{\sigma_{12}, s_{3}}}}{5}
                                            \begin{pmatrix}
                                              0 & 0 & 0 & 0 & 0 \\
                                              0 & -1  & 0 & -1 & 0  \\
                                              0 & 0 & 0 & 0 & 0 \\
                                              0 & 1 & 0 & 1  & 0 \\
                                              0 & 0 & 0 & 0 & 0
                                            \end{pmatrix}
                                                              +\frac{12 g_{V_{34}}}{1+2\Delta_{34}} \cr
                         & \times
                           \begin{pmatrix}
                             -d_r(d_{s_2}+d_{s_3}+d_{\psi})  & 0 & d_rd_{s_2} & -d_rd_{s_3} & -2d_rd_{\psi} \\
                             0 & 0 & 0 & 0 & 0  \\
                             d_rd_{s_2} & 0 & -d_{s_2}(d_r+d_{s_3}+d_{\psi}) & -d_{s_2}d_{s_3} &-2d_{s_2}d_{\psi}  \\
                             d_r d_{s_3} & 0 & d_{s_2}d_{s_3} & d_{s_3}(d_r+d_{s_2}+d_{\psi}) & -2d_{s_3}d_{\psi} \\
                             d_rd_{\psi} & 0 & d_{s_2}d_{\psi} & -d_{s_3}d_{\psi} & 2d_{\psi}(d_r+d_{s_2}+d_{s_2})
                           \end{pmatrix}
\end{align}
where $d_{\psi}=\frac{1}{2}$ and
$d_{\alpha}=\left( \Lambda_{3\alpha}+2\Lambda_{4\alpha} \right)^{2}/2
$.
Also we have $\sum_{\alpha=r,s_2,s_3,\psi}d_{\alpha}=\Delta_{34}$.
See \ref{therm-cond-apndx} for the definition of
$d_{\alpha}$.
$\Lambda$ is the $SO(3,1)$ transformation expressing the
fractional modes $\phi_i$ in terms of
$(\phi_r,\phi_{\sigma_{12}},\phi_{s_2},\phi_{s_{3}})$, i.e., the diagonal
modes of $S_{\Delta_{12}=1}$.

This transformation depends on the velocity matrix in
\eqref{D12-finaction}. We use the velocity matrix in
Eq.~\eqref{V-model} in order to estimate the $d_{\alpha}$. In the
$v^{(0)}_i/w=0$ limit, $\phi_r$ is the total charge mode, and,
consequently, it commutes with the neutral mode $\phi_3+2\phi_4$.
Therefore, in this limit,
$d_{r}=\left( \Lambda_{3,r}+2\Lambda_{4,r} \right)^{2}/2=0 $.  For
finite but small $v^{(0)}/w$, we have
$d_r=O\left( (\frac{v^{(0)}}{w})^2 \right)$ to leading order.

In order to estimate $d_{s_2}$ and $d_{s_{3}}$, we look at the spin of
the operator $\expon{i\phi_3+2i\phi_{4}}$. Generally, for a set of chiral
bosons $\phi_{i}$ with commutation relation
$[\phi_i(x),\phi_j(x')]=\pi i K^{-1}_{ij} \operatorname{sign}(x-x')$,
the spin of the vertex operator $\expon{i\sum_i n_i \phi_i}$ is
\begin{align}\label{eq-spin-def}
  h_- = \frac{1}{2}n_{i} K^{-1}_{ij}n_j = \Delta_R-\Delta_{L},
\end{align}
where $\Delta_R$ ($\Delta_L$) is the scaling dimension of the
right-moving (left-moving) part of $\expon{i\sum_i n_i
  \phi_i}$. Therefore, the spin of the tunneling operator
$\expon{i\phi_3+2i\phi_{4}}$ is
$h_-=-\frac{1}{2}= \Delta_R-\Delta_{L}$.  Also, we have
$\Delta_R=d_r+d_{s_2}$ and $\Delta_L=d_{s_3}$. Along with
$d_r+d_{s_2}+d_{s_3}+d_{\psi}=\Delta_{34}$, to leading order in
$v^{(0)}_i/w$ we find
\begin{subequations}
  \begin{align}
    d_{s_2} &= \frac{\Delta_{34}-1}{2}-d_r =\frac{\Delta_{34}-1}{2}-O\left( (\frac{v^{(0)}}{w})^2 \right) \\
    d_{s_3} &= \frac{\Delta_{34}}{2}.
  \end{align}
\end{subequations}
As we mentioned in Section \ref{D12-section}, we take
$\Delta_{34}\geq \frac{3}{2}$ so that $S_{\Delta_{12}}=1$ describes
the low-energy physics of the $\Delta_{12}=1$ fixed point. On the
other hand, since $g_{V_{34}}L$ is large, based on Eq. \eqref{D12-gs},
we don't expect $\Delta_{34}$ to be very large. This is due to the
fact that \textit{i}) the pre-factor
$\Gamma(\Delta_{34})^{2}/\Gamma(2\Delta_{34})$ vanishes rapidly for
large $\Delta_{34}$ and \textit{ii})
$g_{V_{34}} \sim T^{2(\Delta_{34}-1)}$ and so the equilibration
process corresponding to $g_{V_{34}}$ would have sub-leading
contribution at small temperatures, if $\Delta_{34}$ was large.

We can estimate $\Delta_{34}$ for
$v^{(0)}_i=v^{(0)}$. In this case, using \eqref{chaneut-basis} we can write
\begin{align}\label{eq-samev0-H}
  \nonumber    H &= \frac{1}{4\pi} \int_x
                   V_{ij}\partial_x\phi_i\partial_x\phi_j   \\
                 &= \frac{1}{4\pi} \int_x \Big[
                   (w+\frac{7}{5}v^{(0)})(\partial_x\phi_{\rho})^2 +
                   v^{(0)}(\partial_x\phi_{\sigma_1})^2 \\
  \nonumber    &\qquad \qquad \qquad+v^{(0)}(\partial_x\phi_{\sigma_2})^2
                 +\frac{7}{5} v^{(0)}(\partial_x\phi_{\sigma_3})^2
                 -\frac{4\sqrt{6}}{5} v^{(0)} \partial_x\phi_{\rho}\partial_x\phi_{\sigma_{3}} \Big].
\end{align}
Therefore, for small $v^{(0)}/w$ a small rotation in the
$(\phi_{\rho},\phi_{\sigma_{3}})$ plane would diagonalize the
Hamiltonian. So, using \eqref{chaneut-basis} we find
$\Delta_{34}=\frac{5}{3}$ in the vanishing $v^{(0)}/w$ limit.

We are interested in determining the regime for which this matrix
$G^Q$ leads to a thermal Hall conductance $K=2.5 \kappa_0T$ within the
uncertainties of the experiment.  Quantization of electrical
conductance $G=2.5\sigma_{0}$ implies that $g_{V_{23}}$ is
large. Looking at the last term in \eqref{eq-D12-GQ}, more
specifically, the $(\phi_{s_2},\phi_{s_{3}},\psi)$ block, it appears
that the $\phi_{s_2}$,$\phi_{s_3}$ and $\psi$ modes equilibrate with
each other.  For the moment, let's assume they are completely
equilibrated; we will relax this assumption later. In this case, we
can think of these modes as a single upstream mode with central charge
$c=\frac{1}{2}$.  We call this mode $\tilde{s}$.

If equilibration between the first two modes in \eqref{eq-D12-GQ} and
the $\tilde{s}$ mode is suppressed, the thermal conductance theis sum
of the contributions from the first two modes $K_{r+\sigma_{12}}$ and
from the $\tilde{s}$ mode $K_{\tilde{s}}$. That is
$K= K_{r+\sigma_{12}}+K_{\tilde{s}}= \left(2+\left|-0.5 \right|
\right)\kappa_0T=2.5 \kappa_0T$.  This requires
\begin{align}
  g_{V_{\sigma_{12}, s_2}}L \ll 1, \quad g_{V_{\sigma_{12}, s_3}}L \ll
  1, \quad
  g_{r,\tilde{s}}L \ll 1,
\end{align}
where we defined
$ g_{r,\tilde{s}}= d_{r}(\Delta_{34}-d_r) g_{V_{34}} $. Therefore, we
see that there exists a regime of parameters where the fixed point
$\Delta_{12}=1$ can be consistent with experiments. Using the details
of the experimental measurements, we can gain a more quantitative
estimation of this regime.

We use the above $G^Q$ matrix and boundary conditions given in Section
\ref{conductance-section} to solve for the thermal conductance.
Following our earlier discussion we will take
$\Delta_{34}=\frac{5}{3}$, and consequently
$d_{s_2}=\frac{1}{3},d_{s_3}=\frac{5}{6}$. Later, we will discuss how
our results depend on these values.

We also ignore the first term in \eqref{eq-D12-GQ} in the remainder.
This follows from our discussion in Section
\ref{nu2-transport-section}: we expect $g_{\sigma_{12},r}L$ to be
suppressed both due to the strong Coulomb interaction and small spin
gap. Also, since $g_{\sigma_{12},r}$ quantifies equilibration between
co-propagating modes, its magnitude does not have much effect on the
thermal conductance.

\begin{figure}[h!]
  \centering
  \begin{subfigure}[b]{.45\linewidth}
    \includegraphics[width=\textwidth]{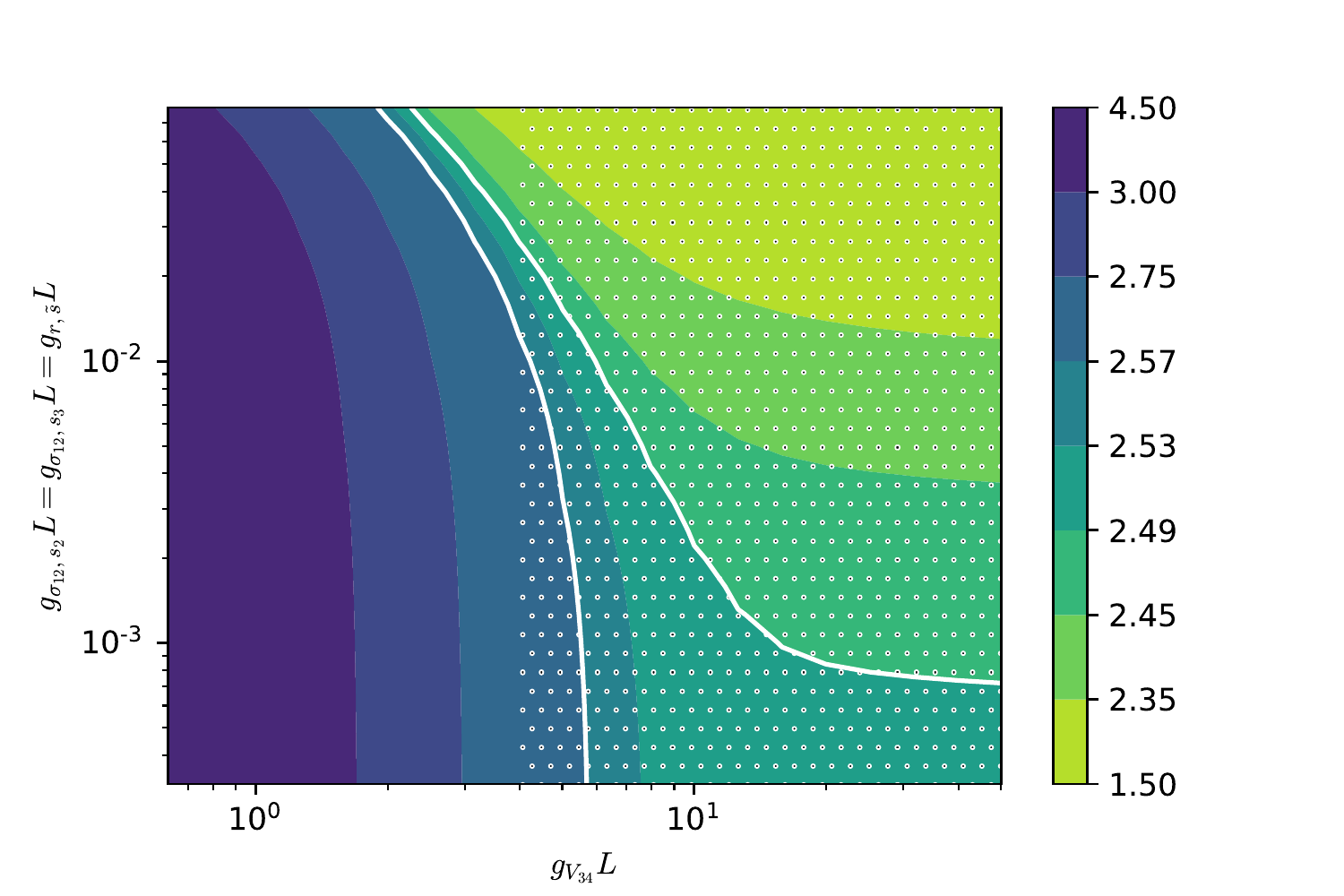}
    \caption{$g_{V_{\sigma_{12},s_2}}=g_{V_{\sigma_{12},s_2}}=g_{r,\tilde{s}}$}
    \label{fig-D12-Ktot-a}
  \end{subfigure}
  \begin{subfigure}[b]{.45\linewidth}
    \includegraphics[width=\textwidth]{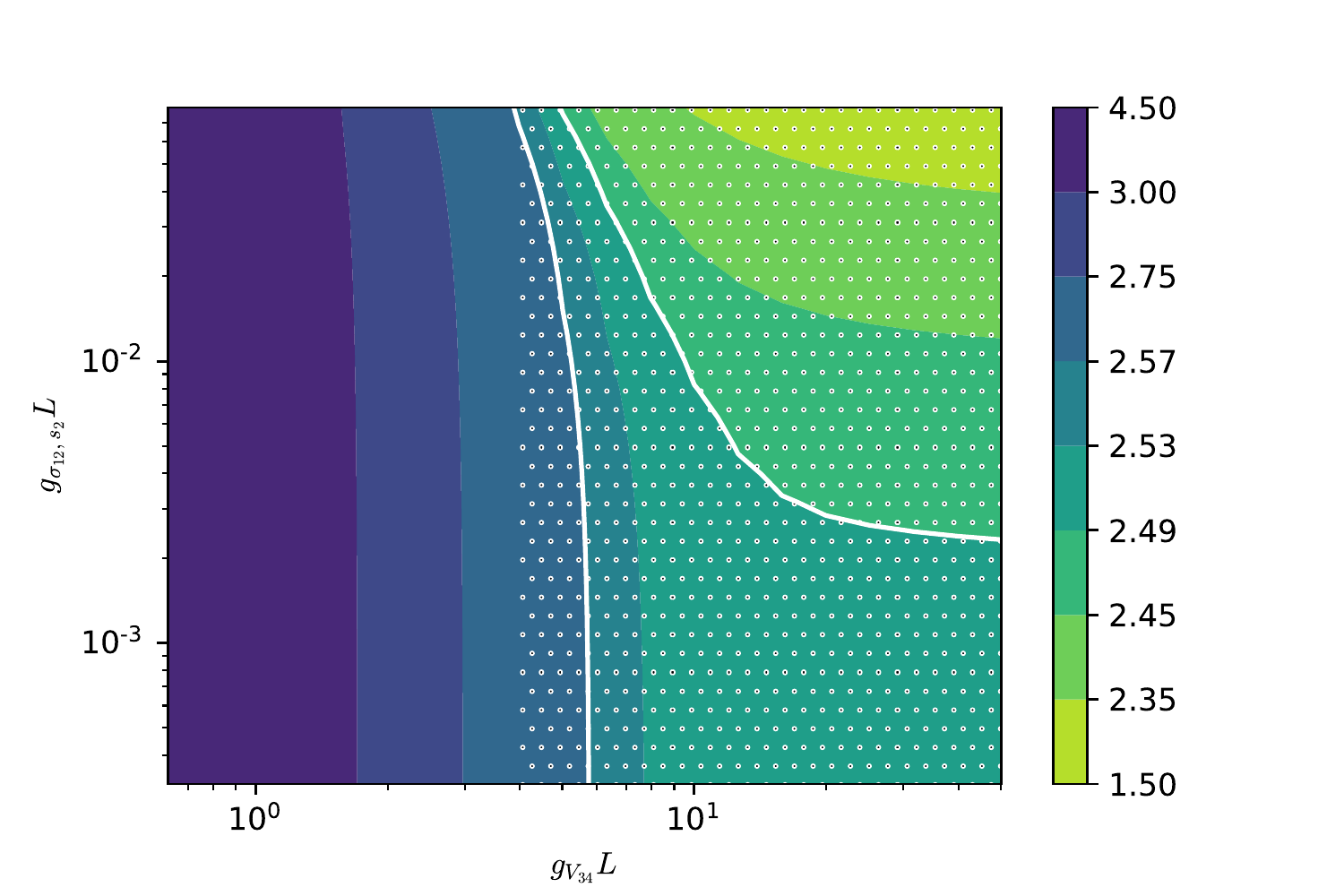}
    \caption{$g_{V_{\sigma_{12},s_3}}=g_{r,\tilde{s}}=0$}
  \end{subfigure}
  \begin{subfigure}[b]{.45\linewidth}
    \includegraphics[width=\textwidth]{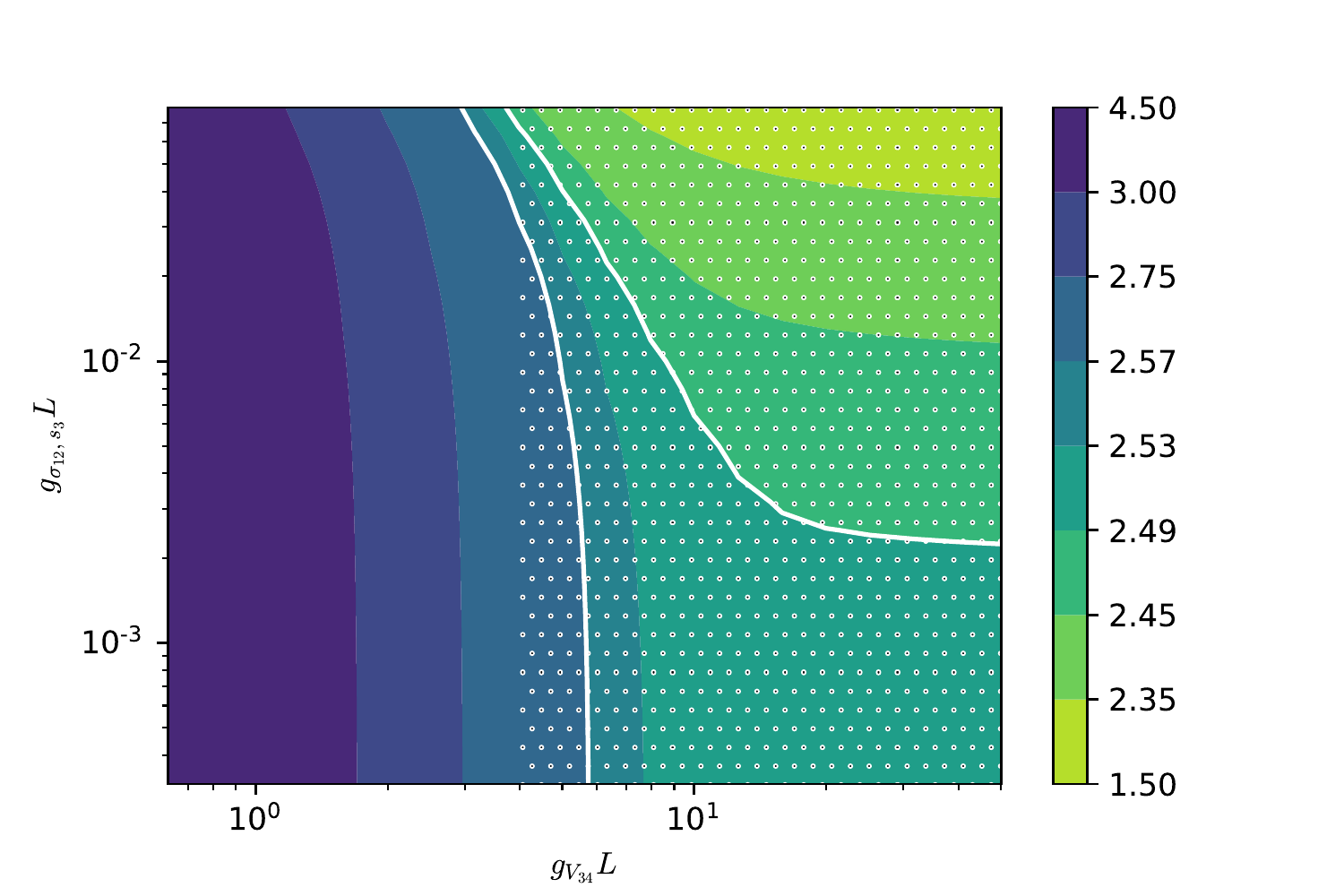}
    \caption{$g_{V_{\sigma_{12},s_2}}=g_{r,\tilde{s}}=0$}
  \end{subfigure}
  \begin{subfigure}[b]{.45\linewidth}
    \includegraphics[width=\textwidth]{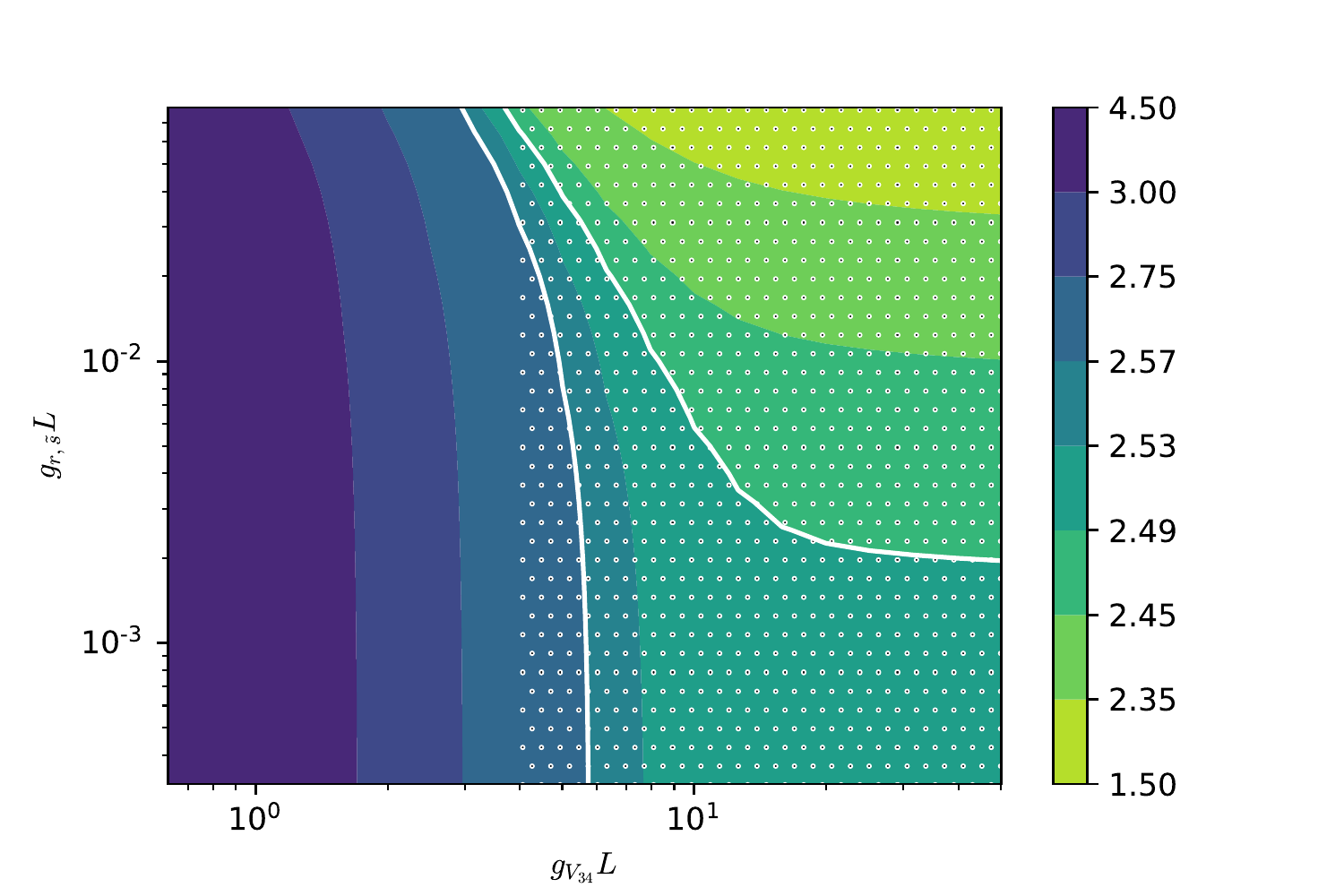}
    \caption{$g_{V_{\sigma_{12},s_2}}=g_{V_{\sigma_{12},s_{3}}}=0$}
  \end{subfigure}
  \caption{Contour plot of thermal conductance about the $\Delta_{12}=1$
    fixed point,
    $K(g_{V_{\sigma_{12},s_2}}L
    ,g_{V_{\sigma_{12},s_3}}L,g_{r,\tilde{s}}L,g_{V_{34}}L)/\kappa_0 T
    $ along several surfaces. $\Delta_{34}=5/3$ for all the sub-plots. 
    The regions within the white contour represent the measured
    thermal conductance $K=(2.53\pm 0.04)\kappa_0T$, while the hatched
    regions represent the regime where $G=(2.50 \pm 0.01) \sigma_0$.}
  \label{fig-D12-Ktot}
\end{figure}

The contour plot of
$K(g_{V_{\sigma_{12},s_2}}L,g_{V_{\sigma_{12},s_2}}L,g_{r,\tilde{s}}L,g_{V_{34}}L)
$ along several surfaces is given in Fig.~\ref{fig-D12-Ktot}. The
thermal conductance observed in the experiments (\cite{banerj-half})
at temperatures ($T\approx18\operatorname{-} 25\ mK$)
$2.49 \kappa_0T < K < 2.57 \kappa_0T $ is enclosed within the white
contours. The hatched region represents the regime where the
electrical conductance $G=(2.50\pm 0.01)\sigma_0$.

We observe that not all of the region observed in the experiment
$2.49 \kappa_0T < K < 2.75 \kappa_0T $ is consistent with the
electrical conductance $G=(2.50\pm 0.01)\sigma_0$: we find that when
$K\gtrsim 2.65 \kappa_0T$, the electrical conductance deviates from
$G=(2.50\pm 0.01)\sigma_0$.  In addition, we can deduce some
information about which point of the region
$2.49 \kappa_0T < K < 2.75 \kappa_0T $ we are at by examining how the
thermal conductance varies as a function of temperature.

The conductivity coefficients have power law dependence on temperature
as Eq.~\eqref{D12-gs}. Therefore, the thermal conductance moves along
straight lines in Fig.~\ref{fig-D12-Ktot}, as the temperature is
varied. From the experimental data, as the temperature is lowered from
$T\approx 18\operatorname{-}25\ mK $ to $T\approx 12\ mK$, i.e., by a
factor of about $2$, the thermal conductance increases from
$K\approx 2.53 \kappa_0T$ to $K\approx 2.75 \kappa_0T$.
It follows that $g_{34}$ would vary by a factor of $2^{(2\Delta_{34}-2)}$
while $g_{\sigma_{12},s_{2}}$ and $g_{\sigma_{12},s_3}$ would vary by
a factor of $4$. We can look for lines in the space of conductivity
coefficients 
where such a variation occurs.

\begin{figure}[h!]
  \centering
  \includegraphics[width=.6\textwidth]{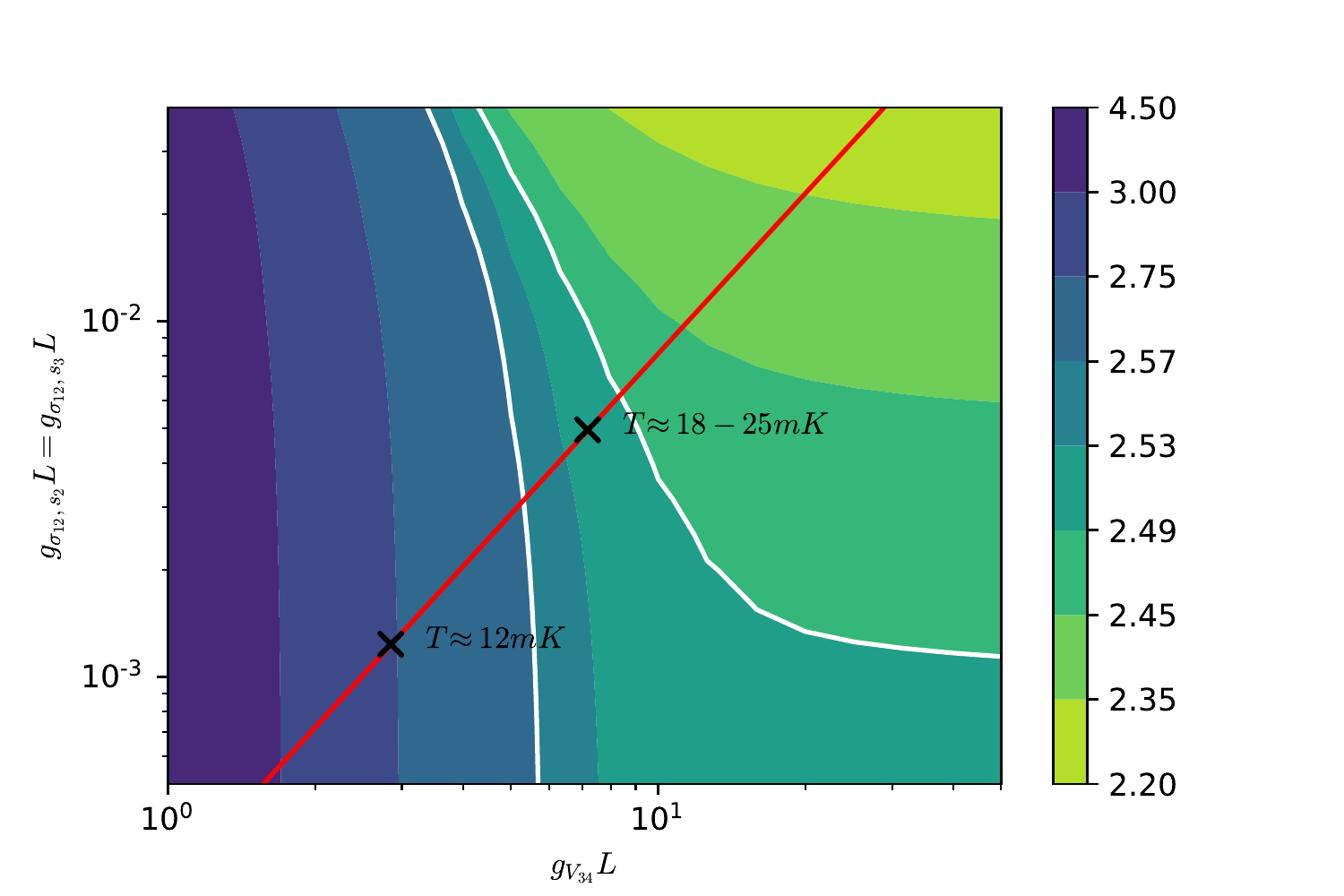}
  \caption{Thermal conductance about the $\Delta_{12}=1$ fixed point
    on the surface $g_{V_{\sigma_{12},s_2}}=g_{V_{\sigma_{12},s_{3}}}$
    and $g_{r,\tilde{s}}=0$. $\Delta_{34}=5/3$. The red line represents
    a typical line along which the thermal conductance varies as a
    function of temperature. This specific red line passes through
    points that are consistent with measurements of thermal
    conductance.
  }
  \label{fig-D12-Ktot-T}
\end{figure}

First, we look at how the thermal conductance varies along the surface
$g_{\sigma_{12},s_{2}}=g_{\sigma_{12},s_3}$ when
$g_{r,\tilde{s}}=0$. This is demonstrated in
Fig.~\ref{fig-D12-Ktot-T}.  The red line showcases a variation of
conductivity coefficients with temperature that is consistent with the
experiments: as the temperature is lowered by a factor of $\sim 2$,
between the cross marks, the thermal conductance increases from
$K\approx 2.53 \kappa_0T$ to $K\approx 2.75 \kappa_0T$.  This gives us
a rough estimate for the value of the conductivity coefficients at
these temperatures. Examining the red line in
Fig.~\ref{fig-D12-Ktot-T} for $T= 18 \operatorname{-}25\ mK$, we find
\begin{align}\label{eq-D12-g-estims}
  g_{V_{34}}L\approx 7 ,\quad g_{\sigma_{12},s_{2}/s_3} L\approx 0.005.
\end{align}
A similar picture also shows $g_{r,\tilde{s}}L\approx 0.005$. Here,
the thermal conductance does not vary much as a function of
$g_{\sigma_{12},s_2/s_3}$ and $g_{r,\tilde{s}}$ when these
coefficients are small. Consequently, the error in the estimate of
$g_{\sigma_{12},s_{2}/s_3}$ and $g_{r,\tilde{s}}$ is large and the
above estimates for $g_{\sigma_{12},s_{2}/s_3}$ and $g_{r,\tilde{s}}$
should be interpreted as upper bounds.

Based on these estimates, we infer
\begin{align}
  \frac{g_{r,\tilde{s}}}{g_{V_{34}}} = (\Delta_{34}-d_r)d_r \sim (\frac{v^{(0)}}{w})^2 \lesssim 0.001.
\end{align}
Since $d_r \sim (\frac{v^{(0)}}{w})^2$ the above bound is not
unexpected for strong short-ranged Coulomb interactions. Our numerical
estimates for $d_r$ based on the velocity matrix in Eq. \ref{V-model}
and sensible choice of $v^{(0)}_i$'s, do satisfy this bound for
$v^{(0)}_i$'s as large as $w/5$.

On the other hand, the coefficients $g_{\sigma_{12},s_2}$ and
$g_{\sigma_{12},s_3}$ in \eqref{D12-gs} are proportional to the square of
$v_{\sigma_{12},s_2}$ and $v_{\sigma_{12},s_3}$.  As we
demonstrated in Eq.~\eqref{eq-vsa-final}, these velocity entries vanish
in the spin-degenerate limit.
Therefore, it is not unexpected that the
bound $g_{\sigma_{12},s_2/s_3} L\lesssim 0.01$ is satisfied when
the spin gap is small. However, we don't have any estimate for these
conductivity coefficients based on the experimental data.

\begin{figure}[h!]
  \centering
  \begin{subfigure}[b]{.45\linewidth}
    \includegraphics[width=\textwidth]{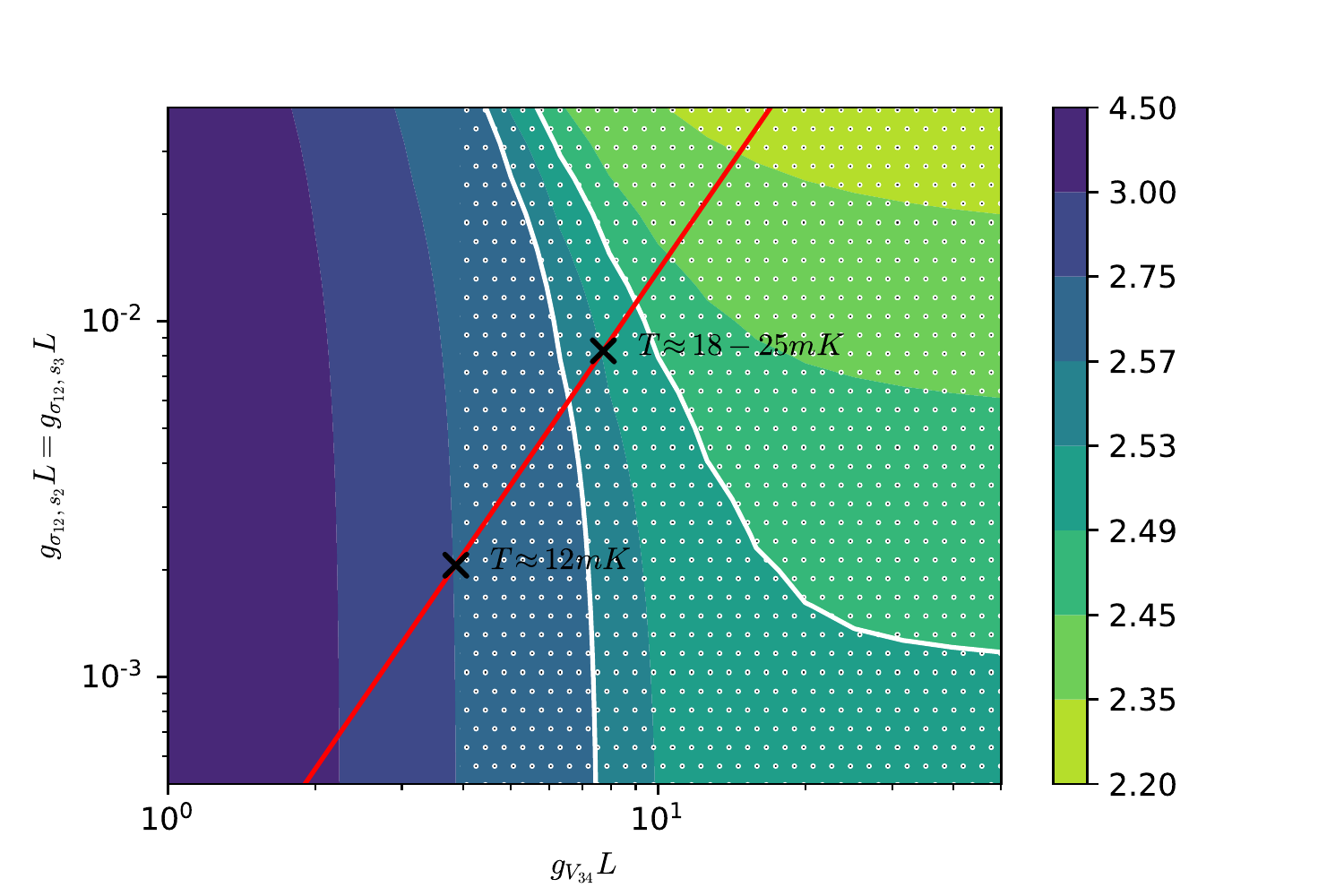}
    \caption{$\Delta_{34}=3/2$}
  \end{subfigure}
  \begin{subfigure}[b]{.45\linewidth}
    \includegraphics[width=\textwidth]{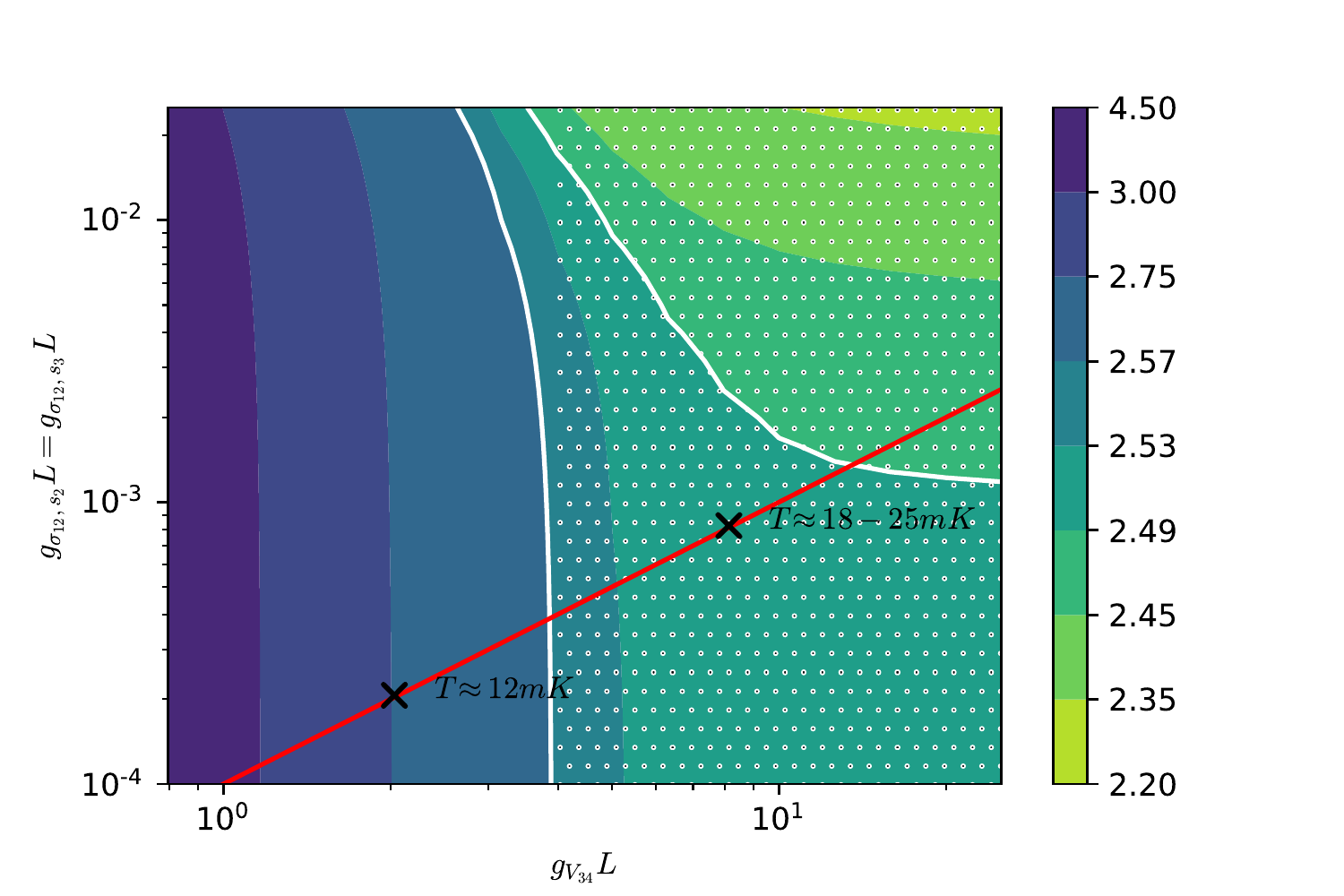}
    \caption{$\Delta_{34}=2$}
  \end{subfigure}
  \caption{Contour plot of thermal conductance about the $\Delta_{12}=1$
    fixed point,
    $K(g_{V_{\sigma_{12},s_2}}L
    ,g_{V_{\sigma_{12},s_3}}L,g_{r,\tilde{s}}L,g_{V_{34}}L)/\kappa_0 T
    $ along the surface
    $g_{V_{\sigma_{12},s_2}}=g_{V_{\sigma_{12},s_2}},g_{r,\tilde{s}}=0$.
    The hatched regions represent the regime where
    $G=(2.50 \pm 0.01) \sigma_0$.}
  \label{fig-D12-Ktot-T-Ds}
\end{figure}

In order to find these results, we used the estimate
$\Delta_{34}=5/3$. In order to see how much our results depend on this
estimate, we look at two other cases: \textit{i}) $\Delta_{34}=3/2$
and \textit{ii}) $\Delta_{34}=2$. For these two values, we plot
$K(g_{V_{\sigma_{12},s_2}}L,g_{V_{\sigma_{12},s_2}}L,g_{r,\tilde{s}}L=0,g_{V_{34}}L)$
along the $g_{V_{\sigma_{12},s_2}}=g_{V_{\sigma_{12},s_3}}$ surface in
Fig. \ref{fig-D12-Ktot-T-Ds}. First, we see that while the observation of
$G=(2.50\pm 0.01)\sigma_0$ is mostly consistent with $2.49 \kappa_0T \leq K
\leq 2.75 \kappa_0T$ for $\Delta_{34}=3/2$, this is not the case for $\Delta_{34}=2$:
in the region $2.57 \kappa_0T \leq K \leq 2.75 \kappa_0T$, the
electrical conductance deviates from
$G=(2.50\pm 0.01)\sigma_0$.
In addition, while for $\Delta_{34}=3/2$ the bounds on the
conductivity coefficients are close to the $\Delta_{34}=5/3$ case,
for $\Delta_{34}=2$ we get
\begin{align}
  g_{V_{34}}L\approx 10 ,\quad g_{\sigma_{12},s_{2}/s_3} L\lesssim 0.001,
\end{align}
which are much stronger bounds.

We conclude that there exists a regime of parameters about
the $\Delta_{12}=1$ fixed point of the anti-Pfaffian edge state where
$K\approx 2.5 \kappa_0 T$ is observed in a range of temperatures
($T\approx 18\operatorname{-}25\ mK$). Our
estimates demonstrate that this regime is possible for realistic
parameters only when $\Delta_{34}\lesssim 5/3$.

\subsection{$\Delta_{12}=\Delta_{34}=1$ fixed point}

\subsubsection{Charge transport}

At this fixed point, the processes that cause equilibration are the
irrelevant terms in \eqref{D12D34-finaction-irrel}. To find the
kinetic equations involving the second Landau level modes, we first
introduce the neutral currents operators
\begin{align}
  J_{34}^a\equiv \frac{i}{8\pi}\bm{\psi}^TL^a\bm{\psi}
\end{align}
where $L^a,a=x,y,z$ are the generators of $SO(3)$. In terms of these
operators we have
$\frac{1}{2\pi}\partial_x\phi_{\sigma_{34}}=J^z_{34} $. Using a
similar set of calculations as in section \ref{sec-rand-dd}, we derive the kinetic
equation for the gauge-transformed density
\begin{align}
  \tilde n_{\sigma_{34}}\equiv \frac{1}{2\pi}\partial_x\tilde \phi_{\sigma_{34}}\equiv \tilde
  J^z_{34} = \frac{i}{8\pi}\bm{\tilde \psi}^TL^z\bm{\tilde \psi}.
\end{align}
We also define the ``slow modes'' basis as
\begin{subequations}
  \begin{align}
    I_3' &=\sqrt{2} I_{\rho_{34}}- \tilde I_{\sigma_{34}} \\
    I'_4 &=-\frac{1}{\sqrt{2}} I_{\rho_{34}}+ \tilde I_{\sigma_{34}}
  \end{align}
\end{subequations}
where $I_{\rho_{34}}$ is the charge current carried by the mode
$\phi_{\rho_{34}}$ and the current neutral current
$\tilde I_{\sigma_{34}}$ is defined by the conservation equation
\begin{align}
  \partial_x\tilde I_{\sigma_{34}} + \partial_t\tilde n_{\sigma_{34}}=0.
\end{align}
It follows that for charge
equilibration in the basis $( I'_{1} , I'_2,
I'_3, I'_4)$ we have
\begin{align}
  G^{e}
  &=-(\sum_{\beta=r,s_2,\sigma_{34}}g_{V_{\sigma_{12}, \beta}})
    \begin{pmatrix}
      1 & -1 & 0 & 0\\
      -1 & 1 & 0 & 0 \\
      0 & 0 & 0 & 0 \\
      0 & 0 & 0& 0
    \end{pmatrix}
                 -(\sum_{\beta=r,\sigma_{12},s_{2}}g_{V_{\sigma_{34},\beta}}) \begin{pmatrix}
                   0 & 0 & 0 & 0\\
                   0 & 0 & 0 & 0 \\
                   0 & 0 & 1 & 2 \\
                   0 & 0 & -1& -2
                 \end{pmatrix}
\end{align}
with
\begin{subequations}\label{eq-D12D34-gs}
  \begin{align}
    g_{V_{\sigma_{12}, \sigma_{34}}}&=  \frac{2 \pi^2v_{\sigma_{12},\sigma_{34}}^2
                                      }{3(v_{\sigma_{12}}^2W_{34}+v_{\sigma_{34}}^2W_{12}) } T^2 \\
    \beta = r, s_2:
    g_{V_{\sigma_{12}, \beta}}&=  \frac{2\pi^2 v_{\sigma_{12},\beta}^2
                                }{3v_{\beta}^2W_{12}} T^2 \\
    \beta = r, s_2:
    g_{V_{\sigma_{34}, \beta}}&=  \frac{2\pi^2 v_{\sigma_{34},\beta}^2
                                T^2}{3v_{\beta}^2W_{34}}.
  \end{align}
\end{subequations}
We can calculate the electrical conductance as in the previous
section. The solution is similar to Eq.~\eqref{eq-D12-Gsol} with
$g_{V_{34}}$ replaced by
$\sum_{\beta=r,\sigma_{12},s_{2}}g_{V_{\sigma_{34},\beta}}$.  An
electrical conductance of $G=(2.50 \pm 0.01)\sigma_0$ implies
$\sum_{\beta=r,\sigma_{12},s_{2}}g_{V_{\sigma_{34},\beta}} L \gtrsim
4$. Looking at Eq.~\eqref{eq-D12D34-gs} we can estimate the relative
magnitude of the terms in
$\sum_{\beta=r,\sigma_{12},s_{2}}g_{V_{\sigma_{34},\beta}}$. We find
\begin{subequations}\label{eq-D1234-g-estim}
  \begin{align}
    \frac{ g_{V_{\sigma_{34},r}} }{ g_{V_{\sigma_{34},s_2}} } &=
                                                                (\frac{ v_{\sigma_{34},r}  v_{s_2}  }{v_{\sigma_{34},s_2}  v_r})^2
                                                                \sim (\frac{v^{(0)}}{\nu w})^2, \\
    \frac{ g_{V_{\sigma_{34},\sigma_{12}}} }{ g_{V_{\sigma_{34},s_2}} }
                                                              &\approx \frac{W_{34}}{W_{12}+W_{34}}.
                                                                (\frac{ v_{\sigma_{34},\sigma_{12}}  v_{s_2}  }{v_{\sigma_{34},s_2}  v_{\sigma_{12}}})^2
                                                                \sim \frac{W_{34}}{W_{12}+W_{34}}.(\frac{v_{\sigma_{34},\sigma_{12}}}{v^{(0)}})^{2}.
  \end{align}
\end{subequations}
Therefore, both $g_{V_{\sigma_{34},\sigma_{12}}}$ and
$g_{V_{\sigma_{34},r}}$ are much smaller than
$g_{V_{\sigma_{34},s_2}}$ for strong Coulomb interactions and small
spin gap, and so we have $g_{V_{\sigma_{34}, s_2} }L \gtrsim 1$ based
on quantization of the electrical conductance.  In the above we used
the estimate that
$v_{s_2},v_{\sigma_{12}},v_{\sigma_{34},r},v_{\sigma_{34},s_2}$ all
have the same order of magnitude $v^{(0)}$. Also, based on the
velocity matrix of Eq. \ref{V-model} and using Eq. \ref{eq-vsa-final}
we should have $v_{\sigma_{34},\sigma_{12}}=0$. However, since we only
take this velocity matrix as an estimation, we allow for finite
$v_{\sigma_{34},\sigma_{12}}$ which vanishes in the spin-symmetric
limit.

\subsubsection{Heat transport}
At this fixed point, since there exists an $SO(3)$
symmetry between the three Majorana modes, we take their
contribution as one upstream mode with central charge
$c=\frac{3}{2}$. We call this mode $\Psi$. Therefore, in the basis
$(r,\sigma_{12},s_2,\Psi)$ we have
\begin{align}
  G^Q=& \frac{12}{5}g_{V_{\sigma_{12}, r}}
        \begin{pmatrix}
          -1 & 1 & 0 & 0 \\
          1 & -1  & 0 & 0   \\
          0 & 0 & 0 & 0  \\
          0 & 0 & 0 & 0
        \end{pmatrix}
                      +\frac{12}{5}g_{V_{\sigma_{12}, s_2}}
                      \begin{pmatrix}
                        0 & 0 & 0 & 0 \\
                        0 & -1  & 1 & 0   \\
                        0 & 1 & -1 & 0  \\
                        0 & 0 & 0 & 0
                      \end{pmatrix}
                                    +\frac{12}{5}g_{V_{\sigma_{12} ,\sigma_{34}}}
                                    \begin{pmatrix}
                                      0 & 0 & 0 & 0 \\
                                      0 & -1  & 0 & -2/3   \\
                                      0 & 0 & 0 & 0  \\
                                      0 & 1 & 0 & 2/3
                                    \end{pmatrix}\\
  \nonumber
      &
        +\frac{12}{5}g_{V_{\sigma_{34}, r}}
        \begin{pmatrix}
          -1 & 0 & 0 & -2/3 \\
          0 & 0  & 0 & 0   \\
          0 & 0 & 0 & 0  \\
          1 & 0 & 0 & 2/3
        \end{pmatrix}
                      +\frac{12}{5}g_{V_{\sigma_{34}, s_2}}
                      \begin{pmatrix}
                        0 & 0 & 0 & 0 \\
                        0 & 0  & 0 & 0   \\
                        0 & 0 & -1 & -2/3  \\
                        0 & 0 & 1 & 2/3
                      \end{pmatrix}.
\end{align}

\begin{figure}[h!]
  \centering
  \includegraphics[width=.6\textwidth]{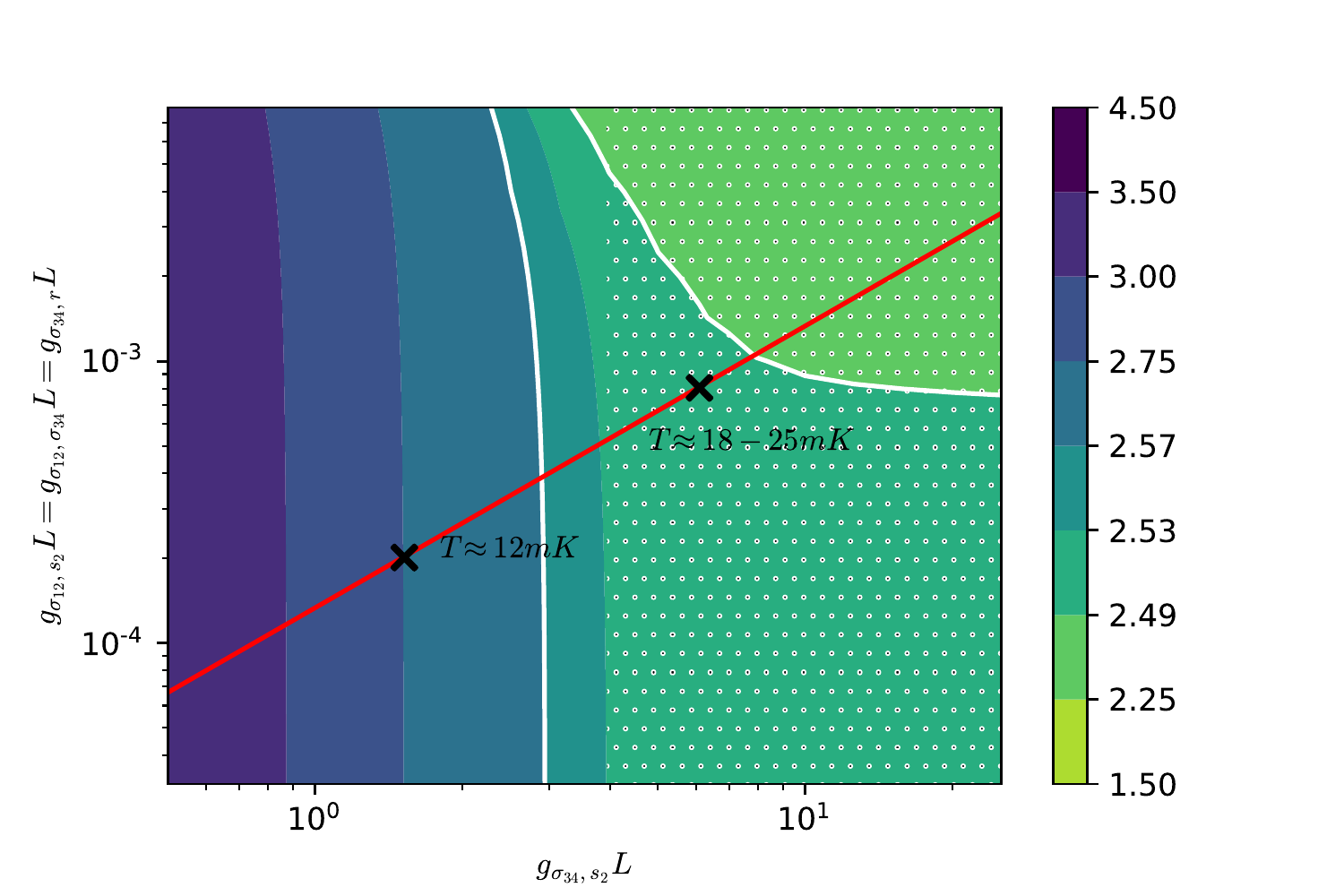}
  \caption{ Thermal conductance about the $\Delta_{12}=\Delta_{34}=1$
    fixed point, along the surface
    $g_{V_{\sigma_{12},s_2}}=g_{V_{\sigma_{12},\sigma_{34}}}=g_{V_{\sigma_{34},r}}$.
    The region within the white contour has
    $2.49\kappa_0T < K < 2.57 \kappa_0T$, while the hatched region
    has electrical conductance $G=(2.50 \pm 0.01)\sigma_0$. The
    thermal conductance varies along lines parallel to the red line
    as the temperature is varied.  }
  \label{fig-D12D34-Ktot-T}
\end{figure}

Since $g_{V_{\sigma_{34},s_2}}L \gtrsim 1$, the modes $s_2$ and $\Psi$
are expected to be well equilibrated.  Therefore, similar to the
$\Delta_{12}=1$ fixed point, the thermal conductance
$K\approx 2.5 \kappa_0T$ is only possible when equilibration between
the modes $\{r,\sigma_{12}\}$ and $\{s_2,\Psi \}$ is suppressed.  In
order to look for such a regime, we solve the heat transport equation
using the above $G^Q$ matrix, and calculate the thermal conductance as
a function of
$g_{V_{\sigma_{12},s_2}},g_{V_{\sigma_{12},\sigma_{34}}},g_{V_{\sigma_{34},r}}$
and $g_{V_{\sigma_{34},r}}$. As before, we ignore the first term in
$G^Q$. Fig.~\ref{fig-D12D34-Ktot-T} shows the contour plot of the
thermal conductance along the surface
$g_{V_{\sigma_{12},s_2}}=g_{V_{\sigma_{12},\sigma_{34}}}=g_{V_{\sigma_{34},r}}$. The
region within the white contour has
$2.49\kappa_0T < K < 2.57 \kappa_0T$, while the hatched region has
electrical conductance $G=(2.50 \pm 0.01)\sigma_0$.

Here, unlike the $\Delta_{12}=1$ fixed point, there exists
a region where $2.49\kappa_0T < K < 2.75 \kappa_0T$ while the
electrical conductance deviates from $G=(2.50 \pm 0.01)\sigma_0$. If,
the electrical conductance is indeed measured to be
$G=(2.50 \pm 0.01)\sigma_0$, even at lowest temperatures $\sim 12 mK$, then
this fixed point is not consistent with the experiments.

We proceed to find estimates for the conductivity coefficients based
on how the thermal conductance varies with temperature. Based on
Fig.~\ref{fig-D12D34-Ktot-T} and following an analysis similar to the
$\Delta_{12}=1$ fixed point, we estimate
\begin{align}
  g_{V_{\sigma_{34},s_2}}L \approx 6, \quad
  g_{V_{\sigma_{12},s_{2}}}L,g_{V_{\sigma_{12},\sigma_{34}}}L,g_{V_{\sigma_{34},r}}L
  \lesssim 10^{-3}
\end{align}
for $T= 18 \operatorname{-}25\ mK$.
Therefore, using Eq.~\eqref{eq-D1234-g-estim},
we require
\begin{subequations}
  \begin{align}\label{eq-D12D34-bounds}
    \frac{ g_{V_{\sigma_{34},r}} }{ g_{V_{\sigma_{34},s_2}} }
    &\sim (\frac{v^{(0)}}{\nu w})^2 \lesssim 2\times 10^{-4},  \\
    \frac{ g_{V_{\sigma_{34},\sigma_{12}}} }{ g_{V_{\sigma_{34},s_2}} }
    &
      \sim
      \frac{W_{34}}{W_{12}+W_{34}}.(\frac{v_{\sigma_{34},\sigma_{12}}}{v^{(0)}})^{2} \lesssim 2\times 10^{-4}, \\
    \frac{ g_{V_{\sigma_{12}, s_2}} }{ g_{V_{\sigma_{34},s_2}} }
    &\sim \frac{W_{34}}{W_{12}}.(\frac{v^{(0)}_1-v^{(0)}_2}{v^{(0)}})^{2} \lesssim 2\times 10^{-4}.
  \end{align}
\end{subequations}
Generally, we expect the conductivity coefficients
$g_{\sigma_{34},r},g_{V_{\sigma_{12}, s_2}}$ and
$g_{V_{\sigma_{12}, \sigma_{34}}}$ to be much smaller than
$g_{\sigma_{34},s_2}$ for strong short-ranged Coulomb interaction
($w \gg v^{(0)}$) and small spin-gap
($v^{(0)}_1-v^{(0)}_2 \ll v^{(0)},v_{\sigma_{34},\sigma_{12}} \ll
v^{(0)}$). However, our estimates for $v^{(0)}/w$ (see Section
\ref{D12-section}) and $(v^{(0)}_1-v^{(0)}_2) / v^{(0)}$ in
Eq.~\eqref{eq-dv0-estim}) only show ratios of about $10^{-1}$.
Therefore, we are not aware of any reason why the bounds in
Eq.~\ref{eq-D12D34-bounds} might be satisfied.

We conclude that the $\Delta_{12}=\Delta_{34}=1$ fixed point of the
anti-Pfaffian state is not consistent with the transport measurements.
This theory predicts that the electrical conductance would deviate
from its quantized value $G=2.5 \sigma_0$ at temperatures
$T \approx 12\ mK$, a feature that does not appear to be observed in
the experiments of Banerjee \textit{et. al.}\cite{banerj-half}.
Furthermore, observation of thermal conductance
$K \approx 2.5 \kappa_0 T$ requires some parameters in this theory
($v^{(0)}/w$ and $(v^{(0)}_1-v^{(0)}_2) / v^{(0)}$) to be fine tuned;
we don't believe such a regime to be realistic.

\section{Quantum point contact tunneling}\label{sec-QPC-tunn}

Tunneling conductance at quantum point contacts (QPC) in the ohmic
regime ($eV\ll k_BT$) scales as $G_{\text{tun}}\sim T^{2g-2}$. Here,
$g$ is the scaling dimension of the tunneling operator that transfers
charge across the Hall bar. Therefore at low temperatures, charge
tunneling is dominated by the operator with the smallest scaling
dimension.  In the case of the anti-Pfaffian state, due to the
physical separation between the lowest and the first Landau level edge modes, this tunneling is dominated by the tunneling of electrons/quasi-particles belonging to the first
Landau level. The most general tunneling operator is then
$\expon{ i(n_3\phi_3+n_4\phi_4/2) }\chi$ where $n_3$ and $n_4$ are
integers and $\chi=1,\psi,\sigma$ \cite{lee-2007,levin-2007}. This
tunneling operator creates an excitation of charge
$q=(n_4/4-n_3)e$. The operator $\sigma$ changes the boundary condition
for the Majorana mode $\psi$ and has scaling dimension
$\Delta_\sigma=1/16$. In addition, $n_4$ is an odd integer when
$\chi=\sigma$.

At the $\Delta_{12}=\Delta_{34}=1$ fixed point, the
charge creation operator with the smallest scaling dimension is
$\sigma e^{i\phi_4/2}$ \cite{lee-2007,levin-2007}, which creates a
quasi-particle of charge $e/4$. A similar operator annihilates this
quasi-particle across the quantum Hall bar. So
\begin{align}
  g=2\Delta(\sigma e^{i\phi_4/2}) =2\Delta_{\sigma}+2\Delta(\expon{i\phi_4/2}) =1/2
\end{align}
where we denote by $\Delta(\mathcal{O})$ the scaling dimension of
operator $\mathcal{O}$.

For the $\Delta_{12}=1$ fixed point, the scaling dimension of the operator
$\expon{ i(n_3\phi_3+n_4\phi_4/2) }$ depends on the velocity matrix in
$S_{\Delta_{12}=1}$ \ref{D12-finaction}, and therefore is
non-universal. In general, the minimum scaling dimension of a vertex
operator is the absolute value of its spin, i.e.,
$\Delta_R+\Delta_L\geq |\Delta_R-\Delta_L|$. See
Eq. \eqref{eq-spin-def}. Therefore, one can check that among all  excitation
operators, $\sigma \expon{i\phi_4/2}$ has the minimum scaling
dimension of $1/8$. Therefore we always have $g\geq 1/4$ for the
anti-Pfaffian state.

We can get a better bound in the limit of strong short-ranged Coulomb
interaction. Using \eqref{chaneut-basis} we can write
\begin{align}
  \expon{ i(n_3\phi_3+n_4\phi_4/2) } = \expon{ i
  \sqrt{\frac{2}{5}}(n_3-n_4/4)\phi_{\rho} }\expon{ -in_3
  \sqrt{\frac{2}{3}}\phi_{\sigma_2}
  -\frac{i}{\sqrt{15}}(n_3+3n_4/2 ) \phi_{\sigma_{3}} }.
\end{align}
Similar to Eq. \eqref{eq-samev0-H}, in the vanishing $v^{(0)}/w$ limit,
$\phi_{\rho}$ is a diagonal mode of $S_{\Delta_{12}=1}$. Therefore in
this limit:
\begin{align}
  \Delta( \expon{ i(n_3\phi_3+n_4\phi_4/2) } )
  \label{eq-comp-1}  =& \Delta( \expon{ i \sqrt{\frac{2}{5}}(n_3-n_4/4)\phi_{\rho} } ) +
                        \Delta( \expon{ -in_3 \sqrt{\frac{2}{3}}\phi_{\sigma_2}
                        -\frac{i}{\sqrt{15}}(n_3+3n_4/2 ) \phi_{\sigma_{3}} }  )\\
                      & \leq \frac{1}{5} (n_3-n_4/4)^2 + \left| \frac{1}{3}n_3^2 -
                        \frac{1}{30}(n_3+3n_4/2 )^2 \right|.
\end{align}
Using this inequality, we can check that the minimum scaling
dimension is $3/20$ for the operator $\sigma e^{i\phi_4/2}$. The next
smallest scaling dimension is $7/20$ for the operator
$e^{i\phi_4}$ which creates an excitation of charge $e/2$.
Therefore, for strong Coulomb
interactions we have $g\geq 3/10$ with the minimum happening for the
operator $\sigma e^{i\phi_4/2}$. Note that this estimate is
independent of the fact that $\Delta_{12}=1$. Therefore, this bound is
also valid for the clean fixed point description of the anti-Pfaffian
edge theory.

Experimental measurements of $g$ give values $g=0.34-0.42$
\cite{radu-2008,lin-tun-2012}, depending on the geometry of the
quantum point contact. So, the fixed points about which the tunneling
term $\expon{i(\phi_3+2\phi_4) }\psi$ is irrelevant can be consistent
with the measured tunneling exponents. These fixed points are realized
only when the short-ranged Coulomb interactions between the Landau
levels is included. This is because, if such interactions are ignored,
the tunneling term $\expon{i(\phi_3+2\phi_4) }\psi$ is always relevant
due to the strong Coulomb interaction within the second Landau level.

\section{Discussion}

We considered equilibration of charge and heat along the edge of the
anti-Pfaffian state realized in the first Landau level at $\nu=5/2$.
We assumed that the
dominant cause of equilibration is due to short-ranged disorder that
allows tunneling of charge between the different edge modes. While tunneling
between edge modes belonging to different Landau levels is ignored in our analysis, a
strong short-ranged Coulomb interaction is assumed. Under these assumptions, we analyzed the conditions under which the
edge modes are not fully in equilibrium.

In the limit of a strong short-ranged Coulomb interaction,
equilibration between the total charge mode and the rest of the edge modes
is suppressed due to the high velocity of the charge mode relative to the
neutral modes. This picture was also considered by Ma and Feldman
in \cite{ma-feldman-2019}.

In the absence of Zeeman splitting between the two
modes in the lowest Landau level, their total spin is independently
conserved. Consequently, heat equilibration between the spin mode
and other modes is suppressed. For finite Zeeman splitting, electron
tunneling between these two modes can drive the edge into the
spin-symmetric fixed point where the spin mode is conserved. At finite
temperature, the irrelevant interactions present due to the spin
asymmetry can bring this spin mode to equilibrium with the other edge modes. For
small enough spin asymmetry, this equilibration processes can be slow
on the length scales of the system size.

Due to these weak equilibration processes, the thermal conductance is
given by
$K=K_{\phi_{\rho}}+K_{\phi_{\sigma_{12}}}+K_{\text{other
    modes}}$, where the nature of  the ``other modes'' depends on the
specific fixed point. Based on the quantization of electrical conductance,
we infer that the ``other modes'' should be in equilibrium with each other. So
$K = (1+1+|-1.5|)\kappa_0T=2.5 \kappa_0T$. This picture relies on the partial equilibration of the fixed points $\Delta_{12}=1$ and
$\Delta_{12}=\Delta_{34}=1$ studied here.
For both of these fixed points,
electron tunneling between the spin-up and spin-down modes
(\textit{i.e.} $\expon{i(\phi_1-\phi_2)}$) drives the edge into a
spin-symmetric fixed point. In contrast, other fixed point theories
where such electron tunnelings are weak do not have such an emergent
symmetry.
However, if the spin asymmetry is small, the spin density
$ \partial_x\phi_{\sigma_{12}}$ is almost conserved and its
equilibration with other modes is suppressed. This situation was
discussed in \cite{ma-feldman-2019} for the $\Delta_{34}=1$ fixed
point.

Therefore, suppressed equilibration of the total charge mode
$\phi_{\rho}$ and the spin mode $\phi_{\sigma_{12}}$ can be realized
for all of the four fixed points mentioned in Ssection
\ref{antipf-setup-section}.  The difference is in the details of the
equilibration process, e.g., the parametric dependence of the
conductivity coefficients and their temperature dependence. We
demonstrated this for the two fixed points: $\Delta_{12}=1$ and
$\Delta_{12}=\Delta_{34}=1$. In light of the existing experimental
data, these two fixed point theories differ in two important ways:
\begin{itemize}
\item About the $\Delta_{12}=\Delta_{34}=1$ fixed point, the
  electrical conductance $G=(2.50\pm 0.01)\sigma_{0}$ and the thermal
  conductance $2.49 \kappa_0T<K<2.75 \kappa_0T$ cannot be
  observed simultaneously. In contrast, these measurements can be
  consistent with the $\Delta_{12}=1$ fixed point when
  $\Delta_{34}\approx 3/2$.
\item About the $\Delta_{12}=\Delta_{34}=1$ fixed point, the range of parameters
  required to have $K\approx 2.5 \kappa_0T$ is not
  compatible with our estimate of these parameters. On the other
  hand, at the $\Delta_{12}=1$ fixed point, there exists a realistic
  regime of parameters (as far as our estimates permit) that results
  in $K\approx 2.5 \kappa_0T$. This regime is possible only when $\Delta_{34}$ is
  small enough $\Delta_{34}\lesssim 5/3$.
\end{itemize}
Therefore, the $\Delta_{12}=1$ fixed point theory of the anti-Pfaffian
state better describes the recent transport measurements
\cite{banerj-half}. About this fixed point the quantum point
contact tunnelings exponents depend on the inter-mode Coulomb
interactions and are, therefore, non-universal.  Nevertheless, the
predictions of this fixed point appear to be consistent with the
existing experimental quantum point contact measurements.
We should point out a limitation in comparing our results with the
experiment: in order to calculate the thermal conductance, we assumed
the temperature difference between the edge modes is small. However,
in the measurements carried out by Banerjee
\textit{et.al.}\cite{banerj-half}, the temperature difference is about
the same order as the average temperature.

From our analysis of the $\Delta_{12}=1$ fixed point, we make the
following predictions for temperatures not reported in \cite{banerj-half}:
\begin{itemize}
\item Based on Fig. \ref{fig-D12-Ktot-T-Ds}, even for the lowest value of
  $\Delta_{34}=3/2$, the electrical conductance would deviate from
  $G=(2.50\pm 0.01)\kappa_0T$ for temperatures lower than $T \approx
  12mK $.
\item Generally at higher temperatures, equilibration between the edge
  modes is improved. Therefore, if the state observed in the
  experiments by Banerjee et al.~\cite{banerj-half} is indeed the
  anti-Pfaffian state, the thermal Hall conductance would decrease
  below $K\approx 2.5 \kappa_0T$ at higher temperatures. Using
  Fig. \ref{fig-D12-Ktot-T}, we can estimate how much of a temperature
  increase is needed in order to observe a measurable decrease from
  the value
  $K\approx 2.5 \kappa_0T$ (\textit{i.e.} to
  $K\approx 2.45 \kappa_0T$): We find the temperature has to increase
  from $T\approx 18-25\ mK$ by at least a factor of $\sim 1.5$, i.e.,
  to $T\approx 35\ mK$.
\end{itemize}

\section*{Acknowledgments}

We thank Dima Feldman, Bert Halperin, Zlatko Papic, and Kirill Shtengel for helpful discussions.
This material is based upon work supported by the U.S. Department of Energy, Office of Science, Office of Basic Energy Sciences under Award Number DE-SC0020007.
This research was supported in part by the National Science Foundation under Grant No.~NSF PHY-1748958.

\appendix

\section{Effective theory of $\Delta_{12}=1$ fixed point}\label{D12-apndx}

After changing variables to the charge mode $\phi_{\rho_{12}}=\frac{1}{\sqrt{2}}
(\phi_1+\phi_{2})$ and the neutral mode $\phi_{\sigma_{12}}=\frac{1}{\sqrt{2}}
(\phi_1-\phi_{2})$ \cite{KF-imp} we have  (we will not write
expressions already defined in \ref{antipf-action})
\begin{subequations}
  \begin{align}
    S &=\sum_{i=\sigma_{12},\rho_{12},3,4 } S_i + \sum_{i,j \in \{
        \sigma_{12},\rho_{12},3,4 \},i\neq j } S_{ij} +S_{\psi}
        +S_{\text{tunneling},34\psi}
    \\
    S_{\rho_{12}} &=
                    -\frac{1}{4\pi}\int_{t,x} \left[ \partial_x
                    \phi_{\rho_{12}}(\partial_t\phi_{\rho_{12}}+v_{\rho_{12}}\partial_x\phi_{\rho_{12}}\right]
    \\ S_{\sigma_{12}} &= -\frac{1}{4\pi}\int_{t,x} \left[ \partial_x
                         \phi_{\sigma_{12}}(\partial_t\phi_{\sigma_{12}}+v_{\sigma_{12}}\partial_x\phi_{\sigma_{12}}\right]
                         + \int_{t,x} \left[\xi_{12}(x) \expon{i\sqrt{ 2} \phi_{\sigma_{12}}}+
                         \text{h.c.} \right] \\
    S_{\rho_{12}, \sigma_{12}} &=
                                 -\frac{2v_{\rho_{12}, \sigma_{12}}}{4\pi}\int_{t,x} \ \partial_x
                                 \phi_{\rho_{12}}
                                 \partial_x\phi_{\sigma_{12}} \\
    i=3,4: S_{\rho_{12},
    i} &= -\frac{2v_{\rho_{12}, i}}{4\pi}\int_{t,x} \ \partial_x
         \phi_{\rho_{12}} \partial_x\phi_{i} \\
    i=3,4: S_{\sigma_{12}, i} &=
                                -\frac{2v_{\sigma_{12}, i}}{4\pi}\int_{t,x} \ \partial_x
                                \phi_{\sigma_{12}} \partial_x\phi_{i}.
  \end{align}
\end{subequations}
When
$v_{\rho_{12} ,\sigma_{12}}=v_{\sigma_{12}, 3}=v_{\sigma_{12},
  4}=W_{34}=0$, $S_{\sigma_{12}}$ has an $SO(3)$
symmetry \cite{KF-imp,mirlin,NAUD2000572}.  To see this, let's define current
operators ($a$ is the short-distance cutoff)
\begin{subequations}\label{current-defs}
  \begin{align}
    J^x &= \frac{1}{2\pi a} \cos(\sqrt{2}\phi_{\sigma_{12}}) \\
    J^y &= \frac{1}{2\pi a} \sin(\sqrt{2}\phi_{\sigma_{12}}) \\
    J^z &= \frac{1}{2\pi \sqrt{2}}\partial_x
          \phi_{\sigma_{12}}.
  \end{align}
\end{subequations}
These operators satisfy a $\mathfrak{su}(2)_1$
current algebra
\begin{equation}
  \label{e4.5} \left[J^{a}(x),
    J^b(x^\prime)\right]=-\frac{i}{4\pi}\eta_{\sigma_{12}}\delta^{ab}\partial_x\delta(x-x^\prime)+i\epsilon^{abc}J^c(x)\delta(x-x^\prime)
\end{equation}
which is preserved under the $SO(3)$ gauge transformation
\begin{align}\label{gauge}
  J^{a}(x)&=O^{ab}(x) \tilde{J}^{b}(x)+h^a(x) , \quad h^{c}(x)=\frac{1}{8\pi}\epsilon^{abc}(O(x)\partial_x O^T)^{ab}.
\end{align}

In terms of these currents, the Hamiltonian of the neutral field is  (we
restore the $\frac{1}{2\pi a}$ coefficient of the tunneling term)
\begin{align}
  H_{\sigma_{12}} &= \int \di x \left[  \frac{2\pi v_{\sigma_{12}}}{3}J^2 + 2\xi^a J^a
                    \right],\quad J^2 = (J^x)^2 + (J^y)^2 + (J^z)^2.
\end{align}
$H_{\sigma_{12}}$ is invariant (up to inconsequential additive constants) under the gauge transformation \ref{gauge} provided the
disorder transforms as
\begin{align}\label{disorder-transf}
  \xi^a(x)\rightarrow \tilde{\xi}^a(x) = \left(\xi^b(x)+\frac{2\pi
  v_{\sigma_{12}}}{3}h^b\right) O^{ba}.
\end{align}

We require $\tilde{\xi}(x)=0$ in order to eliminate the tunneling
term from $H_{\sigma_{12}}$.
This amounts to a specific choice of $O^{ab}$.
After which we express the currents $\tilde{J}^a$ in terms of a new
bosonic field $\tilde{\phi}_{\sigma_{12}}$, similar to \ref{current-defs},
and write $H_{\sigma_{12}}$ as
\begin{align}
  H_{\sigma_{12}} =  \int \di x  \frac{
  v_{\sigma_{12}}}{4\pi} (\partial_x\tilde{\phi})^2.
\end{align}
The resulting action is
\begin{subequations}
  \begin{align}
    S &= \sum_{i=\sigma_{12},\rho_{12},3,4}S_i
        +\sum_{i,j\in \{\sigma_{12}, \rho_{12},3,4\},i\neq j}S_{i j}+S_{\psi} +
        S_{\text{tunneling},34\psi} \\
    S_{\sigma_{12}} &= -\frac{1}{4\pi}\int_{t,x} \left[
                      \partial_x
                      \tilde{\phi}_{\sigma_{12}}(\partial_t\tilde{\phi}_{\sigma_{12}}+v_{\sigma_{12}}\partial_x\tilde{\phi}_{\sigma_{12}}\right]\\
    S_{\sigma_{12},i} &= -\frac{2v_{\sigma_{12}, i}}{4\pi}\int_{t,x} \ \partial_x \phi_{i} \left(
                        \frac{\sqrt{2}}{a}O^{zx}\cos(\sqrt{2}
                        \tilde{\phi}_{\sigma_{12}})+\frac{\sqrt{2}}{a}O^{zy}\sin(\sqrt{2}
                        \tilde{\phi}_{\sigma_{12}})+O^{zz}\partial_x\tilde{\phi}_{\sigma_{12}}
                        \right),
  \end{align}
\end{subequations}
for $ i= \rho_{12},3,4$.
Here, we also used the following transformation in order to eliminate
the terms proportional to $h^z(x)$
\begin{align}
  \phi_{i}(x,t)\rightarrow
  \phi_{i}(x,t) +2\sqrt{2}\pi
  \frac{v_{\sigma_{12},i}}{v_{i}}\int_{-\infty}^x \di x \ h^z(x)
\end{align}
for $i=\rho_{12},3,4$.

\section{Derivation of conductivity coefficients}

\subsection{Electrical conductivity coefficient}\label{elect-cond-apndx}

We want to compute tunneling between a set of
chiral modes described by the free field Hamiltonian $H_{F}=\sum_{\alpha}H_{\alpha}$,
due to interactions of the form
\begin{align}
  V=\int \di x \ \xi(x)
  \prod_{\alpha} X_{\alpha}(x)+\text{h.c.}
\end{align}
in the presence of a
chemical potential bias
\begin{align}
  H_{\mu}=-\int \di x \ \sum_{\alpha} \mu_{\alpha}
  n_{\alpha}(x),\qquad
  n_{\alpha}=\frac{1}{2\pi}\partial_x\phi_{\alpha}.
\end{align}
The bosonic fields $\phi_{\alpha}$ satisfy the commutation relations
$[\phi_{\alpha}(x),\phi_{\beta}(x'')]= \delta_{\alpha \beta}\pi i
\frac{\eta_{\alpha}}{k_{\alpha}}\operatorname{sign}(x-x')$.  Chiral
fermions will be described by chiral bosons.  Here $X_{\alpha}$ is
only a function of $\phi_\alpha$ and $\xi(x)$ is Gaussian-correlated
disorder satisfying $\overline{\xi(x)\xi(x')}=W_V\delta(x-x')$.  The
continuity equation for each number current $I_\alpha$ is
\begin{align}
  -\partial_x I_{\alpha}(x,t)&= \partial_tn_{\alpha}(x,t)=
                               i[H,n_{\alpha}(x)](t).
\end{align}
For the Hamiltonian $H=H_{F}+H_{\mu}+V$
\begin{align}
  -\partial_x I_{\alpha}(t)
  &= -\eta_{\alpha}v_{\alpha}\partial_xn_{\alpha}(x,t)+ i\int \di x' \
    \xi(x') [X_{\alpha}(x'),n_{\alpha}(x)]\prod_{\beta \neq
    \alpha}X_{\alpha}(x')+\text{h.c.}\ .
\end{align}
This equation should be understood as the continuous limit of a series
of point contact tunnelings \cite{KF-contacts,chamon-fradkin}.
Different tunnelings are assumed incoherent so that each mode comes to
local equilibrium between consecutive tunnelings. It follows that
$n_{\alpha} =
\frac{1}{2\pi}\frac{1}{k_{\alpha}v_{\alpha}}\mu_{\alpha}$
so that we drop the $\partial_xn_{\alpha}$ term.

We calculate the expectation value of $\partial_x I_{\alpha}$ using
the Keldysh technique
\begin{align}
  \partial_x \expval{ I_{\alpha}(x,t)} &= \frac{1}{2}\sum_{\sigma_{12}}\expval{T_C \
                                         \partial_x I_{i,H_0}(x,t,s)\expon{i \sum_{s'}s' \int \di t'
                                         V(t',s')_{H_0}}}, \\
  H_0 &\equiv H_{F}+H_{\mu},
\end{align}
where $T_C$ indicates ``time" ordering along the Keldysh contour.
Expanding the exponential to first order in $\xi$ and taking
disorder average
\begin{align}\label{eq-eleckin-genform}
  \partial_x\expval{ I_{\alpha}(x,t)}
  &= \frac{i}{2}\sum_{s s'}s' \int \di t' \expval{T_C \
    \partial_x I_{\alpha}(x,t,s)_{H_{F}}  V(t',s')_{H_0} } \\
  \nonumber  &= \frac{1}{2} W_V \sum_{s s'}s' \int \di x' \int \di t'
               \expval{T_C \
               [X_{\alpha}(x'),n_{\alpha}(x)](t,s) X_{\alpha}^{\dagger}(x',t',s')} \cr
             &  \hspace{10em} \times \prod_{\beta \neq \alpha}
               \expval{ T_C\  X_{\beta}(x',t,s)
               X_{\beta}^{\dagger}(x',t',s')} +\text{h.c}. \ .
\end{align}
We look at two cases separately.

\subsubsection{Random tunneling}

Operators that tunnel electrons/quasiparticles between edge channels
of a fractional quantum Hall state have the form
$\expon{i\sum_im_i\phi_i}$, where $\phi_i$ with Latin index represents
a chiral boson mode carrying charge $\nu_i$ and chirality $\eta_i$
with commutation relation
$[\phi_i(x),\phi_j(x')]= \delta_{ij}\pi i \eta_i
\nu_i\operatorname{sign}(x-x')$. This term also has a coefficient
$\frac{1}{(2\pi a)^{N_e}}$, with $a$ the UV distance cut-off and $N_e$
the number of electrons transferred, which we will retain at the end
of our calculations.  Here conservation of electric charge implies
$\sum_i\eta_i m_i \nu_i=0$. In case there are Coulomb interactions
between these fractional modes we use a transformation
$\phi_i=\Lambda_{i\alpha}\phi_{\alpha}$ to diagonalize the quadratic
part of the action. In terms of the diagonal basis $\phi_{\alpha}$
(which are indexed by Greek letters), the electron/quasi-particle
tunneling operator is
$\expon{i\sum_{\alpha} \lambda_{\alpha}\phi_{\alpha}
}=\prod_{\alpha}X_{\alpha}$ with
$X_{\alpha} \equiv \expon{i\lambda_{\alpha}\phi_{\alpha}}$ and
$\lambda_{\alpha}=\sum_i m_i\Lambda_{i \alpha}$.

From the Heisenberg
equation of motion for $\phi_{\alpha}$, evolved with $H_0$,
\begin{align}
  \partial_t\phi_{\alpha} &= -\eta_{\alpha} v_{\alpha} \partial_x
                            \phi_{\alpha}+\frac{\eta_{\alpha}}{k_{\alpha}}\mu_{\alpha} \\
  \rightarrow X_{\alpha}(x,t)_{H_0} &= \expon{i\eta_{\alpha}
                                      \lambda_{\alpha}\mu_{\alpha}t/k_{\alpha}} X_{\alpha}(x,t)_{H_{F}}.
\end{align}
Also,
\begin{align}
  [X_{\alpha}(x'),n_{\alpha}(x)] = \frac{\eta_{\alpha}\lambda_{\alpha}}{k_{\alpha}} X_{\alpha}(x)\delta(x-x').
\end{align}
So (from now all the time dependencies are with respect to $H_{F}$)
\begin{align}
  \partial_x \expval{ I_{\alpha}(x,t)}
  &= \frac{1}{2}\frac{\eta_{\alpha}\lambda_{\alpha}}{k_{\alpha}} W_V \sum_{s
    s'}s' \int \di t' \prod_{\beta}
    \expon{i\eta_{\beta}\lambda_{\beta}\mu_{\beta} (t-t')/k_{\beta}} \expval{ T_C
    X_{\beta}(x,t,s) X_{\beta}^{\dagger}(x,t',s')} -\text{h.c.}\\
  \nonumber &= i\frac{\eta_{\alpha}\lambda_{\alpha}}{k_{\alpha}} W_V \sum_{s
              s'}s' \int \di t' \sin\left(\Omega (t-t')\right) \prod_{\beta}
              \expval{ T_C
              X_{\beta}(t,s) X_{\beta}^{\dagger}(t',s')}
\end{align}
where $\mu_{\beta}=0$ if $\beta$ is a Majorana mode and we defined
$\Omega\equiv \sum_{\beta}
\frac{\eta_{\beta}\lambda_{\beta}}{k_{\beta}}\mu_{\beta}$.
The Keldysh Green
function of a chiral operator $X_{\alpha}$ is
\begin{align}
  \expval{T_C X_{\alpha}(x,t,s) X_{\alpha}^{\dagger}(0,t',s') } = \left( \frac{
  A_{\alpha} T_{\alpha} }{v_{\alpha} \sin \frac{\pi T_{\alpha}}{v_{\alpha}}(a + i\chi_{ss'}(t-t') (v_{\alpha}(t-t')
  - \eta_{\alpha}x  )  } \right)^{2d_{\alpha}}
\end{align}
where $v_{\alpha}$,$T_{\alpha}$ and $d_{\alpha}$ are the velocity,
temperature, and scaling dimension of operator $X_{\alpha}$.
$A_{\alpha}$ is a constant ($A_{\alpha}=2$ for Majorana fermions,
$A_{\alpha}=\pi a$ for a vertex operator, and
$A_{\alpha}=\frac{\pi}{k_{\alpha}}$ for a boson density operator
$\partial_x\phi_{\alpha}$) and
\begin{align}
  \chi_{s s'}(t) &=
                   \begin{pmatrix}
                     \operatorname{sgn}(t) & -1 \\
                     1 & -\operatorname{sgn}(t)
                   \end{pmatrix}.
\end{align}
Substituting in the appropriate Green functions, assuming all modes are at the same temperature,
\begin{align}
  \partial_x \expval{ I_{\alpha}}
  \nonumber &= i\frac{\eta_{\alpha}\lambda_{\alpha}}{k_{\alpha}}. W_V.\prod_{\beta} (\frac{A_{\beta}
              }{v_{\beta}})^{2d_{\beta}}.T^{2\Delta} \sum_{s
              s'}s' \int \di t'  \frac{
              \sin\left( \Omega (t-t')\right)
              }{
              \prod_{\beta} \sin \left(\frac{\pi
              T_{\beta}}{v_{\beta}}(a + i\chi_{ss'} v_{\beta}(t-t') )
              \right) ^{2d_{\beta}}
              }\\
            &= i\frac{\eta_{\alpha}\lambda_{\alpha}}{k_{\alpha}}. W_V.\prod_{\beta} (\frac{A_{\beta}
              }{v_{\beta}})^{2d_{\beta}}.T^{2\Delta} \sum_{s
              }s \int \di t'  \frac{
              \sin\left( \Omega t'\right)
              }{
              \prod_{\beta} \sin \left(\frac{\pi
              T}{v_{\beta}}(a +is v_{\beta}t' )
              \right) ^{2d_{\beta}}
              },
\end{align}
where in the last equality we dropped the odd terms when
$s=s'$. Changing variables to $t'=-s(t+i/2T)$ and dropping $a$'s
assuming $\forall \beta: aT/v_{\beta}< 1$
\begin{align}
  \partial_x \expval{ I_{\alpha}}
  \nonumber &= i\frac{\eta_{\alpha}\lambda_{\alpha}}{k_{\alpha}}. W_V.\prod_{\beta} (\frac{A_{\beta}
              }{v_{\beta}})^{2d_{\beta}}.T^{2\Delta} \sum_{s
              }s \int \di t  \frac{
              -s\left( \sin\left( \Omega
              t\right)cosh\left( \Omega
              /2T\right) +i \cos\left( \Omega
              t\right)\sinh\left( \Omega /2T
              \right) \right)
              }{
              \cosh \left( \pi T t
              \right) ^{2\Delta}
              } \\
  \nonumber &= 2\frac{\eta_{\alpha}\lambda_{\alpha}}{k_{\alpha}}. W_V.\prod_{\beta} (\frac{A_{\beta}
              }{v_{\beta}})^{2d_{\beta}}.T^{2\Delta}\sinh\left( \frac{\Omega}{2T}
              \right) \int \di t  \frac{
              \cosh\left( i\Omega
              t\right)
              }{
              \cosh \left( \pi T t
              \right) ^{2\Delta}
              }\\
            &= \frac{\eta_{\alpha}\lambda_{\alpha}}{\pi k_{\alpha}}. W_V.\prod_{\beta} (\frac{A_{\beta}
              }{v_{\beta}})^{2d_{\beta}}.2^{2\Delta}T^{2\Delta-1}\sinh\left( \frac{\Omega}{2T}
              \right) B(\Delta+i\frac{\Omega}{2\pi T},\Delta-i\frac{\Omega}{2\pi T}),
\end{align}
where $\Delta=\sum_{\beta} d_{\beta}$ is the scaling
dimension of $\prod_{\beta}X_{\beta}$.
So in the ohmic regime when $\Omega\ll T$ we have (we're also
retaining the $\frac{1}{(2\pi a)^{N_{e}}}$ factor)
\begin{align}\label{elect-kin-final-app}
  \partial_x\expval{ I_{\alpha}} &= \sigma_0 g_V \frac{\eta_{\alpha}\lambda_{\alpha}}{k_{\alpha}}\sum_{\beta}
                                   \frac{\eta_{\beta}\lambda_{\beta}}{k_{\beta}} \mu_{\beta}(x) , \qquad
                                   g_V= W_V. \frac{(2\pi a)^{2(\Delta-N_e)}}{\prod_{\gamma} v_{\gamma}^{2d_{\gamma}}}.\frac{\Gamma(\Delta)^{2}}{\Gamma(2\Delta)}T^{2\Delta-2}.
\end{align}
Assuming local equilibrium we have
$\expval{I_{\beta} }=\eta_{\beta}\sigma_0\mu_{\alpha}/k_{\beta}$.
We can write these set of equations in terms of the original
modes $I_i=\Lambda_{i\alpha}I_{\alpha}$ as
\begin{align}\label{elect-kin-final-app-2}
  \partial_x\expval{ I_{i}} &= - g_V\eta_i\nu_i m_i\sum_{j}
                              m_j \expval{I_j(x)}.
\end{align}

\subsubsection{Random density-density}\label{sec-rand-dd}

For concreteness let's look at the example of the disordered fixed point
of the $\nu=2$ quantum Hall edge state.
This is the theory that we
derived in Appendix \ref{D12-apndx} if we only focus on the $\phi_1$ and
$\phi_2$ modes and ignore the rest:
\begin{subequations}
  \begin{align}
    S &=S_1+S_2+S_{12}+
        S_{\text{tunneling},12}
    \\
    i=1,2: S_i &= -\frac{1}{4\pi} \int_{t,x}
                 \left[\partial_x\phi_i(\eta_i\partial_t\phi_i+v_i \partial_x\phi_i)
                 \right] \\
    S_{12} &= -\frac{v_{12}}{4\pi} \int_{t,x} \
             \partial_x \phi_1 \partial_x\phi_2 \\
    S_{\text{tunneling},12} &= - \int_{t,x} \left[ \xi_{12}(x)
                              \expon{i(\phi_1-\phi_2)}  +\text{h.c.} \right] .
  \end{align}
\end{subequations}
We first change the basis to the charge mode $\phi_{\rho}=\frac{1}{\sqrt{2}}
(\phi_1+\phi_{2})$ and the neutral mode $\phi_{\sigma}=\frac{1}{\sqrt{2}}
(\phi_1-\phi_{2})$ and then perform a gauge transformation $O(x)$ to
eliminate the random tunneling term.
Now, we can write down the Hamiltonian for the disordered
fixed point as
\begin{subequations}
  \begin{align}
    H_F &= H_{\rho}+H_{\sigma} \\
    H_{\rho} &= \frac{v_{\rho}}{4\pi}\int \di x \
               (\partial_x\phi_{\rho})^2 \\
    H_{\sigma} &= \int \di x \  \frac{2\pi
                 v_{\sigma}}{3}\bm{\tilde{J}}_{\sigma}^2
  \end{align}
\end{subequations}
where current operators $\tilde{J}^a$ are defined as in
\eqref{current-defs}.  The residual density-density interaction
between the charge mode $\phi_{\rho}$ and the new neutral mode $\tilde
\phi_{\sigma}$ is
\begin{align}\label{eq-nu2-dd}
  V = H_{\rho \sigma} = \frac{1}{2\pi} \int \di x
  \partial_x\phi_{\rho} (\bm{\xi}_{\sigma}. \bm{\tilde{J}}(x)), \qquad
  \xi_{\sigma}^a\equiv 2\pi\sqrt{2} v_{\rho \sigma} O^{za}(x).
\end{align}
$\xi_{\sigma}^a$ is a quenched random variable, the auto-correlation
of which decays on the length scales of $\sim
v_{\sigma}^2/W_{12}$. This renders $V$ irrelevant. Assuming
$ v_{\sigma}^2/W_{12}$ is small enough, for simplicity we take
$\xi_{\sigma}$ to have Gaussian correlation
$\overline{ \xi_{\sigma}^a(x) \xi_{\sigma}^b(x')
}=\delta^{ab}W_{\sigma}\delta(x-x')$ where
$W_{\sigma}\approx 8\pi^2 v_{\rho \sigma}^2 v_{\sigma}^2/W_{12}$.

In order to find the tunneling current between the charge and neutral
modes we bias the modes with chemical potential by introducing the
interaction
\begin{align}
  H_{\mu}=-\int \di x \left[
  \mu_{\rho}n_{\rho}(x)+ \tilde{\mu}_{\sigma} \tilde{n}_{\sigma}(x) \right]
\end{align}
with the charge density $n_{\rho}=\frac{1}{2\pi}\partial_x \phi_{\rho}$
and the new neutral denstiy
$\tilde{n}_{\sigma}=\frac{1}{2\pi}\partial_x
\tilde{\phi}_{\sigma}=\sqrt{2}\tilde{J}^z$. The charge mode is conserved
\begin{align}
  -\partial_x I_{\rho} =
  \partial_tn_{\rho} = -\eta_{\rho} v_{\rho} \partial_x n_{\rho}
\end{align}
while for the neutral mode we have
\begin{align}
  \nonumber -\partial_x \tilde{I}_{\sigma}(x,t)
  &=\partial_t\tilde{n}_{\sigma}(x,t) =
    -\eta_{\sigma}v_{\sigma}\partial_x\tilde{n}_\sigma + i[H_{\rho
    \sigma},\sqrt{2}\tilde{J}^z(x,t)] \\ &=
                                           -\eta_{\sigma}v_{\sigma}\partial_x\tilde{n}_\sigma +\sqrt{2}n_{\rho}(x,t)
                                           \left( \xi_{\sigma}^x(x) \tilde{J}^y(x,t) - \xi_{\sigma}^y(x)
                                           \tilde{J}^x(x,t) \right).
\end{align}
Similarly as before, we assume the modes are in local equilibrium so
we have
$n_{\rho}(x)=\frac{1}{2\pi v_{\rho}}\mu_{\rho} $
and
$\tilde{n}_{\sigma}(x)=\frac{1}{2\pi v_{\sigma}}\tilde{\mu}_{\sigma}$.
Note that since the density $\tilde{n}_{\sigma}(x)$ decays only due to
the interaction term \ref{eq-nu2-dd}, we expect this density
and its conjugate chemical potential $\tilde
\mu_{\sigma}$  to vary slowly at
low temperatures. Therefore,
we drop the terms $\partial_x n_{\rho}$ and
$\partial_x \tilde{n}_{\sigma}$ in the above equations. Therefore we
drop To leading order in $W_{\sigma}$, the expectation value of this
operator is
\begin{align}
  \partial_x \expval{ \tilde{I}_{\sigma}(x,t)}
  &= -\frac{i}{\sqrt{2}}W_{\sigma}\sum_{s,s'} \int \di t'
    \expval{n_{\rho}(x,t,s)_{H_0} n_{\rho}(x,t',s')_{H_0} }\\
  \nonumber & \hspace{7em} \times \Big[
              \expval{\tilde{J}^y(x,t,s)_{H_0}\tilde{J}^x(x,t',s')_{H_0} }  -
              \expval{\tilde{J}^x(x,t,s)_{H_0}\tilde{J}^y(x,t',s')_{H_0} } \Big].
\end{align}
The equation of motion for $\tilde{J}^a(x)$, evolved with $H_0$, is
\begin{subequations}
  \begin{align}
    \partial_t\tilde{J}^x(x,t)&=-\eta_{\sigma}v_{\sigma}\partial_x\tilde{J}^x(x,t)
                                - \sqrt{2}\tilde{\mu}_{\sigma}\tilde{J}^y(x,t)\\
    \partial_t\tilde{J}^y(x,t)&=-\eta_{\sigma}v_{\sigma}\partial_x\tilde{J}^y(x,t)
                                + \sqrt{2}\tilde{\mu}_{\sigma}\tilde{J}^x(x,t)
  \end{align}
\end{subequations}
with solutions
\begin{subequations}
  \begin{align}
    \tilde{J}^x(x,t) &=
                       \tilde{J}^x(x-\eta_{\sigma}v_{\sigma}t)\cos(\sqrt{2}\tilde{\mu}_{\sigma}t) -
                       \tilde{J}^y(x-\eta_{\sigma}v_{\sigma}t)\sin(\sqrt{2}\tilde{\mu}_{\sigma}t) \\
    \tilde{J}^y(x,t) &=
                       \tilde{J}^y(x-\eta_{\sigma}v_{\sigma}t)\cos(\sqrt{2}\tilde{\mu}_{\sigma}t) +
                       \tilde{J}^x(x-\eta_{\sigma}v_{\sigma}t)\sin(\sqrt{2}\tilde{\mu}_{\sigma}t).
  \end{align}
\end{subequations}
Using this solution we have
\begin{align}
  \partial_x \expval{\tilde{I}_{\sigma}(x,t)}
  &= -\frac{i}{\sqrt{2}}W_{\sigma}\sum_{s,s'} \int \di t'
    \sin(\sqrt{2}\tilde{\mu}_{\sigma}(t-t')) \expval{n_{\rho}(x,t,s)_{H_{F}}
    n_{\rho}(x,t',s')_{H_{F}}  } \\
  \nonumber & \hspace{4em}\bigg[ \expval{\tilde{J}^x(x,t,s)_{H_{F}}\tilde{J}^x(x,t',s')_{H_{F}}
              }
              +\expval{\tilde{J}^y(x,t,s)_{H_{F}}\tilde{J}^y(x,t',s')_{H_{F}}
              }  \bigg] .
\end{align}
We proceed similarly as before to find
\begin{align}
                                           \partial_x \expval{ \tilde{I}_{\sigma}}
  &= -\frac{1}{2\pi}\frac{\sqrt{2} W_{\sigma}}{ v_{\rho}^{2}v_{\sigma}^{2}}  T^{2\Delta-1}\sinh\left( \frac{\sqrt{2}\tilde{\mu}_{\sigma}}{2T}
    \right) B(\Delta+i\frac{\sqrt{2}\tilde{\mu}_{\sigma}}{2\pi T},\Delta-i\frac{\sqrt{2}\tilde{\mu}_{\sigma}}{2\pi T}),
\end{align}
with $\Delta=2$.
To linear order in $\tilde{\mu}_{\sigma}$
\begin{align}\label{elect-cond-final}
  \partial_x \expval{ \tilde{I}_{\sigma}}
  &= - g_{\sigma}\sigma_0 \tilde{\mu}_{\sigma}
    , \qquad
    g_\sigma=
    \frac{ W_\sigma}{12v_{\rho}^{2}v_{\sigma}^{2}}T^{2}=\frac{2\pi^2
    v_{\rho \sigma}^{2}}{3v_{\rho}^2 W_{12}}T^{2}.
\end{align}
We can express this equation along with $\partial_x I_{\rho}=0$ in a
basis similar to the original fractional modes. We define
\begin{subequations}
  \begin{align}
    I'_1 &= \frac{1}{\sqrt{2}}(I_{\rho}+ \tilde I_{\sigma})
    \\
    I'_2 &= \frac{1}{\sqrt{2}}(I_{\rho}- \tilde I_{\sigma}).
  \end{align}
\end{subequations}
These new modes mix only due the irrelevant interactions such as
\ref{eq-nu2-dd} and so are expected to vary slowly at low temperatures.
In this basis the kinetic equations are
\begin{align}
  \partial_x\expval{ I'_{i}} &= -\sigma_0 g_\sigma\eta_i m_i\sum_{j}
                               m_j I'_j(x)
\end{align}
with $m_1=1$ and $m_2=-1$.  While this expression looks similar to
\ref{elect-kin-final-app-2}, the conductivity coefficient is
different and reflects the disordered fixed point.

\subsection{Thermal conductivity coefficient}\label{therm-cond-apndx}
Similarly, we can find the heat currents exchanged between the edge
modes. Here, we work to linear order in the temperature bias and
assume zero chemical potential bias.  From the Heisenberg
equation of motion with total Hamiltonian
$H=\sum_{\alpha}H_{\alpha}+V$:
\begin{align}
  -\partial_x J_{\alpha}(t)&= \partial_t\hamil_{\alpha}(x,t)=
                             i[H,\hamil_{\alpha}(x)](t)\\
                           &= -\eta_{\alpha}v_{\alpha}\partial_x\hamil_{\alpha}(x,t)+ i\int \di x' \ \xi(x') [X_{\alpha}(x'),\hamil_{\alpha}(x)]\prod_{\beta \neq \alpha}X_{\alpha}(x')+\text{h.c.},
\end{align}
where $\hamil_{\alpha}$ is the energy density of mode $\alpha$.
This equation should be understood as change in heat current due
to a series of incoherent tunnelings. Local equilibrium implies
$\expval{\hamil_{\alpha}}=\frac{1}{2v_{\alpha}}\kappa_0T_{\alpha}^2$
so to leading order we can drop the first term on the right hand side.
We will find the expectation value of $\partial_xJ_{\alpha}$ using the
Keldysh technique ($H_F=\sum_{\alpha}H_{\alpha}$),
\begin{align}
  \expval{\partial_x J_{\alpha}(x,t)} &= \frac{1}{2}\sum_{s}\expval{T_C \
                                        \partial_xJ_{\alpha}(x,t,s)_{H_{F}}\expon{i \sum_{s'}s' \int \di t'
                                        V(t',s')_{H_{F}}}}.
\end{align}
Expanding the slow evolution operator to first order
\begin{align}
  \partial_x\expval{ J_{\alpha}(x,t)}
  &= \frac{-i}{2}\sum_{s s'}s' \int \di t' \expval{T_C
    \partial_x J_{\alpha}(x,t,s)_{H_{F}}  V(t',s')_{H_{F}} } \\
  &= \frac{1}{2} W_V \sum_{s s'}s' \int \di x' \int \di t' \expval{T_C
    [X_{\alpha}(x'),\hamil_{\alpha}(x)](t,s) X_{\alpha}^{\dagger}(x',t',s')} \\
  \nonumber & \hspace{12em}  \times \prod_{\beta \neq \alpha}
              \expval{ T_C  X_{\beta}(x',t,s)
              X_{\beta}^{\dagger}(x',t',s')} +\text{h.c}.
\end{align}
where we dropped the $H_{F}$ index after the second equality and also
took the disorder average.  We assume the modes are in local equilibrium
so that the temperatures $T_{\alpha}$ are actually local temperatures
at point $x$.

Using
$[X_{\alpha}(x'),\hamil_{\alpha}(x)](t)_{H_{F}}=i\delta(x-x')\partial_tX_{\alpha}(x,t)_{H_{F}}$
we get
\begin{align}
  \partial_x \expval{ J_{\alpha}}  &= \frac{1}{2} W_V (2\pi d_{\alpha} T_{\alpha})\prod_{\beta} (\frac{A_{\beta}
                                     T_{\beta}}{v_{\beta}})^{2d_{\beta}} \sum_{s s'}s'
                                     \int \di t'  \chi_{ss'}(t-t') \frac{\cot \frac{\pi T_{\alpha}}{v_{\alpha}}(a +
                                     i\chi_{ss'} v_{\alpha}(t-t'))   }{ \prod_{\beta} \sin \left(\frac{\pi
                                     T_{\beta}}{v_{\beta}}(a + i\chi_{ss'} v_{\beta}(t-t') )
                                     \right) ^{2d_{\beta}} }
                                     +\text{h.c}.
\end{align}
$\chi_{ss}(t)$ is an odd function of $t$ so $t \chi_{ss}(t)$ is even
and so the integral vanishes for $s=s'$. Therefore, ($\chi_{s,-s}=-s$)
\begin{align}
  \partial_x \expval{ J_{\alpha}}  &= W_V.\pi d_{\alpha}T_{\alpha}.\prod_{\beta} (\frac{A_{\beta}
                                     T_{\beta}}{v_{\beta}})^{2d_{\beta}}. \sum_{s}
                                     \int \di t' \frac{\cot \frac{\pi T_{\alpha}}{v_{\alpha}}(a +
                                     is v_{\alpha}t' )  }{ \prod_{\beta} \sin \left(\frac{\pi
                                     T_{\beta}}{v_{\beta}}(a + is v_{\beta}t' )
                                     \right) ^{2d_{\alpha}}}
                                     +\text{h.c.}
\end{align}
Ignoring $a$'s (assuming $aT_{\beta}/v_{\beta}<1$) and changing variables to $t'=-s(t+i\frac{1}{2T_{\alpha}})$,
\begin{align}
  \partial_x \expval{ J_{\alpha}}  &= W_V.\pi d_{\alpha}T_{\alpha}.\prod_{\beta} (\frac{A_{\beta}
                                     T_{\beta}}{v_{\beta}})^{2d_{\beta}}. \sum_{s} \int \di t \frac{ i
                                     \sinh (\pi
                                     T_{\alpha} t) }{\cosh(s\pi T_{\alpha} t)^{2d_{\alpha}+1} \prod_{\beta \neq \alpha} \sin
                                     \left(\frac{\pi T_{\beta}}{2T_{\alpha}} - i\pi  T_{\alpha}t ) \right)
                                     ^{2d_{\beta}} } +\text{h.c.}
\end{align}
Expanding the integrand to first order in $\tau_{\beta \alpha}\equiv T_{\beta}-T_{\alpha}$
\begin{align}
  \nonumber \partial_x \expval{ J_{\alpha}} &=  W_V.\pi d_{\alpha}.\prod_{\beta} (\frac{A_{\beta}
                                              T_{\beta}}{v_{\beta}})^{2d_{\beta}}. \sum_{s} \int \di t \frac{i\sinh (\pi
                                              T_{\alpha} t) }{\cosh(\pi T_{\alpha} t)^{\sum_{\beta} 2d_{\beta}+1} }. \\
                                            & \qquad \qquad \qquad  \left(1-i\tanh(\pi
                                              T_{\alpha} t).(\frac{\pi}{2T_{\alpha}}
                                              -i\pi T_{\alpha} t)\sum_{\beta \neq \alpha}2d_{\beta}\tau_{\beta \alpha} \right)
                                              +\text{h.c.}
\end{align}
Dropping the odd terms in the integrand
\begin{align}
  \nonumber \partial_x \expval{ J_{\alpha}}  &=  2\pi d_{\alpha} W_V.\prod_{\beta} (\frac{A_{\beta}
                                               }{v_{\beta}})^{2d_{\beta}}.T^{2\Delta-2}.\frac{\pi}{2T}\sum_{j\neq
                                               i}2d_{\beta}\tau_{\beta \alpha}. \sum_{s} \int \di t \frac{\sinh (\pi
                                               T_{\alpha} t)^2 }{\cosh(s\pi T_{\alpha} t)^{\sum_{\beta} 2d_{\beta}+2} } \\
  \label{therm-kin-final-app} &= \kappa_0\sum_{\beta \neq \alpha}
                                g^Q_{\alpha \beta}\frac{T_{\beta}^{2}-T_{\alpha}^{2}}{2} ,\qquad
                                g^Q_{\alpha \beta}\equiv g_V \frac{12d_{\alpha}d_{\beta}}{1+2\Delta}
\end{align}
with $g_V$ defined in \eqref{elect-kin-final-app}.

\section{Domain of validity of descriptions at weak/strong disorder }\label{apndx-FP-validity}

\subsection*{Weak disorder}
In section \ref{antipf-transport-section}, we observed that the
$\Delta_{12}=1$ fixed point description of the anti-Pfaffian state is
in agreement with experiments only if $\Delta_{34}\approx 3/2$. Since
for $\Delta_{34}<3/2$ the system flows to the $\Delta_{34}=1$ fixed
point \cite{KFP} we might wonder if treating the $W_{34}$ tunneling
term perturbatively is a good description of the anti-Pfaffian
edge. To answer this question we first look at the RG equation for
$W_{34}$. To leading order we have
\begin{align}
  \deriv{W_{34}}{l} = (3-2\Delta_{34})W_{34}.
\end{align}
So, the effective strength of this tunneling term at temperature $T$
is
\begin{align}
  W_{34,\text{eff.}}(T) =W_{34} (\frac{T}{T_0})^{2\Delta_{34}-3}
\end{align}
where $T_0$ is the cutoff temperature, and is related to the
short-distance cutoff $a$ as
\begin{align}
  T_0 = \frac{ v_{\sigma}}{2\pi a}.
\end{align}
Here $v_{\sigma}$ is the typical velocity of the neutral modes. The
reason that we chose the neutral velocity in defining $T_0$ is that
for strong short-ranged Coulomb interactions, tunneling terms only
couple the (``almost'') neutral modes. This can
be seen from the expressions for the conductivity coefficients such as
$g_{V_{34}}$ is Eq. \ref{D12-gs}. We
can write $g_{V_{34}}$ as
\begin{align}\label{eq-g34-rewrt}
  g_{V_{34}} &=
               \frac{\Gamma(\Delta_{34})^{2}}{\Gamma(2\Delta_{34})}
               \frac{W_{34}}{\overline{v}_{V_{34}}^{2}}\left( \frac{T}{T_{0}}  \right)^{2\Delta_{34}-2}
\end{align}
with the above definition of $T_0$ with
$v_{\sigma}=\overline{v}_{V_{34}}\approx v^{(0)}$.

When $\Delta_{34}>3/2$ but is close to $3/2$ and for finite
temperatures, the $W_{34}$ tunneling term might still be strong. A
rough estimate for the range of validity of perturbation in $W_{34}$
can be obtained if we require the length scale associated with the
effective tunneling strength $W_{34,\text{eff}}(T)$ to be larger than the
short-distance cutoff $a$. The length scale
associated with $W_{34,\text{eff.}}$ is
$\ell_{W_{34}}(T)=v_{\sigma}^2/W_{34,\text{eff}}(T)$. So the condition
for the validity of perturbation theory is
\begin{align}
  a \ll \ell_{W_{34}}(T) = \frac{v_{\sigma}^2}{W_{34}}\left( \frac{
  v_{\sigma}}{2\pi a T} \right)^{2\Delta_{34}-3}.
\end{align}
Along with Eq. \ref{eq-g34-rewrt} we can write this condition as
(ignoring numerical factors)
\begin{align}
  g_{V_{34}}^{-1} = \ell_{\text{eq.},V_{34}} \gg L_T.
\end{align}
where $\ell_{\text{eq.},V_{34}}$ is the charge equilibration length
between the modes $\phi_3$ and $\phi_4$ (See Eq. \ref{eq-D12-Gsol}), and $L_T= v_{\sigma}/2\pi T$ is the thermal length.  The last
inequality illustrates a more practical check for the domain of
validity of the incoherent regime.

\subsection*{Strong disorder}
Another question is whether $S_{\Delta_{12}=1}$ is a good description
of modes $\phi_1$ and $\phi_{2}$ at low temperatures, when the
tunnelings between these two modes are weak. The tunnelings between
the $\phi_1$ and $\phi_2$ modes require spin-flipping, and so they are
expected to be weaker than the corresponding spin-conserving
tunnelings. Therefore even for $\Delta_{12}<3/2$ and at finite
temperatures, the tunneling term might not drive the system all the
way to the $\Delta_{12}=1$ fixed point. In order to address such
concerns we first start from the RG equation for $W_{12}$ near the
clean fixed point $W_{12}=1$ (this section follows a similar
estimations as \cite{mirlin}). Solving the RG equation, the effective
tunneling strength at length scale $L$ is
\begin{align}
  W_{12,\text{eff.}}(L) =W_{12} (\frac{L}{a})^{3-2\Delta_{12}}
\end{align}
For weak
$W_{12}$ and small enough lengths $L$ (high enough temperatures) such
that ($\ell_{W_{12},\text{eff.}}(L)=v_{\sigma}^2/W_{12,\text{eff.}}(L)$)
\begin{equation}
\ell_{ W_{12},\text{eff.} }(L) \gg a
\end{equation}
 we
can still treat this tunneling term in perturbation theory. However,
for larger length scales the two modes $\phi_1$ and $\phi_2$ are
strongly mixed and the clean fixed point description is no longer
valid. We can obtain an estimate for the length scale $L_{\text{mix}}$
where such a transition happens by solving
\begin{align}
  \ell_{ W_{12},\text{eff.} }(L_{\text{mix}})=a.
\end{align}
When the velocity of the two modes $\phi_1$ and $\phi_2$ are close to
each other, the mode $\phi_{\sigma_{12}}$ decouples from other modes
(See section \ref{D12-section}) and we have $\Delta_{12}\approx 1$. So we find
\begin{align}
  L_{\text{mix}} \approx \frac{v_{\sigma}^2}{W_{12}}.
\end{align}
This length also serves as the short-distance cutoff for the
$\tilde \phi_{\sigma_{12}}$ mode (See Appendix \ref{D12-apndx}).  For
length scales larger than $L_{\text{mix}}$, i.e. $L_T>L_{\text{mix}}$, we follow the same line of
arguments as before, in order to estimate the the domain of validity of
perturbation theory in the disordered density-density interactions
$S_{\sigma_{12},i}$ in Eq. \ref{D12-finaction-irrel}. We find
\begin{align}
  \beta=r,s_2,s_3:  \ell_{\text{eq.},V_{\sigma_{12},\beta}}\equiv
  g^{-1}_{V_{\sigma_{12},\beta}} & \gg L_T = \frac{v_{\sigma}}{2\pi T}
\end{align}
or
\begin{align}
  W_{12} \gg \frac{v_{\sigma_{12},\beta}^2 v_{\sigma}}{v_{\beta}^2}T.
\end{align}
As we demonstrated in section \ref{D12-section},
$v_{\sigma_{12},\beta}$ goes to zero as the Zeeman gap
vanishes. Therefore, we expect this inequality to be more valid as we
approach the regimes where we expect the thermal conductance
$K=2.5\kappa_0T$.

\bibliographystyle{utphys}
\bibliography{refs}{}

\providecommand{\href}[2]{#2}\begingroup\raggedright\begin{thebibliography}{10}

\bibitem{willet-1987}
R.~Willett, J.~P. Eisenstein, H.~L. St\"ormer, D.~C. Tsui, A.~C. Gossard, and
  J.~H. English, ``Observation of an even-denominator quantum number in the
  fractional quantum hall effect,''
  \href{http://dx.doi.org/10.1103/PhysRevLett.59.1776}{{\em Phys. Rev. Lett.}
  {\bfseries 59} (Oct, 1987) 1776--1779}.
  \url{https://link.aps.org/doi/10.1103/PhysRevLett.59.1776}.

\bibitem{moore-1991}
G.~Moore and N.~Read, ``Nonabelions in the fractional quantum hall effect,''
  \href{http://dx.doi.org/https://doi.org/10.1016/0550-3213(91)90407-O}{{\em
  Nuclear Physics B} {\bfseries 360} no.~2, (1991) 362 -- 396}.
  \url{http://www.sciencedirect.com/science/article/pii/055032139190407O}.

\bibitem{read-2000}
N.~Read and D.~Green, ``Paired states of fermions in two dimensions with
  breaking of parity and time-reversal symmetries and the fractional quantum
  hall effect,'' \href{http://dx.doi.org/10.1103/PhysRevB.61.10267}{{\em Phys.
  Rev. B} {\bfseries 61} (Apr, 2000) 10267--10297}.
  \url{https://link.aps.org/doi/10.1103/PhysRevB.61.10267}.

\bibitem{levin-2007}
M.~Levin, B.~I. Halperin, and B.~Rosenow, ``Particle-hole symmetry and the
  pfaffian state,'' \href{http://dx.doi.org/10.1103/PhysRevLett.99.236806}{{\em
  Phys. Rev. Lett.} {\bfseries 99} (Dec, 2007) 236806}.
  \url{https://link.aps.org/doi/10.1103/PhysRevLett.99.236806}.

\bibitem{lee-2007}
S.-S. Lee, S.~Ryu, C.~Nayak, and M.~P.~A. Fisher, ``Particle-hole symmetry and
  the $\ensuremath{\nu}=\frac{5}{2}$ quantum hall state,''
  \href{http://dx.doi.org/10.1103/PhysRevLett.99.236807}{{\em Phys. Rev. Lett.}
  {\bfseries 99} (Dec, 2007) 236807}.
  \url{https://link.aps.org/doi/10.1103/PhysRevLett.99.236807}.

\bibitem{haldane-num-1988}
F.~D.~M. Haldane and E.~H. Rezayi, ``Spin-singlet wave function for the
  half-integral quantum hall effect,''
  \href{http://dx.doi.org/10.1103/PhysRevLett.60.956}{{\em Phys. Rev. Lett.}
  {\bfseries 60} (Mar, 1988) 956--959}.
  \url{https://link.aps.org/doi/10.1103/PhysRevLett.60.956}.

\bibitem{macdonald-num-1989}
A.~H. MacDonald, D.~Yoshioka, and S.~M. Girvin, ``Comparison of models for the
  even-denominator fractional quantum hall effect,''
  \href{http://dx.doi.org/10.1103/PhysRevB.39.8044}{{\em Phys. Rev. B}
  {\bfseries 39} (Apr, 1989) 8044--8047}.
  \url{https://link.aps.org/doi/10.1103/PhysRevB.39.8044}.

\bibitem{pakrouski-num-2015}
K.~Pakrouski, M.~R. Peterson, T.~Jolicoeur, V.~W. Scarola, C.~Nayak, and
  M.~Troyer, ``Phase diagram of the $\ensuremath{\nu}=5/2$ fractional quantum
  hall effect: Effects of landau-level mixing and nonzero width,''
  \href{http://dx.doi.org/10.1103/PhysRevX.5.021004}{{\em Phys. Rev. X}
  {\bfseries 5} (Apr, 2015) 021004}.
  \url{https://link.aps.org/doi/10.1103/PhysRevX.5.021004}.

\bibitem{rezayi-num-2017}
E.~H. Rezayi, ``Landau level mixing and the ground state of the
  $\ensuremath{\nu}=5/2$ quantum hall effect,''
  \href{http://dx.doi.org/10.1103/PhysRevLett.119.026801}{{\em Phys. Rev.
  Lett.} {\bfseries 119} (Jul, 2017) 026801}.
  \url{https://link.aps.org/doi/10.1103/PhysRevLett.119.026801}.

\bibitem{rezayi-2011}
E.~H. Rezayi and S.~H. Simon, ``Breaking of particle-hole symmetry by landau
  level mixing in the $\ensuremath{\nu}=5/2$ quantized hall state,''
  \href{http://dx.doi.org/10.1103/PhysRevLett.106.116801}{{\em Phys. Rev.
  Lett.} {\bfseries 106} (Mar, 2011) 116801}.
  \url{https://link.aps.org/doi/10.1103/PhysRevLett.106.116801}.

\bibitem{lin-tun-2012}
X.~Lin, C.~Dillard, M.~A. Kastner, L.~N. Pfeiffer, and K.~W. West,
  ``Measurements of quasiparticle tunneling in the
  $\ensuremath{\upsilon}=\frac{5}{2}$ fractional quantum hall state,''
  \href{http://dx.doi.org/10.1103/PhysRevB.85.165321}{{\em Phys. Rev. B}
  {\bfseries 85} (Apr, 2012) 165321}.
  \url{https://link.aps.org/doi/10.1103/PhysRevB.85.165321}.

\bibitem{radu-2008}
I.~P. Radu, J.~B. Miller, C.~M. Marcus, M.~A. Kastner, L.~N. Pfeiffer, and
  K.~W. West, ``Quasi-particle properties from tunneling in the v = 5/2
  fractional quantum hall state,''
  \href{http://dx.doi.org/10.1126/science.1157560}{{\em Science} {\bfseries
  320} no.~5878, (2008) 899--902},
  \href{http://arxiv.org/abs/https://science.sciencemag.org/content/320/5878/899.full.pdf}{{\ttfamily
  https://science.sciencemag.org/content/320/5878/899.full.pdf}}.
  \url{https://science.sciencemag.org/content/320/5878/899}.

\bibitem{5halv-exp-review-2014}
X.~Lin, R.~Du, and X.~Xie, ``{Recent experimental progress of fractional
  quantum Hall effect: 5/2 filling state and graphene},''
  \href{http://dx.doi.org/10.1093/nsr/nwu071}{{\em National Science Review}
  {\bfseries 1} no.~4, (11, 2014) 564--579},
  \href{http://arxiv.org/abs/https://academic.oup.com/nsr/article-pdf/1/4/564/31568876/nwu071.pdf}{{\ttfamily
  https://academic.oup.com/nsr/article-pdf/1/4/564/31568876/nwu071.pdf}}.
  \url{https://doi.org/10.1093/nsr/nwu071}.

\bibitem{yang-feldman-review}
G.~Yang and D.~E. Feldman, ``Influence of device geometry on tunneling in the
  $\ensuremath{\nu}=\frac{5}{2}$ quantum hall liquid,''
  \href{http://dx.doi.org/10.1103/PhysRevB.88.085317}{{\em Phys. Rev. B}
  {\bfseries 88} (Aug, 2013) 085317}.
  \url{https://link.aps.org/doi/10.1103/PhysRevB.88.085317}.

\bibitem{PhysRevB.90.161306}
G.~Yang and D.~E. Feldman, ``Experimental constraints and a possible quantum
  hall state at $\ensuremath{\nu}=5/2$,''
  \href{http://dx.doi.org/10.1103/PhysRevB.90.161306}{{\em Phys. Rev. B}
  {\bfseries 90} (Oct, 2014) 161306}.
  \url{https://link.aps.org/doi/10.1103/PhysRevB.90.161306}.

\bibitem{bid-2010}
A.~Bid, N.~Ofek, H.~Inoue, M.~Heiblum, C.~L. Kane, V.~Umansky, and D.~Mahalu,
  ``Observation of neutral modes in the fractional quantum hall regime,''
  \href{http://dx.doi.org/10.1038/nature09277}{{\em Nature} {\bfseries 466}
  no.~7306, (2010) 585--590}. \url{https://doi.org/10.1038/nature09277}.

\bibitem{gross-2012}
Y.~Gross, M.~Dolev, M.~Heiblum, V.~Umansky, and D.~Mahalu, ``Upstream neutral
  modes in the fractional quantum hall effect regime: Heat waves or coherent
  dipoles,'' \href{http://dx.doi.org/10.1103/PhysRevLett.108.226801}{{\em Phys.
  Rev. Lett.} {\bfseries 108} (May, 2012) 226801}.
  \url{https://link.aps.org/doi/10.1103/PhysRevLett.108.226801}.

\bibitem{KF-Qthermal}
C.~L. Kane and M.~P.~A. Fisher, ``Quantized thermal transport in the fractional
  quantum hall effect,''
  \href{http://dx.doi.org/10.1103/PhysRevB.55.15832}{{\em Phys. Rev. B}
  {\bfseries 55} (Jun, 1997) 15832--15837}.
  \url{https://link.aps.org/doi/10.1103/PhysRevB.55.15832}.

\bibitem{2002NuPhB.636..568C}
A.~{Cappelli}, M.~{Huerta}, and G.~R. {Zemba}, ``{Thermal transport in chiral
  conformal theories and hierarchical quantum Hall states},''
  \href{http://dx.doi.org/10.1016/S0550-3213(02)00340-1}{{\em Nuclear Physics
  B} {\bfseries 636} no.~3, (Aug., 2002) 568--582},
  \href{http://arxiv.org/abs/cond-mat/0111437}{{\ttfamily
  arXiv:cond-mat/0111437 [cond-mat.mes-hall]}}.

\bibitem{banerj-half}
M.~Banerjee, M.~Heiblum, V.~Umansky, D.~E. Feldman, Y.~Oreg, and A.~Stern,
  ``Observation of half-integer thermal hall conductance,''
  \href{http://dx.doi.org/10.1038/s41586-018-0184-1}{{\em Nature} {\bfseries
  559} no.~7713, (2018) 205--210}.
  \url{https://doi.org/10.1038/s41586-018-0184-1}.

\bibitem{son-2015}
D.~T. Son, ``Is the composite fermion a dirac particle?,''
  \href{http://dx.doi.org/10.1103/PhysRevX.5.031027}{{\em Phys. Rev. X}
  {\bfseries 5} (Sep, 2015) 031027}.
  \url{https://link.aps.org/doi/10.1103/PhysRevX.5.031027}.

\bibitem{zucker-feldman-2016}
P.~T. Zucker and D.~E. Feldman, ``Stabilization of the particle-hole pfaffian
  order by landau-level mixing and impurities that break particle-hole
  symmetry,'' \href{http://dx.doi.org/10.1103/PhysRevLett.117.096802}{{\em
  Phys. Rev. Lett.} {\bfseries 117} (Aug, 2016) 096802}.
  \url{https://link.aps.org/doi/10.1103/PhysRevLett.117.096802}.

\bibitem{wang-2018}
C.~Wang, A.~Vishwanath, and B.~I. Halperin, ``Topological order from disorder
  and the quantized hall thermal metal: Possible applications to the
  $\ensuremath{\nu}=5/2$ state,''
  \href{http://dx.doi.org/10.1103/PhysRevB.98.045112}{{\em Phys. Rev. B}
  {\bfseries 98} (Jul, 2018) 045112}.
  \url{https://link.aps.org/doi/10.1103/PhysRevB.98.045112}.

\bibitem{mross-2018}
D.~F. Mross, Y.~Oreg, A.~Stern, G.~Margalit, and M.~Heiblum, ``Theory of
  disorder-induced half-integer thermal hall conductance,''
  \href{http://dx.doi.org/10.1103/PhysRevLett.121.026801}{{\em Phys. Rev.
  Lett.} {\bfseries 121} (Jul, 2018) 026801}.
  \url{https://link.aps.org/doi/10.1103/PhysRevLett.121.026801}.

\bibitem{PhysRevB.97.165124}
B.~Lian and J.~Wang, ``Theory of the disordered $\ensuremath{\nu}=\frac{5}{2}$
  quantum thermal hall state: Emergent symmetry and phase diagram,''
  \href{http://dx.doi.org/10.1103/PhysRevB.97.165124}{{\em Phys. Rev. B}
  {\bfseries 97} (Apr, 2018) 165124}.
  \url{https://link.aps.org/doi/10.1103/PhysRevB.97.165124}.

\bibitem{simon-interp}
S.~H. Simon, ``Interpretation of thermal conductance of the
  $\mathbf{\ensuremath{\nu}}=\mathbf{5}/\mathbf{2}$ edge,''
  \href{http://dx.doi.org/10.1103/PhysRevB.97.121406}{{\em Phys. Rev. B}
  {\bfseries 97} (Mar, 2018) 121406}.
  \url{https://link.aps.org/doi/10.1103/PhysRevB.97.121406}.

\bibitem{Banerjee:2017aa}
M.~Banerjee, M.~Heiblum, A.~Rosenblatt, Y.~Oreg, D.~E. Feldman, A.~Stern, and
  V.~Umansky, ``Observed quantization of anyonic heat flow,''
  \href{http://dx.doi.org/10.1038/nature22052}{{\em Nature} {\bfseries 545}
  no.~7652, (2017) 75--79}. \url{https://doi.org/10.1038/nature22052}.

\bibitem{KF-contacts}
C.~L. Kane and M.~P.~A. Fisher, ``Contacts and edge-state equilibration in the
  fractional quantum hall effect,''
  \href{http://dx.doi.org/10.1103/PhysRevB.52.17393}{{\em Phys. Rev. B}
  {\bfseries 52} (Dec, 1995) 17393--17405}.
  \url{https://link.aps.org/doi/10.1103/PhysRevB.52.17393}.

\bibitem{chamon-fradkin}
C.~de~C.~Chamon and E.~Fradkin, ``Distinct universal conductances in tunneling
  to quantum hall states: The role of contacts,''
  \href{http://dx.doi.org/10.1103/PhysRevB.56.2012}{{\em Phys. Rev. B}
  {\bfseries 56} (Jul, 1997) 2012--2025}.
  \url{https://link.aps.org/doi/10.1103/PhysRevB.56.2012}.

\bibitem{feldman-comment}
D.~E. Feldman, ``Comment on ``interpretation of thermal conductance of the
  $\ensuremath{\nu}=5/2$ edge'',''
  \href{http://dx.doi.org/10.1103/PhysRevB.98.167401}{{\em Phys. Rev. B}
  {\bfseries 98} (Oct, 2018) 167401}.
  \url{https://link.aps.org/doi/10.1103/PhysRevB.98.167401}.

\bibitem{ma-feldman-2019}
K.~K.~W. Ma and D.~E. Feldman, ``Partial equilibration of integer and
  fractional edge channels in the thermal quantum hall effect,''
  \href{http://dx.doi.org/10.1103/PhysRevB.99.085309}{{\em Phys. Rev. B}
  {\bfseries 99} (Feb, 2019) 085309}.
  \url{https://link.aps.org/doi/10.1103/PhysRevB.99.085309}.

\bibitem{simon-rosenow-2019}
S.~H. Simon and B.~Rosenow, ``Partial equilibration of the anti-pfaffian edge
  due to majorana disorder,'' 2019.

\bibitem{PhysRevLett.123.137701}
C.~Sp\aa{}nsl\"att, J.~Park, Y.~Gefen, and A.~D. Mirlin, ``Topological
  classification of shot noise on fractional quantum hall edges,''
  \href{http://dx.doi.org/10.1103/PhysRevLett.123.137701}{{\em Phys. Rev.
  Lett.} {\bfseries 123} (Sep, 2019) 137701}.
  \url{https://link.aps.org/doi/10.1103/PhysRevLett.123.137701}.

\bibitem{PhysRevB.99.161302}
J.~Park, A.~D. Mirlin, B.~Rosenow, and Y.~Gefen, ``Noise on complex quantum
  hall edges: Chiral anomaly and heat diffusion,''
  \href{http://dx.doi.org/10.1103/PhysRevB.99.161302}{{\em Phys. Rev. B}
  {\bfseries 99} (Apr, 2019) 161302}.
  \url{https://link.aps.org/doi/10.1103/PhysRevB.99.161302}.

\bibitem{PhysRevB.101.075308}
C.~Sp\aa{}nsl\"att, J.~Park, Y.~Gefen, and A.~D. Mirlin, ``Conductance plateaus
  and shot noise in fractional quantum hall point contacts,''
  \href{http://dx.doi.org/10.1103/PhysRevB.101.075308}{{\em Phys. Rev. B}
  {\bfseries 101} (Feb, 2020) 075308}.
  \url{https://link.aps.org/doi/10.1103/PhysRevB.101.075308}.

\bibitem{park2020noise}
J.~Park, C.~Sp\aa{}nsl\"att, Y.~Gefen, and A.~D. Mirlin, ``Noise on the
  non-abelian $\nu=5/2$ fractional quantum hall edge,'' 2020.
\newblock \url{https://arxiv.org/abs/2006.06018}.

\bibitem{KF-imp}
C.~L. Kane and M.~P.~A. Fisher, ``Impurity scattering and transport of
  fractional quantum hall edge states,''
  \href{http://dx.doi.org/10.1103/PhysRevB.51.13449}{{\em Phys. Rev. B}
  {\bfseries 51} (May, 1995) 13449--13466}.
  \url{https://link.aps.org/doi/10.1103/PhysRevB.51.13449}.

\bibitem{incoherent-23}
C.~Nosiglia, J.~Park, B.~Rosenow, and Y.~Gefen, ``Incoherent transport on the
  $\ensuremath{\nu}=2/3$ quantum hall edge,''
  \href{http://dx.doi.org/10.1103/PhysRevB.98.115408}{{\em Phys. Rev. B}
  {\bfseries 98} (Sep, 2018) 115408}.
  \url{https://link.aps.org/doi/10.1103/PhysRevB.98.115408}.

\bibitem{wen-1991-edge}
X.-G. Wen, ``Edge transport properties of the fractional quantum hall states
  and weak-impurity scattering of a one-dimensional charge-density wave,''
  \href{http://dx.doi.org/10.1103/PhysRevB.44.5708}{{\em Phys. Rev. B}
  {\bfseries 44} (Sep, 1991) 5708--5719}.
  \url{https://link.aps.org/doi/10.1103/PhysRevB.44.5708}.

\bibitem{KF-thermLL}
C.~L. Kane and M.~P.~A. Fisher, ``Thermal transport in a luttinger liquid,''
  \href{http://dx.doi.org/10.1103/PhysRevLett.76.3192}{{\em Phys. Rev. Lett.}
  {\bfseries 76} (Apr, 1996) 3192--3195}.
  \url{https://link.aps.org/doi/10.1103/PhysRevLett.76.3192}.

\bibitem{muller-1992}
G.~M\"uller, D.~Weiss, A.~V. Khaetskii, K.~von Klitzing, S.~Koch, H.~Nickel,
  W.~Schlapp, and R.~L\"osch, ``Equilibration length of electrons in
  spin-polarized edge channels,''
  \href{http://dx.doi.org/10.1103/PhysRevB.45.3932}{{\em Phys. Rev. B}
  {\bfseries 45} (Feb, 1992) 3932--3935}.
  \url{https://link.aps.org/doi/10.1103/PhysRevB.45.3932}.

\bibitem{wurtz-2002}
A.~W\"urtz, R.~Wildfeuer, A.~Lorke, E.~V. Deviatov, and V.~T. Dolgopolov,
  ``Separately contacted edge states: A spectroscopic tool for the
  investigation of the quantum hall effect,''
  \href{http://dx.doi.org/10.1103/PhysRevB.65.075303}{{\em Phys. Rev. B}
  {\bfseries 65} (Jan, 2002) 075303}.
  \url{https://link.aps.org/doi/10.1103/PhysRevB.65.075303}.

\bibitem{bocq-nature-2013}
E.~Bocquillon, V.~Freulon, J.-.~M. Berroir, P.~Degiovanni, B.~Pla{\c{c}}ais,
  A.~Cavanna, Y.~Jin, and G.~F{\`e}ve, ``Separation of neutral and charge modes
  in one-dimensional chiral edge channels,''
  \href{http://dx.doi.org/10.1038/ncomms2788}{{\em Nature Communications}
  {\bfseries 4} no.~1, (2013) 1839}. \url{https://doi.org/10.1038/ncomms2788}.

\bibitem{kuma-prb-2011}
N.~Kumada, H.~Kamata, and T.~Fujisawa, ``Edge magnetoplasmon transport in gated
  and ungated quantum hall systems,''
  \href{http://dx.doi.org/10.1103/PhysRevB.84.045314}{{\em Phys. Rev. B}
  {\bfseries 84} (Jul, 2011) 045314}.
  \url{https://link.aps.org/doi/10.1103/PhysRevB.84.045314}.

\bibitem{banerj-any}
M.~Banerjee, M.~Heiblum, A.~Rosenblatt, Y.~Oreg, D.~E. Feldman, A.~Stern, and
  V.~Umansky, ``Observed quantization of anyonic heat flow,'' {\em Nature}
  {\bfseries 545} (Apr, 2017) 75 EP --.
  \url{https://doi.org/10.1038/nature22052}.

\bibitem{devyatov-2007}
E.~V. Devyatov, ``Edge states in the regimes of integer and fractional quantum
  hall effects,'' \href{http://dx.doi.org/10.1070/pu2007v050n02abeh006244}{{\em
  Physics-Uspekhi} {\bfseries 50} no.~2, (Feb, 2007) 197--218}.
  \url{https://doi.org/10.1070%2Fpu2007v050n02abeh006244}.

\bibitem{KFP}
C.~L. Kane, M.~P.~A. Fisher, and J.~Polchinski, ``Randomness at the edge:
  Theory of quantum hall transport at filling \ensuremath{\nu}=2/3,''
  \href{http://dx.doi.org/10.1103/PhysRevLett.72.4129}{{\em Phys. Rev. Lett.}
  {\bfseries 72} (Jun, 1994) 4129--4132}.
  \url{https://link.aps.org/doi/10.1103/PhysRevLett.72.4129}.

\bibitem{mirlin}
I.~Protopopov, Y.~Gefen, and A.~Mirlin, ``Transport in a disordered $\nu=5/2$
  fractional quantum hall junction,''
  \href{http://dx.doi.org/https://doi.org/10.1016/j.aop.2017.07.015}{{\em
  Annals of Physics} {\bfseries 385} (2017) 287 -- 327}.
  \url{http://www.sciencedirect.com/science/article/pii/S0003491617302142}.

\bibitem{chamon-wen}
C.~d.~C. Chamon and X.~G. Wen, ``Sharp and smooth boundaries of quantum hall
  liquids,'' \href{http://dx.doi.org/10.1103/PhysRevB.49.8227}{{\em Phys. Rev.
  B} {\bfseries 49} (Mar, 1994) 8227--8241}.
  \url{https://link.aps.org/doi/10.1103/PhysRevB.49.8227}.

\bibitem{aleiner-1994}
I.~L. Aleiner and L.~I. Glazman, ``Novel edge excitations of two-dimensional
  electron liquid in a magnetic field,''
  \href{http://dx.doi.org/10.1103/PhysRevLett.72.2935}{{\em Phys. Rev. Lett.}
  {\bfseries 72} (May, 1994) 2935--2938}.
  \url{https://link.aps.org/doi/10.1103/PhysRevLett.72.2935}.

\bibitem{NAUD2000572}
J.~Naud, L.~P. Pryadko, and S.~Sondhi, ``Quantum hall bilayers and the chiral
  sine-gordon equation,''
  \href{http://dx.doi.org/https://doi.org/10.1016/S0550-3213(99)00658-6}{{\em
  Nuclear Physics B} {\bfseries 565} no.~3, (2000) 572 -- 610}.
  \url{http://www.sciencedirect.com/science/article/pii/S0550321399006586}.

\end{thebibliography}\endgroup

\end{document}